# Weak Flip Codes and their Optimality on the Binary Erasure Channel

Hsuan-Yin Lin, Stefan M. Moser, and Po-Ning Chen*

15 June 2017

**Abstract**

This paper investigates fundamental properties of *nonlinear* binary codes by looking at the codebook matrix not row-wise (codewords), but *column-wise*. The family of *weak flip codes* is presented and shown to contain many beautiful properties. In particular the subfamily *fair weak flip codes*, which goes back to Berlekamp, Gallager, and Shannon and which was shown to achieve the error exponent with a fixed number of codewords $\mathsf{M}$, can be seen as a generalization of linear codes to an arbitrary number of codewords. The fair weak flip codes are related to binary nonlinear Hadamard codes.

Based on the column-wise approach to the codebook matrix, the *$r$-wise Hamming distance* is introduced as a generalization to the well-known and widely used (pairwise) Hamming distance. It is shown that the minimum $r$-wise Hamming distance satisfies a *generalized $r$-wise Plotkin bound*. The $r$-wise Hamming distance structure of the nonlinear fair weak flip codes is analyzed and shown to be superior to many codes. In particular, it is proven that the fair weak flip codes achieve the $r$-wise Plotkin bound with equality for all $r$.

In the second part of the paper, these insights are applied to a *binary erasure channel (BEC)* with an arbitrary erasure probability $0 < \delta < 1$. An exact formula for the average error probability of an arbitrary (linear or nonlinear) code using maximum likelihood decoding is derived and shown to be expressible using only the $r$-wise Hamming distance structure of the code. For a number of codewords $\mathsf{M}$ satisfying $\mathsf{M} \leq 4$ and an arbitrary finite blocklength $n$, the globally optimal codes (in the sense of minimizing the average error probability) are found. For $\mathsf{M} = 5$ or $\mathsf{M} = 6$ and an arbitrary finite blocklength $n$, the optimal codes are conjectured. For larger $\mathsf{M}$, observations regarding the optimal design are presented, e.g., that good codes have a large $r$-wise Hamming distance structure for all $r$. Numerical results validate our code design criteria and show the superiority of our best found nonlinear weak flip codes compared to the best linear codes.

*S. Moser and P.-N. Chen are with the Department of Electrical and Computer Engineering at National Chiao Tung University (NCTU), Hsinchu, Taiwan. S. Moser is also with ETH Zurich, Switzerland. H.-Y. Lin was with NCTU and is now with the Simula@UiB, Bergen, Norway. This work has been supported by National Science Council under NSC 97–2221–E–009–003–MY3 and NSC 100–2221–E–009–068–MY3. This work was presented in parts at the *50th Annual Allerton Conference on Communication, Control, and Computing*, Allerton House, Monticello, IL, USA, Oct. 1–5, 2012 and the *2015 IEEE International Symposium on Information Theory (ISIT)*, Hong Kong, Jun. 14–19, 2015.

# 1 Introduction

A goal in traditional coding theory is to find good codes that operate close to the ultimate limit of the *channel capacity* as introduced by Shannon [1]. Implicitly, by the definition of capacity, such codes are expected to have a large blocklength. Moreover, due to the potential simplifications and because such codes behave well for large blocklength, conventional coding theory often restricts itself to *linear codes*. It is also quite common to use the *minimum Hamming distance* and the *weight enumerating function (WEF)* as a design and quality criterion [2]. This is motivated by the equivalence of Hamming weight and Hamming distance for linear codes, and by the union bound that converts the global error probability into pairwise error probabilities.

In this work we would like to break away from these traditional simplifications and instead focus on an optimal[1] design of codes for finite blocklength. Since for very short blocklength it is not realistic to transmit large quantities of information, we start by looking at codes with only a few codewords, so called *ultrasmall block codes*. Such codes have many practical applications. For example, in the situation of establishing an initial connection in a wireless link, the amount of information that needs to be transmitted during the setup of the link is limited to usually only a couple of bits. However, these bits need to be transmitted in very short time (e.g., blocklength in the range of $n = 20$ to $n = 30$) with the highest possible reliability [3]. Similarly, in the context of 5G wireless communication systems, very reliable codes with very low latency are asked for, which can only be found by restricting oneself to short packets [4].

Also in the area of distributed storage data systems good nonlinear codes are of great interest. Here the nonlinear code constructions presented in this work offer a way to nonlinear code designs that are better compared to the best linear codes of identical given parameters [5].

Another important application of short codes appears in the context of "biological coding", where future digital information storage system designs are attempted based on DNA or DNA-related methods to store data. To that goal very short and simple codes are needed to provide local data integrity. While first architectures relied on a single-parity check code, more advanced systems try more elaborate schemes like simple Reed-Solomon codes [6]–[9]. The code designs presented in this work have the potential to further improve the performance of such systems.

We also would like to mention the emerging field of molecular communication, where short messages are transmitted with the help of molecules that are transported by diffusion. Inherently, in such systems neither the blocklength and nor the number of codewords can be large [10].

Finally, quantum coding is a very strongly growing research area where people are looking for very short codes. So far in that field only some heuristically chosen codes have been applied, thus, a fundamentally new and more systematic way of trying to find good codes is needed. The code designs presented in this paper are very good candidates for such a new approach [11].

While conventional coding theory in the sense of Shannon theory often focuses on stating important fundamental insights and properties like, e.g., at what rates it is possible to transmit information with an error probability that vanishes as the blocklength tends to infinity, we specifically turn our attention to the concrete *code design*, i.e., we are interested in actually finding a globally optimum code for a certain given channel

[1] By *optimal* we always mean *minimizing error probability*.

and a given fixed blocklength.

In this paper, we reintroduce a class of codes, called *fair weak flip codes*, that have many beautiful properties similar to those of binary linear codes. However, while binary linear codes are very much limited since they can only exist if the number of codewords $\mathsf{M}$ happens to be an integer-power of 2, our class of codes exists for arbitrary[2] $\mathsf{M}$. We will investigate these "quasi-linear" codes and show that they satisfy the Plotkin bound.

Fair weak flip codes are related to a class of binary nonlinear codes that are constructed with the help of Hadamard matrices and Levenshtein's theorem [12, Ch. 2]. These *binary nonlinear Hadamard codes* also meet the Plotkin bound. As a matter of fact, if for the parameters $(\mathsf{M}, n)$ of a given fair weak flip code there exists a Hadamard code, then these two codes are equivalent.[3] In this sense we can consider the fair weak flip codes to be a subclass of Hadamard codes. Note, however, that there is no guarantee that for every choice of parameters $(\mathsf{M}, n)$ for which fair weak flip codes exist, there also exists a corresponding Hadamard code.

Moreover, note that while Levenshtein's method is only concerned with an optimal pairwise Hamming distance structure, we will show that fair weak flip codes are *globally* optimal (i.e., they are the best with respect to error probability and not only to pairwise Hamming distance, and they are best among *all* codes, linear or nonlinear). We prove this global optimality in the case of the number of codewords $\mathsf{M} \leq 4$, and conjecture it for $\mathsf{M} \geq 5$.

We introduce a generalization to the Hamming distance, the *$r$-wise Hamming distance*, and we prove that the exact average error probability of an arbitrary binary code on the *binary erasure channel (BEC)* can be fully characterized using the $r$-wise Hamming distances only. Furthermore, we propose a Plotkin-type bound on the $r$-wise Hamming distances for binary codes.

Our definition of the $r$-wise Hamming distance is related to the *$r$th generalized Hamming weight* introduced in [13] and used, e.g., to investigate a code's security performance on the wire-tap channel of Type II. Note, however, that [13] restricts itself to *linear* codes only. Indeed, the $r$th generalized Hamming weight is defined by the minimum support of any $r$-dimensional subcode of a given linear code of dimension $k$ (where a *support* of a linear code is defined as the number of positions where not all codewords are zero), and thus only describes subsets of codewords that form a linear subcode. On the other hand, our $r$-wise Hamming distance is defined for linear and nonlinear codes and characterizes the relation of any subset of $r$ codewords. Since an arbitrary subset of codewords from a linear code can be either linear or nonlinear, this leads to an essential distinction of our work from previous works [14]–[16].

We further define a class of codes called *weak flip codes* that contains the fair weak flip codes as a special case. We prove that some particular weak flip codes are optimal for the BEC for $\mathsf{M} \leq 4$ and for *any* finite blocklength $n$. For $\mathsf{M} \geq 5$, we believe that for certain blocklengths the codes which maximize all the minimum $r$-wise Hamming distances (including the pairwise Hamming distance) are best among all possible codes. Evidence for this claim will be presented for the cases of $\mathsf{M} = 8$ and $\mathsf{M} = 16$. Based on random search, two algorithms are proposed that find nonlinear weak flip code designs that outperform the best linear codes for certain values of $\mathsf{M}$ and many blocklengths $n$.

[2] Note that fair weak flip codes do not exist for all blocklengths $n$.

[3] For a precise definition of *equivalence* see Remark 8 below.

This work is an extension of our previous work [17] and of [18], [19], where we study ultrasmall block codes for the situation of general binary-input binary-output channels and where we derive the optimal code design for the two special cases of the *Z-channel (ZC)* and the *binary symmetric channel (BSC)*. We will also briefly compare our findings here with these channels, especially with the symmetric BSC.

The foundations of our insights lie in a powerful way of creating and analyzing both linear and nonlinear block codes. As is customary, we use the *codebook matrix* containing the codewords in its rows to describe our codes.[4] However, for our code construction and performance analysis, we are looking at this codebook matrix not row-wise, but *column-wise*. All our proofs and also our definitions of the new $r$-wise Hamming distance and the "quasi-linear" codes are fully based on this new approach. (This is another fundamental difference between our results and the binary nonlinear Hadamard codes that are constructed based on Hadamard matrices and Levenshtein's theorem [12].)

The remainder of this paper is structured as follows. After some comments about our notation, we will present the basic setup of this work in Section 2: We review some common definitions in coding, introduce the channel model, and we explain our concept of the column-wise description of general binary codes. We also define several families of binary codes: the family of *weak flip codes* including its subfamily of *fair weak flip codes*, the binary Hadamard codes, and the family of binary linear codes. Section 3 then reviews previous results related to this work. The main results of the paper are summarized and discussed in Sections 4 and 5: Section 4 provides the definition of the $r$-wise Hamming distance and discusses the quasi-linear properties of weak flip codes, and in Section 5 the optimal codes and the best nonlinear codes for the BEC are presented. We conclude in Section 6. Some of the lengthy proofs from Section 5 are postponed to the appendix.

As a convention in coding theory, vectors (denoted by boldface Roman letters, e.g., $\mathbf{x}$) are row-vectors. However, for simplicity of notation and to avoid a large number of transpose-signs, we slightly misuse this notational convention for one special case: any vector $\mathbf{c}$ is a column-vector. It should be always clear from the context because these vectors are used to build codebook matrices and are therefore also conceptually quite different from the transmitted codeword $\mathbf{x}$ or the received sequence $\mathbf{y}$.

Moreover, we use a bar $\bar{\mathbf{x}}$ to denote the flipped version of $\mathbf{x}$, i.e., $\bar{\mathbf{x}} \triangleq \mathbf{x} \oplus \mathbf{1}$ (where $\oplus$ denotes the componentwise XOR operation and where $\mathbf{1}$ is the all-one vector). We use capital letters for random quantities, e.g., $X$, and small letters for their deterministic counterparts, e.g., $x$; constants are depicted by Greek letters, small Romans, or a special font, e.g., $\mathsf{M}$; sets are denoted by calligraphic letters, e.g., $\mathcal{M}$; and $|\mathcal{M}|$ denotes the cardinality of the set $\mathcal{M}$.

# 2 Setup and Definitions

## 2.1 Coding Schemes

**Definition 1.** An $(\mathsf{M}, n)$ *coding scheme* for a discrete memoryless channel (DMC) $(\mathcal{X}, \mathcal{Y}, P_{Y|X})$ consists of the message set $\mathcal{M} \triangleq \{1, 2, \ldots, \mathsf{M}\}$, a codebook $\mathscr{C}^{(\mathsf{M},n)}$ with $\mathsf{M}$ length-$n$ codewords $\mathbf{x}_m = (x_{m,1}, x_{m,2}, \ldots, x_{m,n}) \in \mathcal{X}^n$, $m \in \mathcal{M}$, an encoder that maps every message $m$ into its corresponding codeword $\mathbf{x}_m$, and a decoder that makes a decoding decision $g(\mathbf{y}) \in \mathcal{M}$ for every received $n$-vector $\mathbf{y} \in \mathcal{Y}^n$.

[4] The codebook matrix is not to be confused with a generator matrix that can be used to describe linear codes.

The set of codewords $\mathscr{C}^{(\mathsf{M},n)}$ is called $(\mathsf{M},n)$ *codebook* or simply $(\mathsf{M},n)$ *code*. Sometimes we follow the custom of traditional coding theory and use three parameters:[5] $(\mathsf{M},n,d)$ *code*, where the third parameter $d$ denotes the *minimum Hamming distance*[6] $d_{\min}\bigl(\mathscr{C}^{(\mathsf{M},n)}\bigr)$, i.e., the minimum number of components in which any two codewords differ.

We assume that the $\mathsf{M}$ possible messages are equally likely and $g$ is the *maximum likelihood (ML) decoder*[7]

$$g(\mathbf{y}) \triangleq \operatorname*{argmax}_{1\leq m\leq \mathsf{M}} P_{\mathbf{Y}|\mathbf{X}}(\mathbf{y}|\mathbf{x}_m), \tag{1}$$

where in case that there are several $m$ achieving the maximum, an arbitrary one of them is chosen.

**Definition 2.** For a given code $\mathscr{C}^{(\mathsf{M},n)}$ we define the *decoding region* $\mathcal{D}_m^{(\mathsf{M},n)}$ corresponding to the $m$th codeword $\mathbf{x}_m$ as

$$\mathcal{D}_m^{(\mathsf{M},n)} \triangleq \{\mathbf{y}\colon g(\mathbf{y}) = m\}. \tag{2}$$

Note that in Definition 2, all decoding regions must be disjoint, and their union must be equal to $\mathcal{Y}^n$

$$\mathcal{D}_m^{(\mathsf{M},n)} \cap \mathcal{D}_{m'}^{(\mathsf{M},n)} = \emptyset, \quad 1 \leq m < m' \leq \mathsf{M}, \tag{3}$$

$$\bigcup_{m\in\mathcal{M}} \mathcal{D}_m^{(\mathsf{M},n)} = \mathcal{Y}^n. \tag{4}$$

As mentioned above, there does not necessarily exist a unique $m$ such that for a given $\mathbf{y}$,

$$P_{\mathbf{Y}|\mathbf{X}}(\mathbf{y}|\mathbf{x}_m) = \max_{1\leq m'\leq \mathsf{M}} P_{\mathbf{Y}|\mathbf{X}}(\mathbf{y}|\mathbf{x}_{m'}), \tag{5}$$

i.e., certain received vectors $\mathbf{y}$ could be assigned to different decoding regions without changing the performance of the coding scheme. In the following we define *closed decoding regions* that break the condition (3).

**Definition 3.** The *closed decoding region* $\overline{\mathcal{D}}_m^{(\mathsf{M},n)}$ corresponding to the $m$th codeword $\mathbf{x}_m$ is defined as

$$\overline{\mathcal{D}}_m^{(\mathsf{M},n)} \triangleq \Bigl\{\mathbf{y}\colon P_{\mathbf{Y}|\mathbf{X}}(\mathbf{y}|\mathbf{x}_m) = \max_{1\leq m'\leq \mathsf{M}} P_{\mathbf{Y}|\mathbf{X}}(\mathbf{y}|\mathbf{x}'_m)\Bigr\}, \quad m \in \mathcal{M}. \tag{6}$$

Note that $\mathcal{D}_m^{(\mathsf{M},n)} \subseteq \overline{\mathcal{D}}_m^{(\mathsf{M},n)}$.

[5] Actually, it is usual to have them ordered as $(n,\mathsf{M},d)$, but for consistency and because $\mathsf{M}$ is the more important parameter, we will stick to $(\mathsf{M},n)$ or $(\mathsf{M},n,d)$.

[6] For a definition of *Hamming distance* see Definition 6 below.

[7] Under the assumption of equally likely messages, the ML decoding rule is equivalent to the *maximum a posteriori (MAP)* decoding rule, i.e., for a given code and DMC, it minimizes the average error probability (as defined in (9)) among all possible decoders.

**Definition 4.** For an $(\mathsf{M}, n)$ code, given that message $m$ (and hence the $m$th codeword $\mathbf{x}_m$) has been sent, we define $\lambda_m$ to be the corresponding *probability of a decoding error* under the ML decoder $g$:

$$\lambda_m\bigl(\mathscr{C}^{(\mathsf{M},n)}\bigr) \triangleq \Pr[g(\mathbf{Y}) \neq m \,|\, \mathbf{X} = \mathbf{x}_m] \tag{7}$$

$$= \sum_{\mathbf{y} \notin \mathcal{D}_m^{(\mathsf{M},n)}} P_{\mathbf{Y}|\mathbf{X}}(\mathbf{y}|\mathbf{x}_m). \tag{8}$$

The *average error probability* $P_\mathrm{e}$ of an $(\mathsf{M}, n)$ code is defined as

$$P_\mathrm{e}\bigl(\mathscr{C}^{(\mathsf{M},n)}\bigr) \triangleq \frac{1}{\mathsf{M}} \sum_{m=1}^{\mathsf{M}} \lambda_m\bigl(\mathscr{C}^{(\mathsf{M},n)}\bigr). \tag{9}$$

Sometimes it will be more convenient to focus on the probability of not making any error, denoted *success probability* $\psi_m$:

$$\psi_m\bigl(\mathscr{C}^{(\mathsf{M},n)}\bigr) \triangleq \Pr[g(\mathbf{Y}) = m \,|\, \mathbf{X} = \mathbf{x}_m] \tag{10}$$

$$= \sum_{\mathbf{y} \in \mathcal{D}_m^{(\mathsf{M},n)}} P_{\mathbf{Y}|\mathbf{X}}(\mathbf{y}|\mathbf{x}_m) \tag{11}$$

$$= \Pr\Bigl[\mathbf{Y} \in \mathcal{D}_m^{(\mathsf{M},n)} \,\Big|\, \mathbf{X} = \mathbf{x}_m\Bigr]. \tag{12}$$

The definition of the *average success probability*[8] $P_\mathrm{c}$ follows accordingly.

Our ultimate goal is to find the structure of a code that minimizes the average error probability among all codes based on the ML decoding rule.

**Definition 5.** A code $\mathscr{C}^{(\mathsf{M},n)}$ is called *optimal* and denoted by $\mathscr{C}^{(\mathsf{M},n)*}$ if

$$P_\mathrm{e}\bigl(\mathscr{C}^{(\mathsf{M},n)*}\bigr) \leq P_\mathrm{e}\bigl(\mathscr{C}^{(\mathsf{M},n)}\bigr) \tag{13}$$

for any (linear or nonlinear) code $\mathscr{C}^{(\mathsf{M},n)}$.

## 2.2 The BEC and its Average Error Probability

Regarding a channel model, this work focuses on the well-known *binary erasure channel (BEC)* given in Figure 1. The BEC is a DMC with a binary input alphabet $\mathcal{X} = \{0, 1\}$ and a ternary output alphabet $\mathcal{Y} = \{0, 1, 2\}$, and with a conditional channel law

$$P_{Y|X}(y|x) = \begin{cases} 1-\delta & \text{if } y = x,\; x \in \{0,1\}, \\ \delta & \text{if } y = 2,\; x \in \{0,1\}. \end{cases} \tag{14}$$

Here $0 \leq \delta < 1$ is called the *erasure probability.*

While the focus lies on the BEC, we will sometimes briefly compare our results with the situation of the *binary symmetric channel (BSC)*, particularly in view of [19].

Next we derive a closed-form expression for the average error probability of an arbitrary code used over the BEC, assuming uniformly distributed messages and an optimal ML decoder. To that goal we need the following two definitions.

**Definition 6.** The *Hamming distance* $d_\mathrm{H}(\mathbf{x}_m, \mathbf{x}_{m'})$ between two binary length-$n$ vectors $\mathbf{x}_m$ and $\mathbf{x}_{m'}$ is defined as the number of positions $j$ where $x_{m,j} \neq x_{m',j}$. The *Hamming weight* of a binary length-$n$ vector $\mathbf{x}$ is defined as $w_\mathrm{H}(\mathbf{x}) \triangleq d_\mathrm{H}(\mathbf{x}, \mathbf{0})$.

[8]The subscript "c" stands for "correct."

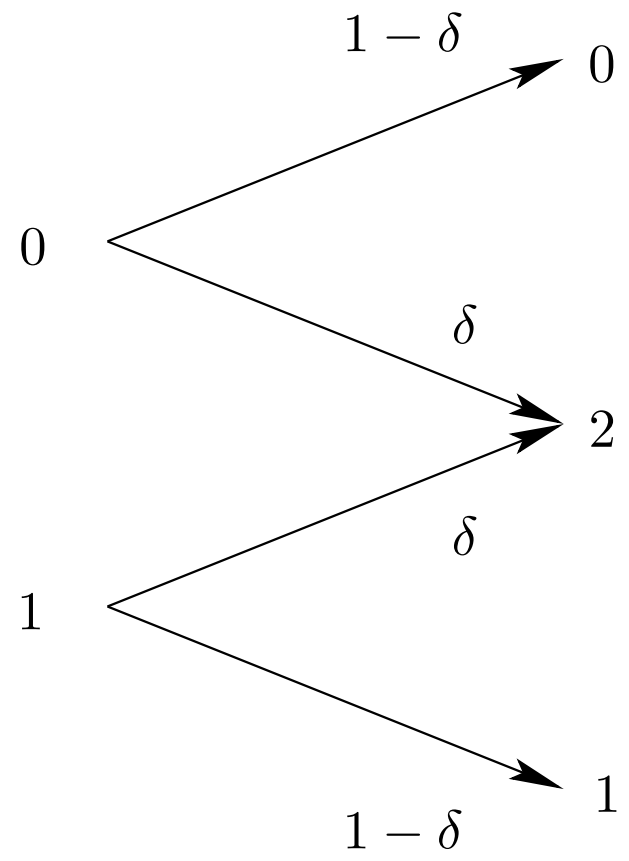

Figure 1: The binary erasure channel (BEC) with erasure probability $\delta$. The channel output 2 corresponds to an erasure.

**Definition 7.** By $\mathsf{N}(\alpha|\mathbf{y})$ we denote the number of occurrences of a symbol $\alpha \in \mathcal{Y}$ in a received vector $\mathbf{y}$, and $\mathcal{I}(\alpha|\mathbf{y})$ is defined as the set of indices $j$ such that $y_j = \alpha$. Thus, $\mathsf{N}(\alpha|\mathbf{y}) = |\mathcal{I}(\alpha|\mathbf{y})|$. Moreover, we use $\mathbf{x}_{m,\mathcal{I}(\alpha|\mathbf{y})}$ (respectively, $\mathbf{y}_{\mathcal{I}(\alpha|\mathbf{y})}$) to describe a vector of length $\mathsf{N}(\alpha|\mathbf{y})$ containing the components $x_{m,j}$ (respectively, $y_j$) where $j \in \mathcal{I}(\alpha|\mathbf{y})$. We also write $\mathbf{x}_{m,\mathcal{I}(\alpha|\mathbf{y})} \cup \mathbf{x}_{m,\mathcal{I}(\mathcal{Y}\setminus\{\alpha\}|\mathbf{y})}$ for the complete vector $\mathbf{x}_m$, where the "union"-operation implicitly reorders the indices in the usual ascending order.

The error probability when transmitting uniformly picked codewords from code $\mathscr{C}^{(\mathsf{M},n)}$ over the BEC can be written as follows:

$$P_{\mathrm{e}}\big(\mathscr{C}^{(\mathsf{M},n)}\big) = \frac{1}{\mathsf{M}} \sum_{m=1}^{\mathsf{M}} \sum_{\substack{\mathbf{y}\in\mathcal{Y}^n \\ g(\mathbf{y})\neq m}} (1-\delta)^{n-\mathsf{N}(2|\mathbf{y})}\, \delta^{\mathsf{N}(2|\mathbf{y})} \mathscr{I}\big\{d_{\mathrm{H}}\big(\mathbf{x}_{m,\mathcal{I}(0|\mathbf{y})}, \mathbf{y}_{\mathcal{I}(0|\mathbf{y})}\big) = 0\big\} \cdot \mathscr{I}\big\{d_{\mathrm{H}}\big(\mathbf{x}_{m,\mathcal{I}(1|\mathbf{y})}, \mathbf{y}_{\mathcal{I}(1|\mathbf{y})}\big) = 0\big\}, \tag{15}$$

where $\mathscr{I}\{\text{STATEMENT}\}$ denotes the indicator function whose value is 1 if the STATEMENT is correct and 0 otherwise.

### 2.3 Column-Wise Description of General Binary Codes

Usually, a general codebook $\mathscr{C}^{(\mathsf{M},n)}$ with $\mathsf{M}$ codewords and with blocklength $n$ is written as an $\mathsf{M} \times n$ codebook matrix where the $\mathsf{M}$ rows correspond to the $\mathsf{M}$ codewords:

$$\mathscr{C}^{(\mathsf{M},n)} = \begin{pmatrix} \text{---}\ \mathbf{x}_1\ \text{---} \\ \vdots \\ \text{---}\ \mathbf{x}_{\mathsf{M}}\ \text{---} \end{pmatrix} = \begin{pmatrix} | & | & & | \\ \mathbf{c}_1 & \mathbf{c}_2 & \cdots & \mathbf{c}_n \\ | & | & & | \end{pmatrix}. \tag{16}$$

In our approach, we prefer to consider the codebook matrix *column-wise* rather than row-wise [19]. We denote the length-$\mathsf{M}$ column-vectors of the codebook by $\mathbf{c}_j$, $j \in \{1, \ldots, n\}$.

**Remark 8.** Since we assume equally likely messages, any permutation of rows only changes the assignment of codewords to messages and has therefore no impact on the performance. We thus consider two codes with permuted rows as being *equal* (this agrees with the concept of a code being a *set* of codewords, where the ordering of the codewords is irrelevant). Furthermore, since we only consider memoryless channels, any permutation of the columns of $\mathscr{C}^{(\mathsf{M},n)}$ will lead to another code with identical error probability. We say that such two codes are *equivalent*. We would like to emphasize that two codes being equivalent is not the same as two codes being equal. However, as we are mainly interested in the performance of a code, we usually treat two equivalent codes as being the same. △

Due to the symmetry of the BEC[9] we have an additional equivalence in the codebook design (compare also with the BSC [19]).

**Lemma 9.** *Consider an arbitrary code $\mathscr{C}^{(\mathsf{M},n)}$ to be used on the BEC and consider an arbitrary $\mathsf{M}$-vector $\mathbf{c}$. Construct a new length-$(n+1)$ code $\mathscr{C}^{(\mathsf{M},n+1)}$ by appending $\mathbf{c}$ to the codebook matrix of $\mathscr{C}^{(\mathsf{M},n)}$ and another new length-$(n+1)$ code $\bar{\mathscr{C}}^{(\mathsf{M},n+1)}$ by appending the flipped vector $\bar{\mathbf{c}} = \mathbf{c} \oplus \mathbf{1}$ to the codebook matrix of $\mathscr{C}^{(\mathsf{M},n)}$. Then the performance of these two new codes are identical:*

$$P_{\mathrm{e}}\big(\mathscr{C}^{(\mathsf{M},n+1)}\big) = P_{\mathrm{e}}\big(\bar{\mathscr{C}}^{(\mathsf{M},n+1)}\big). \tag{17}$$

Note that Lemma 9 cannot be generalized further, i.e., for some $\mathscr{C}^{(\mathsf{M},n)}$, appending a vector $\tilde{\mathbf{c}}$ other than $\bar{\mathbf{c}}$ may result in a length-$(n+1)$ code $\tilde{\mathscr{C}}^{(\mathsf{M},n+1)}$ that is not equivalent to $\mathscr{C}^{(\mathsf{M},n+1)}$.

Next we define a convenient numbering system for the possible columns of the codebook matrix of binary codes.

**Definition 10.** For fixed $\mathsf{M}$ and $b_m \in \{0,1\}$, $m \in \mathcal{M}$, we describe the column vector $(b_1\ b_2\ \cdots\ b_{\mathsf{M}})^{\mathsf{T}}$ by its reverse binary representation of nonnegative integers

$$j = \sum_{m=1}^{\mathsf{M}} b_m\, 2^{\mathsf{M}-m}, \tag{18}$$

and write $\mathbf{c}_j^{(\mathsf{M})} \triangleq (b_1\ b_2\ \cdots\ b_{\mathsf{M}})^{\mathsf{T}}$. For example, $\mathbf{c}_{12}^{(5)} = (0\,1\,1\,0\,0)^{\mathsf{T}}$ and $\mathbf{c}_3^{(5)} = (0\,0\,0\,1\,1)^{\mathsf{T}}$.

Due to Lemma 9, we discard any column starting with a one, i.e., we require $b_1 = 0$. Moreover, as it will never help to improve the performance, we exclude the all-zero column. Hence, the set of all possible *candidate columns* of general binary codes can be restricted to

$$\mathcal{C}^{(\mathsf{M})} \triangleq \Big\{\mathbf{c}_1^{(\mathsf{M})}, \mathbf{c}_2^{(\mathsf{M})}, \ldots, \mathbf{c}_{2^{\mathsf{M}-1}-1}^{(\mathsf{M})}\Big\}. \tag{19}$$

For a given codebook and for any

$$j \in \mathcal{J} \triangleq \{1, \ldots, 2^{\mathsf{M}-1} - 1\}, \tag{20}$$

let $t_j$ denote the number of the corresponding candidate columns $\mathbf{c}_j^{(\mathsf{M})}$ appearing in the codebook matrix of $\mathscr{C}^{(\mathsf{M},n)}$. Because of Remark 8, the ordering of the candidate columns is irrelevant, and any binary code with blocklength

$$n = \sum_{j=1}^{2^{\mathsf{M}-1}-1} t_j \tag{21}$$

[9] The symmetry property here is identical to the symmetry definitions in [20, p. 94]. Hence, it is not surprising that Lemma 9 also holds for general binary-input symmetric channels.

can therefore be fully described by the parameter vector

$$\mathbf{t} \triangleq \left[t_1, t_2, \ldots, t_{2^{\mathsf{M}-1}-1}\right]. \tag{22}$$

We say that such a code has a *type vector* (or simply *type*) $\mathbf{t}$, and write[10] $\mathscr{C}^{(\mathsf{M},n)}_{t_1,\ldots,t_{2^{\mathsf{M}-1}-1}}$ or $\mathscr{C}^{(\mathsf{M},n)}_{\mathbf{t}}$.

**Example 11.** For $\mathsf{M} = 4$, the candidate columns set is

$$\begin{aligned}\mathcal{C}^{(4)} = \Bigg\{ &\mathbf{c}_1^{(4)} = \begin{pmatrix}0\\0\\0\\1\end{pmatrix}, \mathbf{c}_2^{(4)} \triangleq \begin{pmatrix}0\\0\\1\\0\end{pmatrix}, \mathbf{c}_3^{(4)} \triangleq \begin{pmatrix}0\\0\\1\\1\end{pmatrix}, \\ &\mathbf{c}_4^{(4)} = \begin{pmatrix}0\\1\\0\\0\end{pmatrix}, \mathbf{c}_5^{(4)} \triangleq \begin{pmatrix}0\\1\\0\\1\end{pmatrix}, \mathbf{c}_6^{(4)} \triangleq \begin{pmatrix}0\\1\\1\\0\end{pmatrix}, \mathbf{c}_7^{(4)} \triangleq \begin{pmatrix}0\\1\\1\\1\end{pmatrix}\Bigg\}. \end{aligned} \tag{23}$$

A codebook $\mathscr{C}^{(4,7)}_{\mathbf{t}}$ of type $\mathbf{t} = [2, 0, 2, 0, 2, 1, 0]$ is equivalent to all columns permutations of the following codebook:

$$\begin{pmatrix}0&0&0&0&0&0&0\\0&0&0&0&1&1&1\\0&0&1&1&0&0&1\\1&1&1&1&1&1&0\end{pmatrix}. \tag{24}$$

$\Diamond$

## 2.4 Weak Flip Codes

We next introduce some special families of binary codes.

**Definition 12.** Given an integer $\mathsf{M} \geq 2$, a length-$\mathsf{M}$ candidate column is called a *weak flip column* and denoted $\mathbf{c}^{(\mathsf{M})}_{\text{weak}}$ if its first component is 0 and its Hamming weight equals to $\lfloor \frac{\mathsf{M}}{2} \rfloor$ or $\lceil \frac{\mathsf{M}}{2} \rceil$. The collection of all possible weak flip columns is called *weak flip candidate columns set* and is denoted by $\mathcal{C}^{(\mathsf{M})}_{\text{weak}}$. The remaining, nonweak flip candidate columns are collected in $\mathcal{C}^{(\mathsf{M})}_{\text{nonweak}}$, i.e., $\mathcal{C}^{(\mathsf{M})} = \mathcal{C}^{(\mathsf{M})}_{\text{weak}} \cup \mathcal{C}^{(\mathsf{M})}_{\text{nonweak}}$.

We see that a weak flip column contains an almost equal or equal number of zeros and ones. For the remainder of this paper, we introduce the following shorthands:

$$\mathsf{J} \triangleq 2^{\mathsf{M}-1} - 1, \quad \bar{\ell} \triangleq \left\lceil \frac{\mathsf{M}}{2} \right\rceil, \quad \underline{\ell} \triangleq \left\lfloor \frac{\mathsf{M}}{2} \right\rfloor, \quad \mathsf{L} \triangleq \binom{2\bar{\ell} - 1}{\bar{\ell}}. \tag{25}$$

Recall the corresponding sets $\mathcal{M}$ given in Definition 1 and $\mathcal{J}$ given in (20).

**Lemma 13.** *The cardinality of the weak flip candidate columns set is*

$$\left|\mathcal{C}^{(\mathsf{M})}_{\text{weak}}\right| = \mathsf{L}, \tag{26}$$

*and the cardinality of the nonweak flip candidate columns set is*

$$\left|\mathcal{C}^{(\mathsf{M})}_{\text{nonweak}}\right| = \mathsf{J} - \mathsf{L}. \tag{27}$$

[10]Note that sometimes, for the sake of convenience, we will omit the superscripts $(\mathsf{M}, n)$ or $(\mathsf{M})$.

*Proof:* If $\mathsf{M} = 2\bar{\ell}$, then we have $\binom{2\bar{\ell}-1}{\bar{\ell}}$ possible choices of weak flip columns, while if $\mathsf{M} = 2\bar{\ell}-1$, we have $\binom{2\bar{\ell}-2}{\bar{\ell}-1} + \binom{2\bar{\ell}-2}{\bar{\ell}} = \binom{2\bar{\ell}-1}{\bar{\ell}}$ choices. This proves (26). Since in total we have $\mathsf{J}$ candidate columns, (27) follows directly from (26). It can also be computed as

$$\left|\mathcal{C}_{\text{nonweak}}^{(\mathsf{M})}\right| = \sum_{h=1}^{\underline{\ell}-1} \binom{\mathsf{M}-1}{h} + \sum_{h=\bar{\ell}+1}^{\mathsf{M}-1} \binom{\mathsf{M}-1}{h} = \mathsf{J} - \mathsf{L}. \tag{28}$$

□

**Remark 14.** The above lemma assures that the cardinalities of the weak flip candidate columns set for $\mathsf{M} = 2\bar{\ell}-1$ and of the weak flip candidate columns set for $\mathsf{M} = 2\bar{\ell}$ are both the same for any positive integer $\bar{\ell}$ and are both given by $\binom{2\bar{\ell}-1}{\bar{\ell}}$. Actually, if we take $\mathcal{C}_{\text{weak}}^{(2\bar{\ell}-1)}$ and we append as the last bit a one to all its weak flip columns of weight $\underline{\ell} = \bar{\ell}-1$ and a zero to the other weak flip columns of weight $\bar{\ell}$, we obtain $\mathcal{C}_{\text{weak}}^{(2\bar{\ell})}$. Hence, $\mathcal{C}_{\text{weak}}^{(2\bar{\ell}-1)}$ can be obtained from $\mathcal{C}_{\text{weak}}^{(2\bar{\ell})}$ by removing the last bit from all column vectors. △

**Definition 15.** A *weak flip code* $\mathscr{C}_{\text{weak}}^{(\mathsf{M},n)}$ is constructed only by weak flip columns. Since in its type (22) all positions corresponding to nonweak flip columns are zero, we use a reduced type vector:

$$\mathbf{t}_{\text{weak}} \triangleq \left[t_{j_1}, t_{j_2}, \ldots, t_{j_{\mathsf{L}}}\right], \tag{29}$$

where

$$\sum_{w=1}^{\mathsf{L}} t_{j_w} = n \tag{30}$$

with $j_w$, $w = 1, \ldots, \mathsf{L}$, representing the numbers of the candidate columns that are weak flip columns.

For $\mathsf{M} = 2$ or $\mathsf{M} = 3$, all candidate columns are also weak flip columns (note that $2^{\mathsf{M}-1} - 1 = \binom{2\bar{\ell}-1}{\bar{\ell}} = \mathsf{L}$ only when $\mathsf{M} = 2$ or $\mathsf{M} = 3$). For $\mathsf{M} = 4$, $\mathbf{t}_{\text{weak}} = [t_3, t_5, t_6]$. A similar definition can be given also for larger $\mathsf{M}$; however, one needs to be aware that the number of weak flip columns is increasing exponentially fast. For $\mathsf{M} = 5$, we have ten weak flip columns:

$$\mathcal{C}_{\text{weak}}^{(5)} = \left\{ \mathbf{c}_3^{(5)} \triangleq \begin{pmatrix}0\\0\\0\\1\\1\end{pmatrix}, \mathbf{c}_5^{(5)} \triangleq \begin{pmatrix}0\\0\\1\\0\\1\end{pmatrix}, \mathbf{c}_6^{(5)} \triangleq \begin{pmatrix}0\\0\\1\\1\\0\end{pmatrix}, \mathbf{c}_7^{(5)} \triangleq \begin{pmatrix}0\\0\\1\\1\\1\end{pmatrix}, \mathbf{c}_9^{(5)} \triangleq \begin{pmatrix}0\\1\\0\\0\\1\end{pmatrix}, \right.$$
$$\left. \mathbf{c}_{10}^{(5)} \triangleq \begin{pmatrix}0\\1\\0\\1\\0\end{pmatrix}, \mathbf{c}_{11}^{(5)} \triangleq \begin{pmatrix}0\\1\\0\\1\\1\end{pmatrix}, \mathbf{c}_{12}^{(5)} \triangleq \begin{pmatrix}0\\1\\1\\0\\0\end{pmatrix}, \mathbf{c}_{13}^{(5)} \triangleq \begin{pmatrix}0\\1\\1\\0\\1\end{pmatrix}, \mathbf{c}_{14}^{(5)} \triangleq \begin{pmatrix}0\\1\\1\\1\\0\end{pmatrix} \right\}. \tag{31}$$

We will next introduce a special subclass of weak flip codes that, as we will see in Section 4.2, possesses particularly beautiful properties.

**Definition 16.** A weak flip code is called *fair* if it is constructed by an equal number of all possible weak flip columns in $\mathcal{C}_{\text{weak}}^{(\mathsf{M})}$. Note that by definition the blocklength of a fair weak flip code is always an integer-multiple of $\mathsf{L}$.

Fair weak flip codes have been used by Shannon *et al.* [21] for the derivation of error exponents, although the codes were not named at that time. Note that in [21] the error exponents are defined when blocklength $n$ goes to infinity, but here in this work we consider finite $n$.

## 2.5 Hadamard Codes

In this section, we review the family of *Hadamard codes* and investigate its relation to weak flip codes and fair weak flip codes. We follow the definition of [12, Ch. 2].

**Definition 17.** For an even integer $m$, a (*normalized*) *Hadamard matrix* $\mathsf{H}_m$ of order $m$ is an $m \times m$ matrix with entries $+1$ and $-1$ and with the first row and column being all $+1$, such that

$$\mathsf{H}_m \mathsf{H}_m^{\mathsf{T}} = m \mathsf{I}_m, \tag{32}$$

if such a matrix exists. Here $\mathsf{I}_m$ is the identity matrix of size $m$. If the entries $+1$ are replaced by 0 and the entries $-1$ by 1, $\mathsf{H}_m$ is changed into the *binary Hadamard matrix* $\mathsf{A}_m$.

Note that a necessary condition for the existence of $\mathsf{H}_m$ (and the corresponding $\mathsf{A}_m$) is that $m$ is 1, 2, or a multiple of 4 [12, Ch. 2].

**Definition 18.** The binary Hadamard matrix $\mathsf{A}_m$ gives rise to three families of Hadamard codes:[11]

1. The $\left(m, m-1, \frac{m}{2}\right)$ *Hadamard code* $\mathscr{H}_{1,m}$ consists of the rows of $\mathsf{A}_m$ with the first column deleted. Moreover, the codewords in $\mathscr{H}_{1,m}$ that begin with 0 form the $\left(\frac{m}{2}, m-2, \frac{m}{2}\right)$ *Hadamard code* $\mathscr{H}'_{1,m}$ if the initial zero is deleted.
2. The $\left(2m, m-1, \frac{m}{2}-1\right)$ *Hadamard code* $\mathscr{H}_{2,m}$ consists of $\mathscr{H}_{1,m}$ together with the complements of all its codewords.
3. The $\left(2m, m, \frac{m}{2}\right)$ *Hadamard code* $\mathscr{H}_{3,m}$ consists of the rows of $\mathsf{A}_m$ and their complements.

Further Hadamard codes can be created by an arbitrary combination of the codebook matrices of different Hadamard codes.

**Example 19.** Consider the $(8, 7, 4)$ Hadamard code

$$\mathscr{H}_{1,8} = \begin{pmatrix} 0 & 0 & 0 & 0 & 0 & 0 & 0 \\ 0 & 0 & 1 & 0 & 1 & 1 & 1 \\ 0 & 1 & 0 & 1 & 0 & 1 & 1 \\ 0 & 1 & 1 & 1 & 1 & 0 & 0 \\ 1 & 0 & 0 & 1 & 1 & 0 & 1 \\ 1 & 0 & 1 & 1 & 0 & 1 & 0 \\ 1 & 1 & 0 & 0 & 1 & 1 & 0 \\ 1 & 1 & 1 & 0 & 0 & 0 & 1 \end{pmatrix}. \tag{33}$$

[11] Recall that we describe the code parameters as $(\mathsf{M}, n, d)$, where the third parameter denotes the minimum Hamming distance.

From this code, an $(8, 35, 20)$ Hadamard code can be constructed by simply concatenating $\mathscr{H}_{1,8}$ five times. $\diamondsuit$

Note that since the rows of $\mathsf{H}_m$ are orthogonal, so are the columns of $\mathsf{H}_m$, and thus it follows that each column of the corresponding matrix $\mathsf{A}_m$ has a Hamming weight $\frac{m}{2}$. Moreover, by definition the first row of a binary Hadamard matrix is the all-zero row. Hence, we see that all Hadamard codes are weak flip codes, i.e., the family of weak flip codes is a superset of the family of Hadamard codes.

On the other hand, fair weak flip codes can be seen as a "subset" of Hadamard codes because for all parameters $(\mathsf{M}, n)$ for which fair weak flip codes and also Hadamard codes exist, a fair weak flip code can be constructed from a Hadamard code. The problem with this statement lies in the fact that the Hadamard codes rely on the existence of Hadamard matrices, which in general is not guaranteed, i.e., it is difficult to predict whether for a given pair $(\mathsf{M}, n)$, a Hadamard code exists or not. This is in stark contrast to weak flip codes (which exist for all $\mathsf{M}$ and $n$) and fair weak flip codes (which exist for all $\mathsf{M}$ and for all $n$ being a multiple of $\mathsf{L}$).

We also remark that a Hadamard code of parameters $(\mathsf{M}, n)$, for which fair weak flip codes exist, is not necessarily equivalent to a fair weak flip code.

**Example 20.** We continue with Example 19 and note that the $(8, 35, 20)$ Hadamard code that is constructed by five repetitions of the matrix $\mathscr{H}_{1,8}$ given in (33) is actually not a fair weak flip code since we have not used all possible weak flip columns. However, it is possible to find five different $(8, 7, 4)$ Hadamard codes that combine to an $(8, 35, 20)$ fair weak flip code. Recall that the $(8, 35, 20)$ fair weak flip code is composed of all $\binom{7}{4} = 35$ different weak flip columns. $\diamondsuit$

Note that two Hadamard matrices are equivalent if one can be obtained from the other by permuting rows and columns and by multiplying rows and columns by $-1$. In other words, Hadamard codes can actually be constructed from different sets of weak flip columns.

## 2.6 Linear Codes

In conventional coding theory, *linear codes* form an important and well-known class of error correcting codes that have been shown to possess powerful algebraic properties. We refrain from introducing them here in detail, but rather refer to the vast existing literature for more details (e.g., see [2], [12]). Instead we focus briefly on certain properties of linear codes that are important in the context of this work.

We start by categorizing linear codes as a special case of weak flip codes.

**Proposition 21.** *Every linear code is a weak flip code.*

*Proof:* A linear $(\mathsf{M}, n)$ binary code always contains the all-zero codeword, and each column of its codebook matrix has Hamming weight $\frac{\mathsf{M}}{2}$. Thus, it is a weak flip code. $\square$

Note that linear codes only exist if $\mathsf{M} = 2^k$, while weak flip codes are defined for any $\mathsf{M}$. Also note that the converse of Proposition 21 does not necessarily hold, i.e., even if $\mathsf{M} = 2^k$ for some $k \in \mathbb{N} \triangleq \{1, 2, 3, \ldots\}$, a weak flip code $\mathscr{C}^{(\mathsf{M},n)}$ is not necessarily linear. In summary, we have the following relations among linear, weak flip, and arbitrary $(\mathsf{M}, n)$ codes:

$$\left\{\mathscr{C}_{\text{lin}}^{(\mathsf{M},n)}\right\} \subset \left\{\mathscr{C}_{\text{weak}}^{(\mathsf{M},n)}\right\} \subset \left\{\mathscr{C}^{(\mathsf{M},n)}\right\}. \tag{34}$$

Next we recall an important property of linear codes that follows immediately from the fact that linear codes are subspaces of the $n$-dimensional vector space over the channel input alphabet.

**Proposition 22.** *Let $\mathscr{C}_{\text{lin}}$ be linear and let $\mathbf{x}_m \in \mathscr{C}_{\text{lin}}$ be given. Then the code obtained by adding $\mathbf{x}_m$ to each codeword of $\mathscr{C}_{\text{lin}}$ is equal to $\mathscr{C}_{\text{lin}}$.*

Finally, we are going to investigate linear codes from a column-wise perspective. The goal here is to define *fair linear codes.*

Being a subspace, linear codes are usually represented by a generator matrix $\mathsf{G}_{k\times n}$. We now apply our column-wise point-of-view to the construction of generator matrices.[12] The generator matrix $\mathsf{G}_{k\times n}$ consists of $n$ column vectors $\mathbf{c}_j$ of length $k$ similar to (16). Note that in the generator matrix the all-zero column is useless and is therefore excluded. Thus there are totally

$$\mathsf{K} \triangleq 2^k - 1 = \mathsf{M} - 1 \tag{35}$$

possible candidate columns for $\mathsf{G}_{k\times n}$: $\mathbf{c}_j^{(k)} \triangleq (b_1\ b_2\ \cdots\ b_k)^\mathsf{T}$, where $j = \sum_{i=1}^{k} b_i\, 2^{k-i}$ and where $b_1$ is not necessarily equal to zero. Let $\mathsf{U}_k^\mathsf{T}$ be an auxiliary $k \times \mathsf{K}$ matrix consisting of all possible $\mathsf{K}$ candidate columns for the generator matrix: $\mathsf{U}_k^\mathsf{T} = \bigl(\mathbf{c}_1^{(k)}\ \cdots\ \mathbf{c}_\mathsf{K}^{(k)}\bigr)$. This matrix $\mathsf{U}_k^\mathsf{T}$ then allows us to create the set of all possible candidate columns of length $\mathsf{M} = 2^k$ for the codebook matrix of a linear code.

This allows us to derive the set $\mathcal{C}_{\text{lin}}^{(\mathsf{M})}$ of all possible length-$\mathsf{M}$ candidate columns for the codebook matrices of binary linear codes with $\mathsf{M} = 2^k$ codewords:

**Lemma 23.** *Given a dimension $k$, the* candidate columns set $\mathcal{C}_{\text{lin}}^{(\mathsf{M})}$ *for linear codes is given by the columns of the $\mathsf{M} \times (\mathsf{M}-1)$ matrix*

$$\begin{pmatrix} \mathbf{0} \\ \mathsf{U}_k \end{pmatrix} \mathsf{U}_k^\mathsf{T}, \tag{36}$$

*where $\mathbf{0}$ denotes an all-zero row vector of length $k$.*

Thus, the codebook matrix of any linear code can be represented by

$$\mathscr{C}_{\text{lin}}^{(\mathsf{M},n)} = \begin{pmatrix} \mathbf{0} \\ \mathsf{U}_k \end{pmatrix} \mathsf{G}_{k\times n}, \tag{37}$$

which consists of columns taken only from $\mathcal{C}_{\text{lin}}^{(\mathsf{M})}$. Similarly to (29), since in its type all positions corresponding to candidate columns not in $\mathcal{C}_{\text{lin}}^{(\mathsf{M})}$ are zero, we can also use a reduced type vector to describe a $k$-dimensional linear code:

$$\mathbf{t}_{\text{lin}} \triangleq \bigl[t_{j_1}, t_{j_2}, \ldots, t_{j_\mathsf{K}}\bigr], \tag{38}$$

where $\sum_{\ell=1}^{\mathsf{K}} t_{j_\ell} = n$ with $j_\ell$, $\ell = 1, \ldots, \mathsf{K}$, representing the numbers of the corresponding candidate columns in $\mathcal{C}_{\text{lin}}^{(\mathsf{M})}$.

**Definition 24.** A linear code is called *fair* if its codebook matrix is constructed by an equal number of all possible candidate columns in $\mathcal{C}_{\text{lin}}^{(\mathsf{M})}$. Hence the blocklength of a fair linear code[13] $\mathscr{C}_{\text{lin,fair}}^{(\mathsf{M},n)}$ is always a multiple of $\mathsf{K} = \mathsf{M} - 1$.

[12] The authors in [22] have also used this approach to exhaustively examine all possible linear codes.

[13] We point out that a fair linear code actually is a binary simplex code, which is the dual to the well-known Hamming code. However, to remain in sync with the description of fair weak flip codes, throughout this paper we will stick to the name *fair linear codes*.

**Example 25.** Consider the fair linear code with dimension $k = 3$ and blocklength $n = \mathsf{K} = 7$:

$$\mathscr{C}^{(8,7)}_{\text{lin,fair}} = \begin{pmatrix} \mathbf{0} \\ \mathsf{U}_3 \end{pmatrix} \mathsf{U}_3^{\mathsf{T}} = \begin{pmatrix} \mathbf{0} \\ \mathsf{U}_3 \end{pmatrix} \begin{pmatrix} 0 & 0 & 0 & 1 & 1 & 1 & 1 \\ 0 & 1 & 1 & 0 & 0 & 1 & 1 \\ 1 & 0 & 1 & 0 & 1 & 0 & 1 \end{pmatrix} = \begin{pmatrix} 0 & 0 & 0 & 0 & 0 & 0 & 0 \\ 1 & 0 & 1 & 0 & 1 & 0 & 1 \\ 0 & 1 & 1 & 0 & 0 & 1 & 1 \\ 1 & 1 & 0 & 0 & 1 & 1 & 0 \\ 0 & 0 & 0 & 1 & 1 & 1 & 1 \\ 1 & 0 & 1 & 1 & 0 & 1 & 0 \\ 0 & 1 & 1 & 1 & 1 & 0 & 0 \\ 1 & 1 & 0 & 1 & 0 & 0 & 1 \end{pmatrix} \tag{39}$$

with the corresponding type vector

$$\mathbf{t}_{\text{lin}} = [t_{85}, t_{51}, t_{102}, t_{15}, t_{90}, t_{60}, t_{105}] = [1, 1, 1, 1, 1, 1, 1]. \tag{40}$$

Note that the fair linear code with $k = 3$ and $n = 7$ is an $(8, 7, 4)$ Hadamard linear code with all pairwise Hamming distances equal to 4. ♢

## 2.7 Plotkin Bound

Finally, we recall an important bound that holds for any $(\mathsf{M}, n)$ code.

**Lemma 26 (Plotkin Bound [12]).** *The minimum distance of an $(\mathsf{M}, n)$ binary code $\mathscr{C}^{(\mathsf{M},n)}$ always satisfies*

$$d_{\min}\bigl(\mathscr{C}^{(\mathsf{M},n)}\bigr) \le \begin{cases} \frac{n \cdot \frac{\mathsf{M}}{2}}{\mathsf{M}-1} & \text{if } \mathsf{M} \text{ is even,} \\ \frac{n \cdot \frac{\mathsf{M}+1}{2}}{\mathsf{M}} & \text{if } \mathsf{M} \text{ is odd.} \end{cases} \tag{41}$$

Note that from the proof[14] of Lemma 26, one can actually find that a necessary condition for a codebook to meet the Plotkin Bound with equality is that the codebook is composed of weak flip columns. Furthermore, Levenshtein [12, Ch. 2] proved that the Plotkin bound can be achieved provided that Hadamard matrices exist for orders divisible by 4.

# 3 Previous Results

## 3.1 SGB Bounds on the Average Error Probability

In [21], Shannon, Gallager, and Berlekamp derive upper and lower bounds on the average error probability of a given code used on a DMC. We quickly summarize their results.

**Theorem 27 (SGB Bounds on Average Error Probability [21]).** *For an arbitrary DMC, the average error probability $P_{\mathrm{e}}\bigl(\mathscr{C}^{(\mathsf{M},n)}\bigr)$ of a given code $\mathscr{C}^{(\mathsf{M},n)}$ with $\mathsf{M}$ codewords and blocklength $n$ is upper- and lower-bounded as follows:*

$$\frac{1}{4\mathsf{M}}\, e^{-n\left(\mathsf{D}^{(\text{DMC})}_{\min}(\mathscr{C}^{(\mathsf{M},n)}) + \sqrt{\frac{2}{n}} \log \frac{1}{P_{\min}}\right)} \le P_{\mathrm{e}}\bigl(\mathscr{C}^{(\mathsf{M},n)}\bigr) \le (\mathsf{M}-1)\, e^{-n \mathsf{D}^{(\text{DMC})}_{\min}(\mathscr{C}^{(\mathsf{M},n)})} \tag{42}$$

[14] We omit this proof, but instead refer to our generalization of the Plotkin Bound in Theorem 43 in Section 4.3.

*where* $\mathsf{D}_{\min}^{(\mathrm{DMC})}(\mathscr{C}^{(\mathsf{M},n)})$ *is the* minimum discrepancy *for a codebook* $\mathscr{C}^{(\mathsf{M},n)}$ *and where* $P_{\min}$ *denotes the smallest nonzero transition probability of the DMC (cf. [19, Sec. VI] and [21] for detailed explanations). Here we use a superscript "(DMC)" to indicate the channel to which the discrepancy refers.*

Note that these bounds are specific to a given code design (via $\mathsf{D}_{\min}^{(\mathrm{DMC})}$). Therefore, the upper bound is a generally valid upper bound on the optimal performance, while the lower bound may not bound the optimal performance from below unless we apply it to the optimal code or to a suboptimal code that achieves the optimal $\mathsf{D}_{\min}^{(\mathrm{DMC})}$.

### 3.2 PPV Bounds for the BEC

In [23], Polyanskiy, Poor, and Verdú present upper and lower bounds on the optimal average error probability for finite blocklength for general DMCs. For some special cases like the BSC or the BEC, these bounds can be expressed explicitly by closed-form formulas. The upper bound is based on *random coding.*

**Theorem 28 (PPV Upper Bound [23, Th. 36]).** *For the BEC with erasure probability $\delta$, if the codebook $\mathscr{C}^{(\mathsf{M},n)}$ is created at random based on a uniform distribution, the expected average error probability (averaged over all codewords and all codebooks) satisfies*

$$\begin{aligned}&\mathsf{E}\Big[P_{\mathrm{e}}\big(\mathscr{C}^{(\mathsf{M},n)}\big)\Big]\\&\quad = 1-\sum_{j=0}^{n}\binom{n}{j}(1-\delta)^{j}\delta^{n-j}\sum_{m=0}^{\mathsf{M}-1}\frac{1}{m+1}\binom{\mathsf{M}-1}{m}(2^{-j})^{m}(1-2^{-j})^{\mathsf{M}-1-m}.\end{aligned} \tag{43}$$

Note that there must exist a codebook whose average error probability achieves (43), so Theorem 28 provides a general achievable upper bound on the error probability, although we do not know the concrete code structure.

Polyanskiy, Poor, and Verdú also provide a new general converse for the average error probability, based on which a closed-form formula can be derived for the BEC.

**Theorem 29 (PPV Lower Bound [23, Th. 38]).** *For the BEC with erasure probability $\delta$, any codebook $\mathscr{C}^{(\mathsf{M},n)}$ satisfies*

$$P_{\mathrm{e}}\big(\mathscr{C}^{(\mathsf{M},n)}\big) \geq \sum_{e=\lfloor n-\log_2 \mathsf{M}\rfloor+1}^{n}\binom{n}{e}\delta^{e}(1-\delta)^{n-e}\left(1-\frac{2^{n-e}}{\mathsf{M}}\right). \tag{44}$$

Note that (44) was first derived based on an "*ad hoc*" (i.e., BEC specific) argument in [23]. It is then shown in [24] that the same result can also be obtained using the so-called *meta-converse* methodology.

## 4 Column-Wise Analysis of Codes

### 4.1 $r$-Wise Hamming Distance and $r$-Wise Hamming Match

The minimum Hamming distance is a well-known and widely used quality criterion of a code. Unfortunately, a design solely based on the minimum Hamming distance can be strictly suboptimal even for a very symmetric channel like the BSC and even for linear codes [19], [25].[15] In order to remedy this, we start by defining a slightly more general and more concise description of a code: the *pairwise Hamming distance vector*.

**Definition 30.** The *pairwise Hamming distance vector* $\mathbf{d}^{(\mathsf{M},n)}$ *of a code* $\mathscr{C}^{(\mathsf{M},n)}$ is defined as the length-$\bigl(\frac{1}{2}(\mathsf{M}-1)\mathsf{M}\bigr)$ vector containing as components the Hamming distances of all possible codeword pairs:

$$\mathbf{d}^{(\mathsf{M},n)} \triangleq \Bigl(d_{12}^{(n)},\;\; d_{13}^{(n)}, d_{23}^{(n)},\;\; d_{14}^{(n)}, d_{24}^{(n)}, d_{34}^{(n)}, \ldots, \\ d_{1\mathsf{M}}^{(n)}, d_{2\mathsf{M}}^{(n)}, \ldots, d_{(\mathsf{M}-1)\mathsf{M}}^{(n)}\Bigr) \tag{45}$$

with $d_{mm'}^{(n)} \triangleq d_{\mathrm{H}}(\mathbf{x}_m, \mathbf{x}_{m'})$, $1 \leq m < m' \leq \mathsf{M}$. We remind the reader of our convention to number the codewords according to rows in the codebook matrix, see (16).

The *minimum Hamming distance* $d_{\min}$ is then the minimum component of the pairwise Hamming distance vector $\mathbf{d}^{(\mathsf{M},n)}$.

Note that for this definition it is completely irrelevant whether the code is linear or not.

While the pairwise Hamming distance vector already contains more information about a particular code than simply the minimum Hamming distance, it is still not sufficient to describe the exact performance of a code. We will therefore next provide an extension of the pairwise Hamming distance: the so-called *$r$-wise Hamming distance of a code*. We will see that this generalization (in combination with the type vector $\mathbf{t}$) allows a precise formulation of the exact error probability of the code over a BEC.

**Definition 31** ($r$-Wise Hamming Distance and $r$-Wise Hamming Match)**.** For a given general codebook $\mathscr{C}^{(\mathsf{M},n)}$ and an arbitrary integer $2 \leq r \leq \mathsf{M}$, we fix some integers $1 \leq i_1 < i_2 < \cdots < i_r \leq \mathsf{M}$ and define the *$r$-wise Hamming match* $a_{i_1 i_2 \cdots i_r}\bigl(\mathscr{C}^{(\mathsf{M},n)}\bigr)$ to be the number of codebook columns $\mathbf{c}$ whose $i_1$th, $i_2$th, ..., $i_r$th coordinates are all identical:

$$a_{i_1 i_2 \cdots i_r}\bigl(\mathscr{C}^{(\mathsf{M},n)}\bigr) \triangleq \bigl|\bigl\{j \in \{1,\ldots,n\} \colon c_{j,i_1} = c_{j,i_2} = \cdots = c_{j,i_r}\bigr\}\bigr|, \\ 1 \leq i_1 < i_2 < \cdots < i_r \leq \mathsf{M}. \tag{46}$$

The *$r$-wise Hamming distance* $d_{i_1 i_2 \cdots i_r}\bigl(\mathscr{C}^{(\mathsf{M},n)}\bigr)$ is accordingly defined as

$$d_{i_1 i_2 \cdots i_r}\bigl(\mathscr{C}^{(\mathsf{M},n)}\bigr) \triangleq n - a_{i_1 i_r \cdots i_r}\bigl(\mathscr{C}^{(\mathsf{M},n)}\bigr), \qquad 1 \leq i_1 < i_2 < \cdots < i_r \leq \mathsf{M}. \tag{47}$$

It is straightforward to verify that the 2-wise Hamming distances according to Definition 31 are identical to the pairwise Hamming distances given in the pairwise Hamming distance vector (45).

The $r$-wise Hamming distances can be written elegantly with the help of the type vector:

$$d_{i_1 i_2 \cdots i_r}\bigl(\mathscr{C}_{\mathbf{t}}^{(\mathsf{M},n)}\bigr) = n - \sum_{\substack{j \in \mathcal{J} \text{ s.t.} \\ c_{j,i_1} = c_{j,i_2} = \cdots = c_{j,i_r}}} t_j, \qquad 1 \leq i_1 < i_2 < \cdots < i_r \leq \mathsf{M}. \tag{48}$$

Here $t_j$ denotes the $j$th component of the type vector $\mathbf{t}$ of length $\mathsf{J} = 2^{\mathsf{M}-1} - 1$, and $c_{j,i_\ell}$ is the $i_\ell$th component of the $j$th candidate column $\mathbf{c}_j^{(\mathsf{M})}$ as given in Definition 10, and $\mathcal{J} \triangleq \{1,\ldots,2^{\mathsf{M}-1}-1\} = \{1,\ldots,\mathsf{J}\}$ was defined in (20).

[15] This is in spite of the fact that the error probability performance of a BSC is completely specified by the Hamming distances between codewords and received vectors!

When the considered type-$\mathbf{t}$ code is unambiguous from the context, we will usually omit the explicit specification of the code and abbreviate (46) and (47) as $a^{(\mathsf{M},n)}_{i_1 i_2 \cdots i_r}$ and $d^{(\mathsf{M},n)}_{i_1 i_2 \cdots i_r}$ or, even shorter, as $a^{(\mathsf{M},n)}_{\mathcal{I}}$ and $d^{(\mathsf{M},n)}_{\mathcal{I}}$ for some given $\mathcal{I} = \{i_1, i_2, \ldots, i_r\}$. Note that there are $\binom{\mathsf{M}}{r}$ different choices of parameters $1 \le i_1 < i_2 < \cdots < i_r \le \mathsf{M}$, i.e., there are $\binom{\mathsf{M}}{r}$ different $r$-wise Hamming distances per code.

**Example 32.** For $\mathsf{M} = 4$ and $r = 3$, there are $\binom{\mathsf{M}}{r} = \binom{4}{3} = 4$ different 3-wise Hamming distances:

$$d^{(4,n)}_{123} = n - t_1, \quad d^{(4,n)}_{124} = n - t_2, \quad d^{(4,n)}_{134} = n - t_4, \quad d^{(4,n)}_{234} = n - t_7, \tag{49}$$

and there is only one 4-wise Hamming distance: $d^{(4,n)}_{1234} = n$. ♢

The definition of the $r$-wise Hamming distances leads to a natural extension of the minimum Hamming distance.

**Definition 33** (Minimum $r$-Wise Hamming Distance)**.** For a given $r \in \{2, \ldots, \mathsf{M}\}$, the *minimum $r$-wise Hamming distance $d_{\min;r}$ of a code $\mathscr{C}^{(\mathsf{M},n)}$* is defined as the minimum of all possible $r$-wise Hamming distances of this $(\mathsf{M}, n)$ code:

$$d_{\min;r}\bigl(\mathscr{C}^{(\mathsf{M},n)}\bigr) \triangleq \min_{\substack{\mathcal{I} \subseteq \{1,\ldots,\mathsf{M}\}:\\ |\mathcal{I}|=r}} d_{\mathcal{I}}\bigl(\mathscr{C}^{(\mathsf{M},n)}\bigr), \tag{50}$$

where the minimization is over all size-$r$ subsets $\mathcal{I} \subseteq \{1, \ldots, \mathsf{M}\}$.

Correspondingly, the *maximum $r$-wise Hamming match* $a_{\max;r}$ is defined as the maximum of all possible $r$-wise Hamming matches $a_{\mathcal{I}}\bigl(\mathscr{C}^{(\mathsf{M},n)}\bigr)$ and is given by

$$a_{\max;r}\bigl(\mathscr{C}^{(\mathsf{M},n)}\bigr) = n - d_{\min;r}\bigl(\mathscr{C}^{(\mathsf{M},n)}\bigr). \tag{51}$$

Recall that in traditional coding theory it is customary to specify a code with three parameters $(\mathsf{M}, n, d_{\mathrm{H,min}})$, where the third parameter specifies the minimum pairwise Hamming distance. We follow this tradition but replace the minimum pairwise Hamming distance by a vector containing all minimum $r$-wise Hamming distances for $r = 2, \ldots, \bar{\ell}$:

$$\mathbf{d}_{\min} \triangleq \bigl(d_{\min;2},\ d_{\min;3},\ \ldots,\ d_{\min;\bar{\ell}}\bigr). \tag{52}$$

The reason why we restrict ourselves to $r \le \bar{\ell}$ lies in the fact that for weak flip codes the minimum $r$-wise Hamming distance is only relevant for $2 \le r \le \bar{\ell}$; see the remark after Theorem 43 below.

**Example 34.** We continue with Example 25. The fair linear code with $k = 3$ and $n = 7$ given in (39) is an $(8, 7, \mathbf{d}_{\min})$ Hadamard linear code with $\mathbf{d}_{\min} = (4, 6, 6)$. Similarly, the fair linear code with $k = 3$ and $n = 35$ that is created by concatenating the codebook matrix (39) five times is an $\bigl(8, 35, (20, 30, 30)\bigr)$ Hadamard linear code.

Both codes are obviously not fair weak flip codes for $\mathsf{M} = 8$. Later in Theorem 45 we will show that the fair weak flip code with $\mathsf{M} = 8$ codewords is actually an $\bigl(8, 35, (20, 30, 34)\bigr)$ code. ♢

In [13], Wei defines the *sth generalized Hamming weight* of a $k$-dimensional linear code as the minimum support of any $s$-dimensional linear subcode, where the *support* is the number of codebit positions at which not all codewords are zero. Obviously, this definition is strongly restricted because firstly it is only defined for a *linear* code, and because secondly in general an arbitrarily picked subset of codewords of a linear code is not a linear subcode, i.e., Wei only considers a very much limited number of subsets of codewords taken from the given linear code. Nevertheless, it can be shown that if we pick $2^s$ codewords ($s \le k$) from a $k$-dimensional linear code in such a way that these $2^s$ codewords form a linear subcode, then the $s$th generalized Hamming weight is equal to the smallest $r$-wise Hamming distance among all $r$ satisfying $2^{s-1} < r \le 2^s$ [26], [27].

Following the classical definition of an *equidistant code* being a code whose pairwise Hamming distance between all codewords is the same, we extend this definition to the $r$-wise Hamming distance and define *$r$-wise equidistant codes.*

**Definition 35** ($r$-Wise Equidistant Codes)**.** For a given integer $2 \le r \le \mathsf{M}$, an $(\mathsf{M}, n)$ code $\mathscr{C}^{(\mathsf{M},n)}$ is called *$r$-wise equidistant* if all $r$-wise Hamming distances are equal, i.e., if for all choices of integers $1 \le i_1 < i_2 < \cdots < i_r \le \mathsf{M}$

$$d_{i_1 \cdots i_r}\left(\mathscr{C}^{(\mathsf{M},n)}\right) = \text{constant}. \tag{53}$$

We end this section with a relation between the $r$-wise Hamming distance and the type vector of a code. To that goal, we first state a property regarding the number of candidate columns with $r$ equal components.

**Lemma 36.** *For any integer $2 \le r \le \mathsf{M}$ and any choice $1 \le i_1 < i_2 < \cdots < i_r \le \mathsf{M}$, the cardinality of the index set*[16]

$$\mathcal{J}_{i_1 i_2 \cdots i_r} \triangleq \left\{ j \in \mathcal{J} : c_{j,i_1} = c_{j,i_2} = \cdots = c_{j,i_r} \right\} \tag{54}$$

*is equal to $2^{\mathsf{M}-r} - 1$. In other words, there are totally $2^{\mathsf{M}-r} - 1$ candidate columns in $\mathcal{C}^{(\mathsf{M})}$ that have identical components at the given positions $i_1, i_2, \ldots, i_r$.*

*Proof:* First, consider the case when $i_1 = 1$. Since the first position of each candidate column is always equal to zero, we only need to consider those $j \in \mathcal{J}$ such that $c_{j,i_1} = c_{j,i_2} = \cdots = c_{j,i_r} = 0$. There are in total $2^{\mathsf{M}-r}$ such columns, but we need to subtract 1 because we exclude the all-zero column.

Second, consider the case when $i_1 > 1$. Since the first position is fixed to zero, we ignore it. There are $2^{\mathsf{M}-1-r}$ columns with $c_{j,i_1} = c_{j,i_2} = \cdots = c_{j,i_r} = 0$ and the same number with $c_{j,i_1} = c_{j,i_2} = \cdots = c_{j,i_r} = 1$. Once again excluding the all-zero column, we have in total $2 \cdot 2^{\mathsf{M}-1-r} - 1$ possible columns. □

**Corollary 37.** *The $r$-wise Hamming distance $d_{1\,2\cdots r}\left(\mathscr{C}_{\mathbf{t}}^{(\mathsf{M},n)}\right)$ of the first $r$ codewords is given by*

$$d_{1\,2\cdots r}^{(\mathsf{M},n)} = \sum_{j=2^{\mathsf{M}-r}}^{\mathsf{J}} t_j. \tag{55}$$

*If every candidate column in $\mathcal{C}^{(\mathsf{M})}$ is used exactly once in $\mathscr{C}_{\mathbf{t}}^{(\mathsf{M},n)}$, i.e., $t_j = 1$ for $1 \le j \le \mathsf{J}$, then all $r$-wise Hamming distances $d_{i_1\cdots i_r}^{(\mathsf{M},n)}$ have an identical value:*

$$d_{i_1\cdots i_r}^{(\mathsf{M},n)} = 2^{\mathsf{M}-1} - 2^{\mathsf{M}-r}, \quad 1 \le i_1 < \cdots < i_r \le \mathsf{M}. \tag{56}$$

[16] Here again, $c_{j,i_\ell}$ denotes the $i_\ell$th component of the $j$th candidate column $\mathbf{c}_j^{(\mathsf{M})}$.

*Proof:* By the numbering system in Definition 10, together with Definition 31 and Lemma 36, we have

$$d_{12\cdots r}^{(\mathsf{M},n)} = n - \sum_{j\in\mathcal{J}_{1,\ldots,r}} t_j = n - \sum_{j=1}^{2^{\mathsf{M}-r}-1} t_j = \sum_{j=2^{\mathsf{M}-r}}^{\mathsf{J}} t_j. \tag{57}$$

If $t_1 = t_2 = \cdots = t_{\mathsf{J}} = 1$ (see (19)), we obtain again by Lemma 36 for arbitrary $1 \le i_1 < \cdots < i_r \le \mathsf{M}$,

$$d_{i_1\cdots i_r}^{(\mathsf{M},n)} = 2^{\mathsf{M}-1} - 1 - \sum_{j=1}^{2^{\mathsf{M}-r}-1} 1 = 2^{\mathsf{M}-1} - 2^{\mathsf{M}-r} = d_{12\cdots r}^{(\mathsf{M},n)}. \tag{58}$$

□

### 4.2 Characteristics of Weak Flip Codes

In this section, we concentrate on the analysis of the family of weak flip codes.

First, we pose the question which of the many powerful algebraic properties of linear codes are retained in weak flip codes.

**Theorem 38.** *Consider a weak flip code $\mathscr{C}_{\text{weak}}^{(\mathsf{M},n)}$ and fix some codeword $\mathbf{x}_m \in \mathscr{C}_{\text{weak}}^{(\mathsf{M},n)}$. If we add this codeword to all codewords in $\mathscr{C}_{\text{weak}}^{(\mathsf{M},n)}$, then the resulting code*

$$\tilde{\mathscr{C}}^{(\mathsf{M},n)} \triangleq \left\{\mathbf{x} \oplus \mathbf{x}_m : \mathbf{x} \in \mathscr{C}_{\text{weak}}^{(\mathsf{M},n)}\right\} \tag{59}$$

*is still a weak flip code; however, it is not necessarily the same one.*

*Proof:* Let $\mathscr{C}_{\text{weak}}^{(\mathsf{M},n)}$ be a weak flip code according to Definition 15. We have to prove that

$$\begin{pmatrix} \mathbf{x}_1 \\ \mathbf{x}_2 \\ \vdots \\ \mathbf{x}_{\mathsf{M}} \end{pmatrix} \oplus \begin{pmatrix} \mathbf{x}_m \\ \mathbf{x}_m \\ \vdots \\ \mathbf{x}_m \end{pmatrix} = \begin{pmatrix} \mathbf{x}_1 \oplus \mathbf{x}_m \\ \vdots \\ \mathbf{x}_m \oplus \mathbf{x}_m = \mathbf{0} \\ \vdots \\ \mathbf{x}_{\mathsf{M}} \oplus \mathbf{x}_m \end{pmatrix} \triangleq \tilde{\mathscr{C}}^{(\mathsf{M},n)} \tag{60}$$

is a weak flip code. Let $\mathbf{c}_j$, $1 \le j \le n$, denote the $j$th column vector of the code matrix of $\mathscr{C}_{\text{weak}}^{(\mathsf{M},n)}$. Then $\tilde{\mathscr{C}}^{(\mathsf{M},n)}$ has the column vectors

$$\tilde{\mathbf{c}}_j = \begin{cases} \mathbf{c}_j & \text{if } x_{m,j} = 0, \\ \bar{\mathbf{c}}_j & \text{if } x_{m,j} = 1. \end{cases} \tag{61}$$

Since $\mathbf{c}_j$ is a weak flip column, either $w_{\text{H}}(\mathbf{c}_j) = \lfloor \frac{\mathsf{M}}{2} \rfloor$ or $w_{\text{H}}(\mathbf{c}_j) = \lceil \frac{\mathsf{M}}{2} \rceil$, which implies that either $w_{\text{H}}(\bar{\mathbf{c}}_j) = \lceil \frac{\mathsf{M}}{2} \rceil$ or $w_{\text{H}}(\bar{\mathbf{c}}_j) = \lfloor \frac{\mathsf{M}}{2} \rfloor$. Now it only remains to interchange the first codeword of $\tilde{\mathscr{C}}^{(\mathsf{M},n)}$ and the all-zero codeword in the $m$th row in $\tilde{\mathscr{C}}^{(\mathsf{M},n)}$ (which is always possible, see Remark 8). As a result, $\tilde{\mathscr{C}}^{(\mathsf{M},n)}$ is also a weak flip code. □

Theorem 38 is a beautiful property of weak flip codes; however, it still represents a considerable weakening of the powerful property of linear codes given in Proposition 22. This can be fixed by considering the subfamily of *fair* weak flip codes.

**Theorem 39 (Quasi-Linear Codes).** *Let* $\mathscr{C}_{\text{fair}}^{(\mathsf{M},n)}$ *be a fair weak flip code and let* $\mathbf{x}_m \in \mathscr{C}_{\text{fair}}^{(\mathsf{M},n)}$ *be given. Then the code*

$$\tilde{\mathscr{C}}^{(\mathsf{M},n)} \triangleq \left\{\mathbf{x} \oplus \mathbf{x}_m : \mathbf{x} \in \mathscr{C}_{\text{fair}}^{(\mathsf{M},n)}\right\} \tag{62}$$

*is equivalent to* $\mathscr{C}_{\text{fair}}^{(\mathsf{M},n)}$.

*Proof:* We divide the weak flip candidate columns in $\mathcal{C}_{\text{weak}}^{(\mathsf{M})}$ into two subfamilies: one subfamily consists of the columns with the $m$th component being zero, and the columns in the other subfamily have their $m$th component equal to one. Next we add the $m$th codeword to the codewords in $\mathscr{C}_{\text{fair}}^{(\mathsf{M},n)}$ and then interchange the first and $m$th components of each column in the code matrix of $\mathscr{C}_{\text{fair}}^{(\mathsf{M},n)}$ to form a new code $\tilde{\mathscr{C}}^{(\mathsf{M},n)}$. It is apparent that the columns in the first subfamily are unchanged by such code-addition-and-interchanging manipulation. However, when $\mathsf{M}$ is odd, the weights of columns in the second subfamily change either from $\underline{\ell}$ to $\bar{\ell}$, or from $\bar{\ell}$ to $\underline{\ell}$, while these weights stay the same when $\mathsf{M}$ is even. As a result, after such code-addition-and-interchanging manipulation, the columns belonging to the second subfamily remain distinct weak flip columns and are still contained in the second subfamily (since their $m$th components are still equal to one). Thus, all the weak flip columns remain to be used equally in $\tilde{\mathscr{C}}^{(\mathsf{M},n)}$, showing that $\tilde{\mathscr{C}}^{(\mathsf{M},n)}$ is fair. □

Comparing Theorem 39 with Proposition 22 and recalling Proposition 21 and the discussion after it, we realize that the family of fair weak flip codes is a considerable enlargement of the family of linear codes.

The following corollary is a direct consequence of Lemma 36.

**Corollary 40.** *For any integer* $2 \le r \le \mathsf{M}$*, the* $r$*-wise Hamming distances* $d_{i_1\cdots i_r}^{(\mathsf{M},n)}$ *of a fair weak flip code* $\mathscr{C}_{\text{fair}}^{(\mathsf{M},n)}$ *for any choice* $1 \le i_1 < i_2 < \cdots < i_r \le \mathsf{M}$*, are all identical and are given by*

$$d_{i_1\cdots i_r}^{(\mathsf{M},n)} = \frac{n}{\mathsf{L}}\, d_{i_1\cdots i_r}^{(\mathsf{M},\mathsf{L})} = \frac{n}{\mathsf{L}}\left[\mathsf{L} - \binom{2\bar{\ell}-r}{\bar{\ell}}\right]. \tag{63}$$

*Proof:* By definition of a fair weak flip code, we observe that the $r$-wise Hamming distance of arbitrary $r$ codewords is a fixed integer multiple (i.e., $n/\mathsf{L}$) of the $r$-wise Hamming distance $d_{i_1\cdots i_r}^{(\mathsf{M},\mathsf{L})}$ of a fair weak flip code of blocklength $n = \mathsf{L}$.

We apply the proof idea of Lemma 36. When $\mathsf{M} = 2\bar{\ell} - 1$ is odd, first consider the case of $i_1 = 1$. Since the first position of each weak flip column is always equal to zero, the number of weak flip columns with weight $\underline{\ell}$ such that $c_{j,i_1} = c_{j,i_2} = \cdots = c_{j,i_r} = 0$ equals $\binom{\mathsf{M}-r}{\bar{\ell}-1}$, and the number of weak flip columns with weight $\bar{\ell}$ is $\binom{\mathsf{M}-r}{\bar{\ell}}$. In total, we have the $r$-wise Hamming match

$$a_{1\,i_2\cdots,i_r}^{(\mathsf{M},n)} = \frac{n}{\mathsf{L}}\left[\binom{\mathsf{M}-r}{\bar{\ell}-1} + \binom{\mathsf{M}-r}{\bar{\ell}}\right] = \frac{n}{\mathsf{L}}\binom{2\bar{\ell}-r}{\bar{\ell}}, \tag{64}$$

where we take $\mathsf{M} = 2\bar{\ell} - 1$ in the last equality.

Second, consider the case when $i_1 > 1$. Since the first position is fixed to zero, we ignore it. There are $\binom{\mathsf{M}-r-1}{\bar{\ell}-1}$ columns with weight $\bar{\ell} - 1$ such that $c_{j,i_1} = c_{j,i_2} = \cdots = c_{j,i_r} = 0$, and there are $\binom{\mathsf{M}-r-1}{\bar{\ell}}$ columns with weight $\bar{\ell}$ such that $c_{j,i_1} = c_{j,i_2} = \cdots = c_{j,i_r} = 0$. Similarly, there are $\binom{\mathsf{M}-r-1}{\bar{\ell}-1-r}$ columns with weight $\bar{\ell} - 1$ such that $c_{j,i_1} = c_{j,i_2} = \cdots = c_{j,i_r} = 1$, and there are $\binom{\mathsf{M}-r-1}{\bar{\ell}-r}$ columns with weight $\bar{\ell}$ such that $c_{j,i_1} = c_{j,i_2} = \cdots = c_{j,i_r} = 1$. In total we have the $r$-wise Hamming match

$$a^{(\mathsf{M},n)}_{i_1 i_2 \cdots i_r} = \frac{n}{\mathsf{L}}\left[\binom{\mathsf{M}-r-1}{\bar{\ell}-1} + \binom{\mathsf{M}-r-1}{\bar{\ell}} + \binom{\mathsf{M}-r-1}{\bar{\ell}-1-r} + \binom{\mathsf{M}-r-1}{\bar{\ell}-r}\right] \tag{65}$$

$$= \frac{n}{\mathsf{L}}\left[\binom{\mathsf{M}-r}{\bar{\ell}} + \binom{\mathsf{M}-r}{\bar{\ell}-r}\right] \tag{66}$$

$$= \frac{n}{\mathsf{L}}\binom{2\bar{\ell}-r}{\bar{\ell}}, \tag{67}$$

where in the last equality we use $\mathsf{M} = 2\bar{\ell} - 1$.

In a similar way, given $\mathsf{M} = 2\bar{\ell}$ is even, we obtain

$$a^{(\mathsf{M},n)}_{i_1 \cdots i_r} = \begin{cases} \frac{n}{\mathsf{L}}\binom{\mathsf{M}-r}{\bar{\ell}} & \text{if } i_1 = 1, \\ \frac{n}{\mathsf{L}}\left[\binom{\mathsf{M}-r-1}{\bar{\ell}} + \binom{\mathsf{M}-r-1}{\bar{\ell}-r}\right] & \text{if } i_1 > 1. \end{cases} \tag{68}$$

The proof is completed by combining all possible cases using that $d^{(\mathsf{M},n)}_{i_1 \cdots i_r} = n - a^{(\mathsf{M},n)}_{i_1 \cdots i_r}$. □

Recall that for a given choice of $r$ column positions $1 \le i_1 < i_2 < \cdots < i_r \le \mathsf{M}$, the $r$-wise Hamming match counts how many columns exist in the codebook matrix that have identical entries in these $r$ positions. Now we would like to look at this the other way around: for a fixed candidate column, we would like to count how many different choices of $r$ positions $1 \le i_1 < i_2 < \cdots < i_r \le \mathsf{M}$ exist such that all these positions have identical entries.

Since a candidate column with Hamming weight equal to $h$ has $h$ components of value 1 and $\mathsf{M} - h$ components of value 0, it is easy to see that the following lemma always holds.

**Lemma 41.** *For given an integer $2 \le r \le \bar{\ell}$ and an arbitrary candidate column $\mathbf{c}_j$, $j = 1, \ldots, \mathsf{J}$, the cardinality of the set*

$$\bigl\{(i_1, i_2, \ldots, i_r) : 1 \le i_1 < i_2 < \cdots < i_r \le \mathsf{M}, \quad c_{j,i_1} = c_{j,i_2} = \cdots = c_{j,i_r}\bigr\} \tag{69}$$

*is equal to $\binom{h}{r} + \binom{\mathsf{M}-h}{r}$, where $h = w_{\mathrm{H}}(\mathbf{c}_j)$.*

Finally, we illustrate an example of Lemma 41.

**Example 42.** For $\mathsf{M} = 4$, the pairwise Hamming distance vector of a weak flip code of type $\mathbf{t}_{\text{weak}}$ can be listed as follows:

$$\mathbf{d}^{(4,n)} = (n - t_3, n - t_5, n - t_6, n - t_6, n - t_5, n - t_3), \tag{70}$$

i.e., each $t_{j_w}$, $w = 1, 2, 3$, shows up exactly twice. ◊

### 4.3 Generalized Plotkin Bound for the $r$-wise Hamming Distance

The $r$-wise Hamming distance (together with the type vector $\mathbf{t}$) plays an important role in the closed-form expression of the average error probability for an arbitrary code $\mathscr{C}^{(\mathsf{M},n)}_{\mathbf{t}}$ over a BEC. It is therefore interesting to find some bounds on the $r$-wise Hamming distance. We start with a generalization of the Plotkin bound for the minimum pairwise Hamming distance to the situation of the minimum $r$-wise Hamming distance.

**Theorem 43 (Plotkin Bound for the Minimum $r$-wise Hamming Distance).** *The minimum $r$-wise Hamming distance with $2 \le r \le \mathsf{M}$ of an $(\mathsf{M}, n)$ binary code satisfies*

$$d_{\min;r}\bigl(\mathscr{C}^{(\mathsf{M},n)}\bigr) \le \begin{cases} n\left(1 - \frac{\binom{\bar{\ell}-1}{r-1}}{\binom{2\bar{\ell}-1}{r-1}}\right) & \text{if } 2 \le r \le \bar{\ell}, \\ n & \text{if } \bar{\ell} < r \le \mathsf{M}. \end{cases} \tag{71}$$

*Proof:* The bound for $r > \bar{\ell}$ is trivial and therefore needs no proof. We focus on $2 \le r \le \bar{\ell}$. Note that because there are $\mathsf{M}(\mathsf{M}-1)\cdots(\mathsf{M}-r+1)$ different choices for $1 \le i_1 < \cdots < i_r \le \mathsf{M}$, we have

$$\sum_{\substack{\mathcal{I} \subseteq \{1,\ldots,\mathsf{M}\}: \\ |\mathcal{I}|=r}} a_{\mathcal{I}}\bigl(\mathscr{C}^{(\mathsf{M},n)}\bigr) \le \mathsf{M}(\mathsf{M}-1)\cdots(\mathsf{M}-r+1) \cdot a_{\max;r}\bigl(\mathscr{C}^{(\mathsf{M},n)}\bigr). \tag{72}$$

On the other hand, if we look at the codebook matrix $\mathscr{C}^{(\mathsf{M},n)}$ from a column-wise point of view and define $h_j$ to be the number of zeros in the $j$th column (and hence $\mathsf{M} - h_j$ to be the number of ones in the $j$th column), we see that the $j$th column contributes $h_j(h_j-1)\cdots(h_j-r+1)$ possible choices of picking $r$ different components that all are zero and $(\mathsf{M}-h_j)(\mathsf{M}-h_j-1)\cdots(\mathsf{M}-h_j-r+1)$ choices of picking $r$ different components that all are one. Hence,[17]

$$\begin{aligned} \sum_{\substack{\mathcal{I} \subseteq \{1,\ldots,\mathsf{M}\}: \\ |\mathcal{I}|=r}} a_{\mathcal{I}}\bigl(\mathscr{C}^{(\mathsf{M},n)}\bigr) &= \sum_{j=1}^{n} \Bigl[h_j(h_j-1)\cdots(h_j-r+1) \\ &\qquad + (\mathsf{M}-h_j)(\mathsf{M}-h_j-1)\cdots(\mathsf{M}-h_j-r+1)\Bigr] \qquad (75) \\ &\ge n\Bigl[\underline{\ell}(\underline{\ell}-1)\cdots(\underline{\ell}-r+1) + \bar{\ell}(\bar{\ell}-1)\cdots(\bar{\ell}-r+1)\Bigr], \qquad (76) \end{aligned}$$

where the lower bound is achieved if $h_j = \bar{\ell}$ or $\underline{\ell}$ for all $j = 1, \ldots, n$, i.e., if the columns are weak flip columns. Note that when $r = \bar{\ell}$ and $\mathsf{M}$ is odd, the first term in the bracket in (76) is zero because $(\underline{\ell} - r + 1) = (\underline{\ell} - \bar{\ell} + 1) = 0$.

---

[17] Under $r \le h_j \le \mathsf{M} - h_j$, (75) can be lower bounded as follows:

$$\begin{aligned} &h_j(h_j-1)\cdots(h_j-r+1) + (\mathsf{M}-h_j)(\mathsf{M}-h_j-1)\cdots(\mathsf{M}-h_j-r+1) \\ &= r!\left[\binom{h_j}{r} + \binom{M-h_j}{r}\right] \ge r!\left[\binom{h_j+1}{r} + \binom{M-h_j-1}{r}\right] \ge \cdots \ge r!\left[\binom{\underline{\ell}}{r} + \binom{\bar{\ell}}{r}\right], \end{aligned} \tag{73}$$

where the first inequality holds as long as $\mathsf{M} - h_j - 1 \ge h_j$ because

$$\left[\binom{h_j}{r} + \binom{M-h_j}{r}\right] - \left[\binom{h_j+1}{r} + \binom{M-h_j-1}{r}\right] = \binom{\mathsf{M}-h_j-1}{r-1} - \binom{h_j}{r-1} \ge 0 \tag{74}$$

and we can continue the process of adding one to the top number in the first binomial coefficient and meanwhile subtracting one from the top number in the second binomial coefficient until the last inequality in (73) is reached. The same argument can be used to validate (76) under $r \le \mathsf{M} - h_j \le h_j$. In the special case that $h_j < r \le \mathsf{M} - h_j$ (or $\mathsf{M} - h_j < r \le h_j$), which occurs definitely when $r = \bar{\ell}$ and $\mathsf{M}$ odd, (75) should be refined to

$$\begin{aligned} &\max\{h_j(h_j-1)\cdots(h_j-r+1), 0\} + \max\{(\mathsf{M}-h_j)(\mathsf{M}-h_j-1)\cdots(\mathsf{M}-h_j-r+1), 0\} \\ &\ge \max\{(h_j+1)(h_j)\cdots(h_j-r+2), 0\} + \max\{(\mathsf{M}-h_j-1)(\mathsf{M}-h_j-2)\cdots(\mathsf{M}-h_j-r), 0\} \\ &\ge \max\{(h_j+2)(h_j+1)\cdots(h_j-r+3), 0\} + \max\{(\mathsf{M}-h_j-2)(\mathsf{M}-h_j-3)\cdots(\mathsf{M}-h_j-r-1), 0\} \\ &\ge \cdots \end{aligned}$$

for which the process can be repeated $(r - h_j)$ times to reach the case considered in (73); hence (76) still holds.

Combining (72) and (76) (and separately calculating the cases where $\mathsf{M}$ is even or odd), we obtain

$$a_{\max;r}\left(\mathscr{C}^{(\mathsf{M},n)}\right) \geq \begin{cases} n\,\frac{2\cdot\bar{\ell}(\bar{\ell}-1)(\bar{\ell}-2)\cdots(\bar{\ell}-r+1)}{(2\bar{\ell})(2\bar{\ell}-1)(2\bar{\ell}-2)\cdots(2\bar{\ell}-r+1)} & \text{if } \mathsf{M}=2\bar{\ell}, \\ n\,\frac{\bar{\ell}(\bar{\ell}-1)(\bar{\ell}-2)\cdots(\bar{\ell}-r+1)+(\bar{\ell}-1)(\bar{\ell}-2)\cdots(\bar{\ell}-r)}{(2\bar{\ell}-1)(2\bar{\ell}-2)\cdots(2\bar{\ell}-r)} & \text{if } \mathsf{M}=2\bar{\ell}-1 \end{cases} \tag{77}$$

$$= n\,\frac{\binom{\bar{\ell}-1}{r-1}}{\binom{2\bar{\ell}-1}{r-1}}. \tag{78}$$

□

The above theorem only provides absorbing bounds to the $r$-wise Hamming distance for $2 \leq r \leq \bar{\ell}$, while further increasing the parameter $r$ only renders trivially $d_{\min;r} \leq n$. Since the minimum $r$-wise Hamming distance of a weak flip code for $r > \bar{\ell}$ is always equal to this trivial bound $n$ and therefore is irrelevant for the exact error performance, the vector (52) contains the minimum $r$-wise Hamming distances for $2 \leq r \leq \bar{\ell}$ only.

It is well-known that Hadamard codes achieve the Plotkin bound (Lemma 26) with equality, i.e., they achieve the largest minimum pairwise Hamming distance (or equivalently, the largest minimum 2-wise Hamming distance) [12, Ch. 2]. Moreover, Hadamard codes are also (pairwise) equidistant.[18] In the following we will investigate generalizations of these two properties for weak flip codes. We will show the following:

1. If a weak flip code (of a certain blocklength $n$) is $r$-wise equidistant, then it is also $s$-wise equidistant for all $s = 2, \ldots, r-1$.

2. If in addition to be $r$-wise equidistant, it also achieves the $r$-wise Plotkin bound (Theorem 43), then it also achieves the $s$-wise Plotkin bound for all $s = 2, \ldots, r-1$.

3. Fair weak flip codes are $r$-wise equidistant and achieve the $r$-wise Plotkin bound for all $2 \leq r \leq \mathsf{M}$.

The proof will make use of *$s$-designs* [28] from combinatorial design theory:

**Definition 44** ( [28, Ch. 9])**.** Let $v$, $\kappa$, $\lambda_s$, and $s$ be positive integers such that $v > \kappa \geq s$. An $s$-$(v, \kappa, \lambda_s)$ *design* or simply *$s$-design* is a pair $(\mathcal{X}, \mathscr{B})$, where $\mathcal{X}$ is a set of size $v$ and $\mathscr{B}$ is a collection of subsets of $\mathcal{X}$ (called *blocks*), such that the following properties are satisfied:

1. each block $\mathcal{B} \in \mathscr{B}$ contains exactly $\kappa$ points, and

2. every set of $s$ distinct points is contained in exactly $\lambda_s$ blocks.

We now claim that some specific weak flip codes (for an arbitrary $\mathsf{M}$ and for certain blocklengths) can be seen as $r$-designs with $2 \leq r \leq \bar{\ell}$ and achieve the generalized Plotkin upper bound (71) with equality (again, it is trivial to see that their $d_{\mathrm{min;r}}$ for $r > \bar{\ell}$ are equal to $n$).

[18]Note that the two properties of a code being equidistant and a code achieving the Plotkin bound do not imply each other. There exist Plotkin-bound achieving codes that are not equidistant, and there also exist equidistant codes that do not achieve the Plotkin bound.

**Theorem 45.** *Fix some* $\mathsf{M}$*, a blocklength* $n$ *with* $n \bmod \mathsf{L} = 0$*, and some* $2 \leq r \leq \bar{\ell}$*. Then if a weak flip code is* $r$*-wise equidistant, then it is also* $s$*-wise equidistant for all* $2 \leq s < r$*. Moreover, if this* $r$*-wise equidistant weak flip code* $\mathscr{C}^{(\mathsf{M},n)}_{\text{equidist}}$ *also achieves the generalized Plotkin bound (and hence achieves the largest minimum* $r$*-wise Hamming distance), i.e., it satisfies*

$$d_{\min;r}\left(\mathscr{C}^{(\mathsf{M},n)}_{\text{equidist}}\right) = n\left(1 - \frac{\binom{\bar{\ell}-1}{r-1}}{\binom{2\bar{\ell}-1}{r-1}}\right), \tag{79}$$

*then* $\mathscr{C}^{(\mathsf{M},n)}_{\text{equidist}}$ *must also achieve the largest minimum* $s$*-wise Hamming distances for all* $2 \leq s < r$*.*

*Proof:* We start by explaining how we connect the $r$-wise Hamming distance with $2 \leq r \leq \bar{\ell}$ of an $r$-wise equidistant weak flip code to the $s$-$(v, \kappa, \lambda_s)$ design. Consider an $r$-wise equidistant weak flip code with a certain blocklength $n$. Let $\mathcal{M} \triangleq \{1, 2, \ldots, \mathsf{M}\}$. Denote by $\mathscr{B}$ the collection containing all $\bar{\ell}$-size subsets $\mathcal{B} \triangleq \{i_1, i_2, \ldots, i_{\bar{\ell}}\} \subseteq \mathcal{M}$ such that $c_{j,i_1} = c_{j,i_2} = \cdots = c_{j,i_{\bar{\ell}}}$, $1 \leq j \leq n$. It can then be verified from the definition of an $r$-wise equidistant weak flip code that this completes the construction of an $r$-$(\mathsf{M}, \bar{\ell}, \lambda_r)$ design, where $\lambda_r$ is by definition equal to $n - d^{(\mathsf{M},n)}_{\mathcal{I}}$ with $\mathcal{I}$ being any size-$r$ subset of $\mathcal{M}$.

Using a fundamental theorem in combinatorial design theory [28, Thm. 9.4], we next infer that an $r$-$(\mathsf{M}, \bar{\ell}, \lambda_r)$ design is also an $s$-$(\mathsf{M}, \bar{\ell}, \lambda_s)$ design with $2 \leq s < r$ and

$$\lambda_s = \lambda_r \frac{\binom{\mathsf{M}-s}{r-s}}{\binom{\bar{\ell}-s}{r-s}}. \tag{80}$$

Since an $s$-$(\mathsf{M}, \bar{\ell}, \lambda_s)$ design corresponds to an $s$-wise equidistant weak flip code, this proves the first statement.

If we additionally assume that the parameter $\lambda_r$ is equal to the maximum $r$-wise Hamming match $a_{\max;r}$ satisfying (79), we then obtain for $\mathsf{M} = 2\bar{\ell}$:

$$\begin{aligned}
a_{\max;s} &= a_{\max;r} \frac{\binom{\mathsf{M}-s}{r-s}}{\binom{\bar{\ell}-s}{r-s}} && (81)\\
&= n \frac{\binom{\bar{\ell}-1}{r-1}}{\binom{2\bar{\ell}-1}{r-1}} \frac{\binom{2\bar{\ell}-s}{r-s}}{\binom{\bar{\ell}-s}{r-s}} && (82)\\
&= n \frac{\frac{(\bar{\ell}-1)!}{(\bar{\ell}-r)!\,(r-1)!}}{\frac{(2\bar{\ell}-1)!}{(2\bar{\ell}-r)!\,(r-1)!}} \frac{\frac{(2\bar{\ell}-s)!}{(r-s)!\,(2\bar{\ell}-r)!}}{\frac{(\bar{\ell}-s)!}{(\bar{\ell}-r)!\,(r-s)!}} && (83)\\
&= n \frac{\frac{(\bar{\ell}-1)!}{(\bar{\ell}-s)!\,(s-1)!}}{\frac{(2\bar{\ell}-1)!}{(2\bar{\ell}-s)!\,(s-1)!}} && (84)\\
&= n \frac{\binom{\bar{\ell}-1}{s-1}}{\binom{2\bar{\ell}-1}{s-1}}. && (85)
\end{aligned}$$

We thus confirm that $\mathscr{C}^{(\mathsf{M},n)}_{\text{equidist}}$ also meets the smallest maximum $s$-wise Hamming matches (i.e., the largest minimum $s$-wise Hamming distances) for $2 \leq s < r$.

In the case of $\mathsf{M} = 2\bar{\ell} - 1$, the definition of weak flip codes indicates that all codewords of $\mathscr{C}_{\text{equidist}}^{(2\bar{\ell}-1,n)}$ are contained in $\mathscr{C}_{\text{equidist}}^{(2\bar{\ell},n)}$. Hence,

$$d_{\min;r}\left(\mathscr{C}_{\text{equidist}}^{(2\bar{\ell}-1,n)}\right) \geq d_{\min;r}\left(\mathscr{C}_{\text{equidist}}^{(2\bar{\ell},n)}\right) = n - n\frac{\binom{\bar{\ell}-1}{r-1}}{\binom{2\bar{\ell}-1}{r-1}}, \tag{86}$$

which again achieves the Plotkin upper bound for $r$-wise Hamming distances in Theorem 43. □

The following corollary follows directly from Theorem 45 and Corollary 40.

**Corollary 46.** *The fair weak flip code $\mathscr{C}_{\text{fair}}^{(\mathsf{M},n)}$ achieves the largest minimum $r$-wise Hamming distance for all $2 \leq r \leq \bar{\ell}$ among all $(\mathsf{M}, n)$ codes.*

*Proof:* The proof is completed by observing that the smallest maximum $\bar{\ell}$-wise Hamming matches of (71) is equal to

$$n\,\frac{\binom{\bar{\ell}-1}{\bar{\ell}-1}}{\binom{2\bar{\ell}-1}{\bar{\ell}-1}} = n\,\frac{1}{\mathsf{L}}, \tag{87}$$

which, according to Corollary 40 with $r$ there replaced by $\bar{\ell}$, is achieved by $\mathscr{C}_{\text{fair}}^{(\mathsf{M},n)}$. □

We make the following remark to Corollary 46: The fair *linear* code always meets the Plotkin bound for the 2-wise Hamming distance; however, in contrast to the fair weak flip code $\mathscr{C}_{\text{fair}}^{(\mathsf{M},n)}$, it does not necessarily meet the Plotkin bound for $r$-wise Hamming distances for $r > 2$. This gives rise to our suspicion that a fair linear code may perform strictly worse than the optimal fair weak flip code even if it is the best linear code. Proper evidence for this claim will be given in Section 5.7.

# 5 Performance Analysis of the BEC

In Section 2.3 we have shown that any codebook can be described by the type vector $\mathbf{t}$. Therefore the minimization of the average error probability among all possible codebooks turns into an optimization problem on the discrete vector $\mathbf{t}$, subject to the condition that $\sum_{j=1}^{\mathsf{J}} t_j = n$. Consequently, the $r$-wise Hamming distance and the properties of the type vector play an important role in our analysis.

## 5.1 Exact Average Error Probability of a Code with an Arbitrary Number of Codewords $\mathsf{M}$

We firstly derive a useful result that gives the exact average error probability as a function of the type vector $\mathbf{t}$.

**Lemma 47 (Inclusion–Exclusion Principle in Probability Theory [29]).** *Let $\mathcal{A}_1, \mathcal{A}_2, \ldots, \mathcal{A}_{\mathsf{M}}$ be $\mathsf{M}$ (not necessarily independent) events in a probability space. The inclusion–exclusion principle states that*

$$\Pr\left(\bigcup_{m=1}^{\mathsf{M}} \mathcal{A}_m\right) = \sum_{r=1}^{\mathsf{M}} (-1)^{r-1} \sum_{\substack{\mathcal{I}\subseteq\{1,2,\ldots,\mathsf{M}\}:\\ |\mathcal{I}|=r}} \Pr\left(\bigcap_{i\in\mathcal{I}} \mathcal{A}_i\right). \tag{88}$$

We will next apply the idea of the inclusion–exclusion principle to the closed decoding regions given in Definition 3. To simplify our notation, we define the following shorthands:

$$\Pr\Bigl(\overline{\mathcal{D}}_m^{(\mathsf{M},n)}\Bigm|\mathbf{x}_m\Bigr) \triangleq \sum_{\mathbf{y}\in\overline{\mathcal{D}}_m^{(\mathsf{M},n)}} P_{\mathbf{Y}|\mathbf{X}}(\mathbf{y}|\mathbf{x}_m), \tag{89}$$

$$\Pr\Biggl(\bigcap_{i\in\mathcal{I}}\overline{\mathcal{D}}_i^{(\mathsf{M},n)}\Biggm|\mathbf{x}_\ell\Biggr) \triangleq \sum_{\mathbf{y}\in\bigcap_{i\in\mathcal{I}}\overline{\mathcal{D}}_i^{(\mathsf{M},n)}} P_{\mathbf{Y}|\mathbf{X}}(\mathbf{y}|\mathbf{x}_{\ell,\ell\in\mathcal{I}}), \quad \mathcal{I}\subseteq\mathcal{M}, \tag{90}$$

where for every $\mathbf{y}$ in $\bigcap_{i\in\mathcal{I}=\{i_1,i_2,\ldots,i_r\}}\overline{\mathcal{D}}_i^{(\mathsf{M},n)}$, we note according to Definition 3 that

$$\max_{1\le m'\le\mathsf{M}} P_{\mathbf{Y}|\mathbf{X}}(\mathbf{y}|\mathbf{x}_{m'}) = P_{\mathbf{Y}|\mathbf{X}}(\mathbf{y}|\mathbf{x}_{i_1}) = P_{\mathbf{Y}|\mathbf{X}}(\mathbf{y}|\mathbf{x}_{i_2}) = \cdots = P_{\mathbf{Y}|\mathbf{X}}(\mathbf{y}|\mathbf{x}_{i_r}), \tag{91}$$

and hence the exact choice of $\ell$ is irrelevant in (90).

**Theorem 48.** *Consider an $(\mathsf{M},n)$ coding scheme with its corresponding closed ML decoding regions $\overline{\mathcal{D}}_m$ as given in Definition 3, where we drop the superscript "$(\mathsf{M},n)$" for notational convenience. Defining*

$$\mathcal{D}_m \triangleq \overline{\mathcal{D}}_m \setminus \Biggl(\overline{\mathcal{D}}_m \cap \Biggl(\bigcup_{i\in\{1,\ldots,m-1\}}\overline{\mathcal{D}}_i\Biggr)\Biggr) \tag{92}$$

*we have*

$$\Pr(\mathcal{D}_m\,|\,\mathbf{x}_m) = \Pr\bigl(\overline{\mathcal{D}}_m\,\big|\,\mathbf{x}_m\bigr) - \sum_{r=1}^{m-1}(-1)^{r-1}\sum_{\substack{\mathcal{I}\subseteq\{1,\ldots,m-1\}:\\|\mathcal{I}|=r}}\Pr\Biggl(\bigcap_{i\in\mathcal{I}}\bigl(\overline{\mathcal{D}}_i\cap\overline{\mathcal{D}}_m\bigr)\Biggm|\mathbf{x}_m\Biggr) \tag{93}$$

*and the exact average success probability can be expressed as*

$$P_{\mathrm{c}}\bigl(\mathscr{C}^{(\mathsf{M},n)}\bigr) = \frac{1}{\mathsf{M}}\sum_{r=1}^{\mathsf{M}}(-1)^{r-1}\sum_{\substack{\mathcal{I}\subseteq\{1,\ldots,\mathsf{M}\}:\\|\mathcal{I}|=r}}\Pr\Biggl(\bigcap_{i\in\mathcal{I}}\overline{\mathcal{D}}_i\Biggm|\mathbf{x}_{\ell,\ell\in\mathcal{I}}\Biggr). \tag{94}$$

*Proof:* By Definition 3, a possible choice of ML decoding regions is given as follows:

$$\begin{aligned}
\mathcal{D}_1 &\triangleq \overline{\mathcal{D}}_1, && (95)\\
\mathcal{D}_2 &\triangleq \overline{\mathcal{D}}_2\setminus\overline{\mathcal{D}}_1 && (96)\\
&= \overline{\mathcal{D}}_2\setminus\bigl(\overline{\mathcal{D}}_2\cap\overline{\mathcal{D}}_1\bigr), && (97)\\
\mathcal{D}_3 &\triangleq \overline{\mathcal{D}}_3\setminus\bigl(\overline{\mathcal{D}}_1\cup\overline{\mathcal{D}}_2\bigr) && (98)\\
&= \overline{\mathcal{D}}_3\setminus\bigl(\overline{\mathcal{D}}_3\cap\bigl(\overline{\mathcal{D}}_1\cup\overline{\mathcal{D}}_2\bigr)\bigr), && (99)\\
&\ \ \vdots
\end{aligned}$$

i.e., we obtain (92). We rewrite

$$\overline{\mathcal{D}}_m \cap \Biggl(\bigcup_{i\in\{1,\ldots,m-1\}}\overline{\mathcal{D}}_i\Biggr) = \bigcup_{i\in\{1,\ldots,m-1\}}\bigl(\overline{\mathcal{D}}_m\cap\overline{\mathcal{D}}_i\bigr) \tag{100}$$

and use Lemma 47 to obtain

$$\Pr(\mathcal{D}_m \,|\, \mathbf{x}_m) = \Pr\left(\overline{\mathcal{D}}_m \setminus \left(\bigcup_{i\in\{1,\ldots,m-1\}} (\overline{\mathcal{D}}_m \cap \overline{\mathcal{D}}_i)\right) \middle|\, \mathbf{x}_m\right) \tag{101}$$

$$= \Pr\bigl(\overline{\mathcal{D}}_m \,\big|\, \mathbf{x}_m\bigr) - \Pr\left(\bigcup_{i\in\{1,\ldots,m-1\}} (\overline{\mathcal{D}}_m \cap \overline{\mathcal{D}}_i) \,\middle|\, \mathbf{x}_m\right) \tag{102}$$

$$= \Pr\bigl(\overline{\mathcal{D}}_m \,\big|\, \mathbf{x}_m\bigr) - \sum_{r=1}^{m-1} (-1)^{r-1} \sum_{\substack{\mathcal{I}\subseteq\{1,\ldots,m-1\}:\\ |\mathcal{I}|=r}} \Pr\left(\bigcap_{i\in\mathcal{I}} (\overline{\mathcal{D}}_m \cap \overline{\mathcal{D}}_i) \,\middle|\, \mathbf{x}_m\right), \tag{103}$$

which proves (93).

The average success probability can now be expressed as follows:

$$P_{\mathrm{c}}\bigl(\mathscr{C}^{(\mathsf{M},n)}\bigr)$$

$$= \frac{1}{\mathsf{M}} \sum_{m=1}^{\mathsf{M}} \Pr(\mathcal{D}_m \,|\, \mathbf{x}_m) \tag{104}$$

$$= \frac{1}{\mathsf{M}} \sum_{m=1}^{\mathsf{M}} \left( \Pr\bigl(\overline{\mathcal{D}}_m \,\big|\, \mathbf{x}_m\bigr) - \sum_{r=1}^{m-1} (-1)^{r-1} \sum_{\substack{\mathcal{I}\subseteq\{1,\ldots,m-1\}:\\ |\mathcal{I}|=r}} \Pr\left(\bigcap_{i\in\mathcal{I}} (\overline{\mathcal{D}}_m \cap \overline{\mathcal{D}}_i) \,\middle|\, \mathbf{x}_m\right)\right) \tag{105}$$

$$= \frac{1}{\mathsf{M}} \sum_{m=1}^{\mathsf{M}} \left( \Pr\bigl(\overline{\mathcal{D}}_m \,\big|\, \mathbf{x}_m\bigr) + \sum_{r=1}^{m-1} (-1)^{r} \sum_{\substack{\mathcal{I}\subseteq\{1,\ldots,m-1\}:\\ |\mathcal{I}|=r}} \Pr\left(\bigcap_{i\in\mathcal{I}} (\overline{\mathcal{D}}_m \cap \overline{\mathcal{D}}_i) \,\middle|\, \mathbf{x}_{\ell,\ell\in\mathcal{I}\cup\{m\}}\right)\right) \tag{106}$$

$$= \frac{1}{\mathsf{M}} \sum_{m=1}^{\mathsf{M}} \left( \sum_{r=0}^{m-1} (-1)^{r} \sum_{\substack{\mathcal{I}\subseteq\{1,\ldots,m-1\}:\\ |\mathcal{I}|=r}} \Pr\left(\bigcap_{i\in\mathcal{I}} (\overline{\mathcal{D}}_m \cap \overline{\mathcal{D}}_i) \,\middle|\, \mathbf{x}_{\ell,\ell\in\mathcal{I}\cup\{m\}}\right)\right) \tag{107}$$

$$= \frac{1}{\mathsf{M}} \sum_{r=0}^{\mathsf{M}-1} (-1)^{r} \sum_{m=r+1}^{\mathsf{M}} \left( \sum_{\substack{\mathcal{I}\subseteq\{1,\ldots,m-1\}:\\ |\mathcal{I}|=r}} \Pr\left(\bigcap_{i\in\mathcal{I}} (\overline{\mathcal{D}}_m \cap \overline{\mathcal{D}}_i) \,\middle|\, \mathbf{x}_{\ell,\ell\in\mathcal{I}\cup\{m\}}\right)\right) \tag{108}$$

$$= \frac{1}{\mathsf{M}} \sum_{r=0}^{\mathsf{M}-1} (-1)^{r} \sum_{\substack{\mathcal{I}\subseteq\{1,\ldots,\mathsf{M}\}:\\ |\mathcal{I}|=r+1}} \Pr\left(\bigcap_{i\in\mathcal{I}} \overline{\mathcal{D}}_i \,\middle|\, \mathbf{x}_{\ell,\ell\in\mathcal{I}}\right) \tag{109}$$

$$= \frac{1}{\mathsf{M}} \sum_{r=1}^{\mathsf{M}} (-1)^{r-1} \sum_{\substack{\mathcal{I}\subseteq\{1,\ldots,\mathsf{M}\}:\\ |\mathcal{I}|=r}} \Pr\left(\bigcap_{i\in\mathcal{I}} \overline{\mathcal{D}}_i \,\middle|\, \mathbf{x}_{\ell,\ell\in\mathcal{I}}\right). \tag{110}$$

Here, (105) follows from (103); in (106) we allow different choices of the conditioning argument, which does not change the expression because of (91); in (107) we include the empty set into the sum to take care of the first term; and in (108) and (109) we exchange the two outer sums and then combine the resulting two inner sums. This completes the proof. □

By the $r$-wise Hamming distance and Theorem 48, we are now able to give a closed-form expression for the exact average error probability of an arbitrary code $\mathscr{C}_{\mathbf{t}}^{(\mathsf{M},n)}$ used on a BEC.

**Theorem 49 (Average Error Probability on the BEC).** *Consider a BEC with arbitrary erasure probability $0 \le \delta < 1$ and an arbitrary code $\mathscr{C}_{\mathbf{t}}^{(\mathsf{M},n)}$ with $\mathsf{M} \ge 2$. The average ML error probability can be expressed using the type vector $\mathbf{t}$ as follows:*

$$P_{\mathrm{e}}\left(\mathscr{C}_{\mathbf{t}}^{(\mathsf{M},n)}\right) = \frac{1}{\mathsf{M}} \sum_{r=2}^{\mathsf{M}} (-1)^r \sum_{\substack{\mathcal{I} \subseteq \{1,\ldots,\mathsf{M}\}: \\ |\mathcal{I}|=r}} \delta^{d_{\mathcal{I}}^{(\mathsf{M},n)}}, \tag{111}$$

*where $d_{\mathcal{I}}^{(\mathsf{M},n)}$ denotes the $r$-wise Hamming distance as given in Definition 31.*

*Proof:* Comparing (94) and (111), we see that the theorem can be proved by showing that

$$\Pr\left(\overline{\mathcal{D}}_m \,\middle|\, \mathbf{x}_m\right) = 1, \quad \forall\, m \in \mathcal{M}, \tag{112}$$

$$\Pr\left(\bigcap_{i\in\mathcal{I}} \overline{\mathcal{D}}_i \,\middle|\, \mathbf{x}_{\ell,\ell\in\mathcal{I}}\right) = \delta^{d_{\mathcal{I}}^{(\mathsf{M},n)}}, \quad \forall\, \mathcal{I} \subseteq \mathcal{M} \text{ with } |\mathcal{I}| \ge 2. \tag{113}$$

By definition and because the channel is a BEC,

$$\begin{aligned} \overline{\mathcal{D}}_m &= \left\{\mathbf{y}\colon d_{\mathrm{H}}\left(\mathbf{x}_{m,\mathcal{I}(0|\mathbf{y})}, \mathbf{y}_{\mathcal{I}(0|\mathbf{y})}\right) = d_{\mathrm{H}}\left(\mathbf{x}_{m,\mathcal{I}(1|\mathbf{y})}, \mathbf{y}_{\mathcal{I}(1|\mathbf{y})}\right) = 0\right\} && (114) \\ &= \bigcup_{\mathsf{N}=0}^{n} \bigcup_{\substack{\mathcal{N}\subseteq\mathbb{N}_n: \\ |\mathcal{N}|=\mathsf{N}}} \left\{\mathbf{y}\colon d_{\mathrm{H}}(\mathbf{2}_{\mathcal{N}}, \mathbf{y}_{\mathcal{N}}) = d_{\mathrm{H}}\left(\mathbf{x}_{m,\mathbb{N}_n\setminus\mathcal{N}}, \mathbf{y}_{\mathbb{N}_n\setminus\mathcal{N}}\right) = 0\right\} && (115) \end{aligned}$$

where we abbreviate $\mathbb{N}_n \triangleq \{1,\ldots,n\}$ and $\mathbf{2}$ denotes the all-2 vector. Therefore, the conditional success probability of the closed decoding region $\overline{\mathcal{D}}_m$ is

$$\Pr\left(\overline{\mathcal{D}}_m \,\middle|\, \mathbf{x}_m\right) = \sum_{\mathbf{y}\in\overline{\mathcal{D}}_m} P_{\mathbf{Y}|\mathbf{X}}\left(\mathbf{y}\middle|\mathbf{x}_m\right) = \sum_{\mathsf{N}=0}^{n} \binom{n}{\mathsf{N}} \delta^{\mathsf{N}} (1-\delta)^{n-\mathsf{N}} = 1. \tag{116}$$

Similarly,

$$\begin{aligned} \bigcap_{i\in\mathcal{I}} \overline{\mathcal{D}}_i &= \left\{\mathbf{y}\colon d_{\mathrm{H}}\left(\mathbf{x}_{i,\mathcal{I}(0|\mathbf{y})}, \mathbf{y}_{\mathcal{I}(0|\mathbf{y})}\right) = d_{\mathrm{H}}\left(\mathbf{x}_{i,\mathcal{I}(1|\mathbf{y})}, \mathbf{y}_{\mathcal{I}(1|\mathbf{y})}\right) = 0 \quad \forall\, i \in \mathcal{I}\right\} && (117) \\ &= \bigcup_{\mathsf{N}=d_{\mathcal{I}}^{(\mathsf{M},n)}}^{n} \bigcup_{\substack{\mathcal{N}\supseteq\mathbb{N}_n\setminus\mathbb{N}_{\mathcal{I}}: \\ |\mathcal{N}|=\mathsf{N}}} \left\{\mathbf{y}\colon d_{\mathrm{H}}(\mathbf{2}_{\mathcal{N}}, \mathbf{y}_{\mathcal{N}}) = d_{\mathrm{H}}\left(\mathbf{x}_{i_1,\mathbb{N}_n\setminus\mathcal{N}}, \mathbf{y}_{\mathbb{N}_n\setminus\mathcal{N}}\right) = 0\right\}, && (118) \end{aligned}$$

where for convenience, we set $\mathcal{I} = \{i_1,\ldots,i_r\}$ and $\mathbb{N}_{\mathcal{I}} \triangleq \{j \in \mathbb{N}_n\colon x_{i_1,j} = x_{i_2,j} = \cdots = x_{i_r,j}\}$. This implies

$$\begin{aligned} \Pr\left(\bigcap_{i\in\mathcal{I}} \overline{\mathcal{D}}_i \,\middle|\, \mathbf{x}_{\ell,\ell\in\mathcal{I}}\right) &= \sum_{\mathbf{y}\in\bigcap_{i\in\mathcal{I}}\overline{\mathcal{D}}_i} P_{\mathbf{Y}|\mathbf{X}}(\mathbf{y}|\mathbf{x}_{i_1}) && (119) \\ &= \sum_{\mathsf{N}=d_{\mathcal{I}}^{(\mathsf{M},n)}}^{n} \binom{n-d_{\mathcal{I}}^{(\mathsf{M},n)}}{\mathsf{N}-d_{\mathcal{I}}^{(\mathsf{M},n)}} \delta^{\mathsf{N}} (1-\delta)^{n-\mathsf{N}} && (120) \\ &= \delta^{d_{\mathcal{I}}^{(\mathsf{M},n)}}. && (121) \end{aligned}$$

□

## 5.2 Optimal Codes with Three or Four Codewords ($\mathsf{M} = 3, 4$)

We start to investigate the optimal codes for $\mathsf{M} = 3$, since the optimal code for $\mathsf{M} = 2$ on a BEC is quite trivially the repetition code.

Even though we know the exact average error probability for a code with an arbitrary number of codewords $\mathsf{M}$ on a BEC, the optimal code structure is not obvious. We are now trying to shed more light on this problem.

We start with the following lemma.

**Lemma 50 ( [19, Lem. 32]).** *Fix the number of codewords $\mathsf{M}$ and a DMC. The success probability $P_c(\mathscr{C}^{(\mathsf{M},n)})$ for a sequence of codes $\{\mathscr{C}^{(\mathsf{M},n)}\}_{n\geq 1}$, where each code is generated by appending a column to the code of smaller blocklength, is nondecreasing with respect to the blocklength $n$.*

*Proof:* See [19, Sec. VIII-B]. □

Lemma 50 suggests a recursive code construction that guarantees the largest *total success probability increase*,[19] i.e., we can find some locally optimal code type.

**Theorem 51.** *For a BEC with arbitrary erasure probability $0 \leq \delta < 1$, an optimal code with three codewords $\mathsf{M} = 3$ or four codewords $\mathsf{M} = 4$ and with a blocklength $n = 2$ is*

$$\mathscr{C}_{\text{BEC}}^{(\mathsf{M},2)*} = \begin{cases} \begin{pmatrix} \mathbf{c}_1^{(\mathsf{M})} & \mathbf{c}_2^{(\mathsf{M})} \end{pmatrix} & \text{if } \mathsf{M} = 3, \\ \begin{pmatrix} \mathbf{c}_3^{(\mathsf{M})} & \mathbf{c}_5^{(\mathsf{M})} \end{pmatrix} & \text{if } \mathsf{M} = 4. \end{cases} \tag{122}$$

*If we recursively construct a locally optimal codebook with three codewords $\mathsf{M} = 3$ or four codewords $\mathsf{M} = 4$ and with a blocklength $n \geq 3$ by appending a new column to $\mathscr{C}_{\text{BEC}}^{(\mathsf{M},n-1)\diamond}$, where we append a "$\diamond$" to $(\mathsf{M}, n)$ to denote a locally optimal recursive-constructed code of size $\mathsf{M}$ and length $(n-1)$, the increase in average success probability is maximized by the following choice of appended columns:*

$$\begin{cases} \mathbf{c}_3^{(\mathsf{M})} & \text{if } n \bmod 3 = 0, \\ \mathbf{c}_1^{(\mathsf{M})} & \text{if } n \bmod 3 = 1, \\ \mathbf{c}_2^{(\mathsf{M})} & \text{if } n \bmod 3 = 2, \end{cases} \quad \text{when } \mathsf{M} = 3 \tag{123}$$

*and*

$$\begin{cases} \mathbf{c}_6^{(\mathsf{M})} & \text{if } n \bmod 3 = 0, \\ \mathbf{c}_3^{(\mathsf{M})} & \text{if } n \bmod 3 = 1, \\ \mathbf{c}_5^{(\mathsf{M})} & \text{if } n \bmod 3 = 2, \end{cases} \quad \text{when } \mathsf{M} = 4. \tag{124}$$

*Proof:* See Appendix A. □

This theorem suggests that for a given fixed code size $\mathsf{M}$, a sequence of good codes can be generated by appending the correct columns to a code of smaller blocklength. For a given DMC and code of blocklength $n$, we ask the question what is the optimal improvement (i.e., the maximum reduction of error probability) when increasing the blocklength from $n$ to $n+1$ when $\mathsf{M} = 3$ or 4. (Note that in general one might achieve better results if we design a sequence of codes that increases from blocklength $n$ to $n + \gamma$ with a step-size $\gamma > 1$; however, as we will see below, for $\mathsf{M} = 3$ or $\mathsf{M} = 4$,

[19] See [19, Def. 33].

$\gamma = 1$ turns out to be optimal.) The answer to this question then leads to the recursive construction of (123) and (124).

While Theorem 51 only guarantees local optimality for the given recursive construction, further investigation shows that the given construction is actually globally optimum.

**Theorem 52.** *For a BEC and for any $n \geq 2$, an optimal codebook with $\mathsf{M} = 3$ or $\mathsf{M} = 4$ codewords is the weak flip code of type $\mathbf{t}^*_{\text{weak}}$, where for $\mathsf{M} = 3$*

$$t_1^* = \left\lfloor \frac{n+2}{3} \right\rfloor, \quad t_2^* = \left\lfloor \frac{n+1}{3} \right\rfloor, \quad t_3^* = \left\lfloor \frac{n}{3} \right\rfloor \tag{125}$$

*and for $\mathsf{M} = 4$*

$$t_3^* = \left\lfloor \frac{n+2}{3} \right\rfloor, \quad t_5^* = \left\lfloor \frac{n+1}{3} \right\rfloor, \quad t_6^* = \left\lfloor \frac{n}{3} \right\rfloor. \tag{126}$$

*Note that the recursively constructed code of Theorem 51 is equivalent to the optimal code given here:*

$$\mathscr{C}_{\text{BEC}}^{(\mathsf{M},n)\diamond} \equiv \mathscr{C}_{\mathbf{t}^*_{\text{weak}}}^{(\mathsf{M},n)}. \tag{127}$$

*Proof:* See Appendix B. □

Using the shorthand

$$k \triangleq \left\lfloor \frac{n}{3} \right\rfloor, \tag{128}$$

the code parameters of these optimal codes can be summarized as

$$\mathbf{t}^*_{\text{weak}} = \begin{cases} [t_1^*, t_2^*, t_3^*] & \text{for } \mathsf{M} = 3, \\ [t_3^*, t_5^*, t_6^*] & \text{for } \mathsf{M} = 4 \end{cases} \tag{129}$$

$$= \begin{cases} [k, k, k] & \text{if } n \bmod 3 = 0, \\ [k+1, k, k] & \text{if } n \bmod 3 = 1, \\ [k+1, k+1, k] & \text{if } n \bmod 3 = 2. \end{cases} \tag{130}$$

From (123) and (124), or from (125) and (126), or from (129), we confirm again that $\mathscr{C}_{\mathbf{t}^*_{\text{weak}}}^{(3,n)}$ can be obtained by simply removing the last codeword of $\mathscr{C}_{\mathbf{t}^*_{\text{weak}}}^{(4,n)}$ (compare with Remark 14).

The corresponding optimal average error probabilities are given as

$$P_{\text{e}}\left(\mathscr{C}_{\mathbf{t}^*_{\text{weak}}}^{(\mathsf{M},n)}\right) = \begin{cases} \frac{1}{3}\left(\delta^{n-t_1^*} + \delta^{n-t_2^*} + \delta^{n-t_3^*} - \delta^n\right) & \text{if } \mathsf{M} = 3, \\ \frac{1}{4}\left(2\delta^{n-t_3^*} + 2\delta^{n-t_5^*} + 2\delta^{n-t_6^*} - 3\delta^n\right) & \text{if } \mathsf{M} = 4. \end{cases} \tag{131}$$

### 5.3 A Brief Comparison between BSC and BEC

In [19], it has been shown that the optimal codes for $\mathsf{M} = 3$ or $\mathsf{M} = 4$ for the BSC are weak flip codes with type

$$\mathbf{t}^*_{\text{weak}} = \begin{cases} [k+1, k, k-1] & \text{if } n \bmod 3 = 0, \\ [k+1, k, k] & \text{if } n \bmod 3 = 1, \\ [k+1, k+1, k] & \text{if } n \bmod 3 = 2. \end{cases} \tag{132}$$

which by (130) immediately gives the following corollary.

**Corollary 53.** *For* $\mathsf{M} = 3$ *or* $\mathsf{M} = 4$ *and for* $n \bmod 3 \neq 0$*, the weak flip codes with type* $\mathbf{t}^*_{\text{weak}}$ *defined in* (132) *(equivalently,* (130)*) are optimal for both BSC and BEC.*

The corresponding pairwise Hamming distance vectors of the BSC optimal codes for $\mathsf{M} = 3$ and $\mathsf{M} = 4$ are respectively[20]

$$\mathbf{d}^{(3,n)*} = \begin{cases} (2k-1, 2k, 2k+1) & \text{if } n \bmod 3 = 0, \\ (2k, 2k+1, 2k+1) & \text{if } n \bmod 3 = 1, \\ (2k+1, 2k+1, 2k+2) & \text{if } n \bmod 3 = 2, \end{cases} \tag{133}$$

and

$$\mathbf{d}^{(4,n)*} = \begin{cases} (2k-1, 2k, 2k+1, 2k+1, 2k, 2k-1) & \text{if } n \bmod 3 = 0, \\ (2k, 2k+1, 2k+1, 2k+1, 2k+1, 2k) & \text{if } n \bmod 3 = 1, \\ (2k+1, 2k+1, 2k+2, 2k+2, 2k+1, 2k+1) & \text{if } n \bmod 3 = 2. \end{cases} \tag{134}$$

Comparing these to the corresponding pairwise Hamming distance vectors of the BEC optimal codes (Theorem 52),

$$\mathbf{d}^{(3,n)*} = \begin{cases} (2k, 2k, 2k) & \text{if } n \bmod 3 = 0, \\ (2k, 2k+1, 2k+1) & \text{if } n \bmod 3 = 1, \\ (2k+1, 2k+1, 2k+2) & \text{if } n \bmod 3 = 2 \end{cases} \tag{135}$$

and

$$\mathbf{d}^{(4,n)*} = \begin{cases} (2k, 2k, 2k, 2k, 2k, 2k) & \text{if } n \bmod 3 = 0, \\ (2k, 2k+1, 2k+1, 2k+1, 2k+1, 2k) & \text{if } n \bmod 3 = 1, \\ (2k+1, 2k+1, 2k+2, 2k+2, 2k+1, 2k+1) & \text{if } n \bmod 3 = 2, \end{cases} \tag{136}$$

we note that when $n \bmod 3 = 0$, the optimal codes for the BEC are fair and therefore maximize the minimum Hamming distance, while this is not the case for the very symmetric BSC (i.e, on the BSC, an optimal code of length $n \bmod 3 = 0$ does not maximize the minimum Hamming distance among all code designs of the same size and length!). In fact, for $\mathsf{M} = 3$ or 4 and for every $n$, a code maximizes the minimum Hamming distance if, and only if, it is an optimal code for the BEC. However, when $\mathsf{M} > 4$, numerical evidence can be created to disprove the statement that a code maximizing the minimum Hamming distance is an optimal code for the BEC! As we will see in the cases of $\mathsf{M} = 8$ and 16, the pairwise Hamming distance vector (2-wise Hamming distance) is not sufficient for determining global optimality, but the $r$-wise Hamming distances with $r > 2$ have to be taken into account.

## 5.4 Application to Known Bounds on the Error Probability for a Finite Blocklength ($\mathsf{M} = 3, 4$)

Since we now know the optimal code structure, we can compare its performance to the known bounds in Section 3.

[20]For weak flip codes with $\mathsf{M} = 3$ or $\mathsf{M} = 4$ codewords, we only need to compare the pairwise Hamming distances because the 3-wise and 4-wise Hamming distances are all equal to $n$ and hence are identical.

Note that for $\mathsf{M} = 3, 4$,

$$\mathrm{D}_{\min}^{(\mathrm{BEC})}\left(\mathscr{C}_{\mathbf{t}^*_{\mathrm{weak}}}^{(\mathsf{M},n)}\right) = \begin{cases} -\frac{2}{3}\log\delta & \text{if } n \bmod 3 = 0, \\ -\frac{\lfloor\frac{n}{3}\rfloor + \lfloor\frac{n+1}{3}\rfloor}{n}\log\delta & \text{if } n \bmod 3 = 1, \\ -\frac{\lfloor\frac{n}{3}\rfloor + \lfloor\frac{n+1}{3}\rfloor}{n}\log\delta & \text{if } n \bmod 3 = 2. \end{cases} \tag{137}$$

Figures 2 and 3 compare the exact optimal performance for $\mathsf{M} = 3$ and $\mathsf{M} = 4$, respectively, with the following bounds: the SGB upper and lower bounds based on the optimal code as used by Shannon *et al.* for a blocklength $n \bmod 3 = 0$ (thereby confirming that this lower bound is valid generally), the Gallager upper bound, and also the PPV upper and lower bounds.

We can see that the SGB upper bound is closer to the exact optimal performance (and hence tighter) than the PPV upper bound and the Gallager upper bound. Note that the PPV upper bound is not exactly the same as the Gallager upper bound, even though for $\mathsf{M} = 3$ their curves look almost identical. Also note that the SGB upper bound does exhibit the correct error exponent. It is shown in [23] that when $n$ goes to infinity under fixed $\mathsf{M}$, the PPV upper bound only tends to the suboptimal Gallager exponent [20]; this fact is also confirmed by the two figures.

Regarding the lower bounds we see that the PPV lower bound is much better for finite $n$ than the SGB lower bound. However, the exponential growth rate of the PPV lower bound only approaches that of the sphere-packing bound [24], and does not equal the optimal exponent either [21].

Once more we would like to emphasize that even though for $\mathsf{M} = 3, 4$, the fair weak flip codes are optimal for the BEC and achieve the optimal error exponent for both the BEC and the BSC, they are strictly suboptimal for every $n \bmod 3 = 0$ for the BSC.

### 5.5 Optimal Codes with Five or Six Codewords ($\mathsf{M} = 5, 6$)

The idea of recursively designing a locally optimal code turned out to be a powerful approach to obtain globally optimal codes for $\mathsf{M} = 3, 4$. Unfortunately, for larger values of $\mathsf{M}$, we might need a recursion from $n$ to $n+\gamma$ with a step-size $\gamma > 1$, and—according to our numerical examination— this step-size $\gamma$ might be a function of the blocklength $n$. Since the exact average error probability expression becomes involved as $\mathsf{M}$ grows, we only succeeded in investigating a locally optimal code construction subject to the recursive design approach when the blocklength $n$ is a multiple of $\mathsf{L}$. Based on our definition of fair weak flip codes and on Conjecture 54 below, we conjecture[21] that the necessary step-size for global optimality satisfies $\gamma \leq \mathsf{L}$.

**Conjecture 54.** *For a BEC and for any $n$ being a multiple of $\mathsf{L} = 10$, an optimal codebook with $\mathsf{M} = 5$ or $\mathsf{M} = 6$ codewords is the corresponding fair weak flip code.*

Note that the restriction on $n$ stems from the fact that fair weak flip codes are only defined for blocklengths satisfying $n \bmod \mathsf{L} = 0$ (the code uses each weak flip column $\tau$ times, where $\tau = n/\mathsf{L}$ is an integer). We can show that if we relax the error minimization problem by allowing noninteger values for the type $\mathbf{t}$, the optimal type will be equally distributed among all possible weak flip columns also when $n$ mod $\mathsf{L} \neq 0$. Unfortunately, a block code always must use an integer number of candidate columns, and the globally optimal choice of an integer in the neighborhood of the optimal noninteger value is rather involved. Based on this observation and on our extensive numerical examinations, we give the following conjecture.

[21] Note that in the following conjectures, despite of Conjecture 55, we actually can prove local optimality of the proposed type vector by verifying the Karush–Kuhn–Tucker (KKT) conditions. However, since the discrete multivariate average error probability function is not convex, we did not succeed in confirming global optimality.

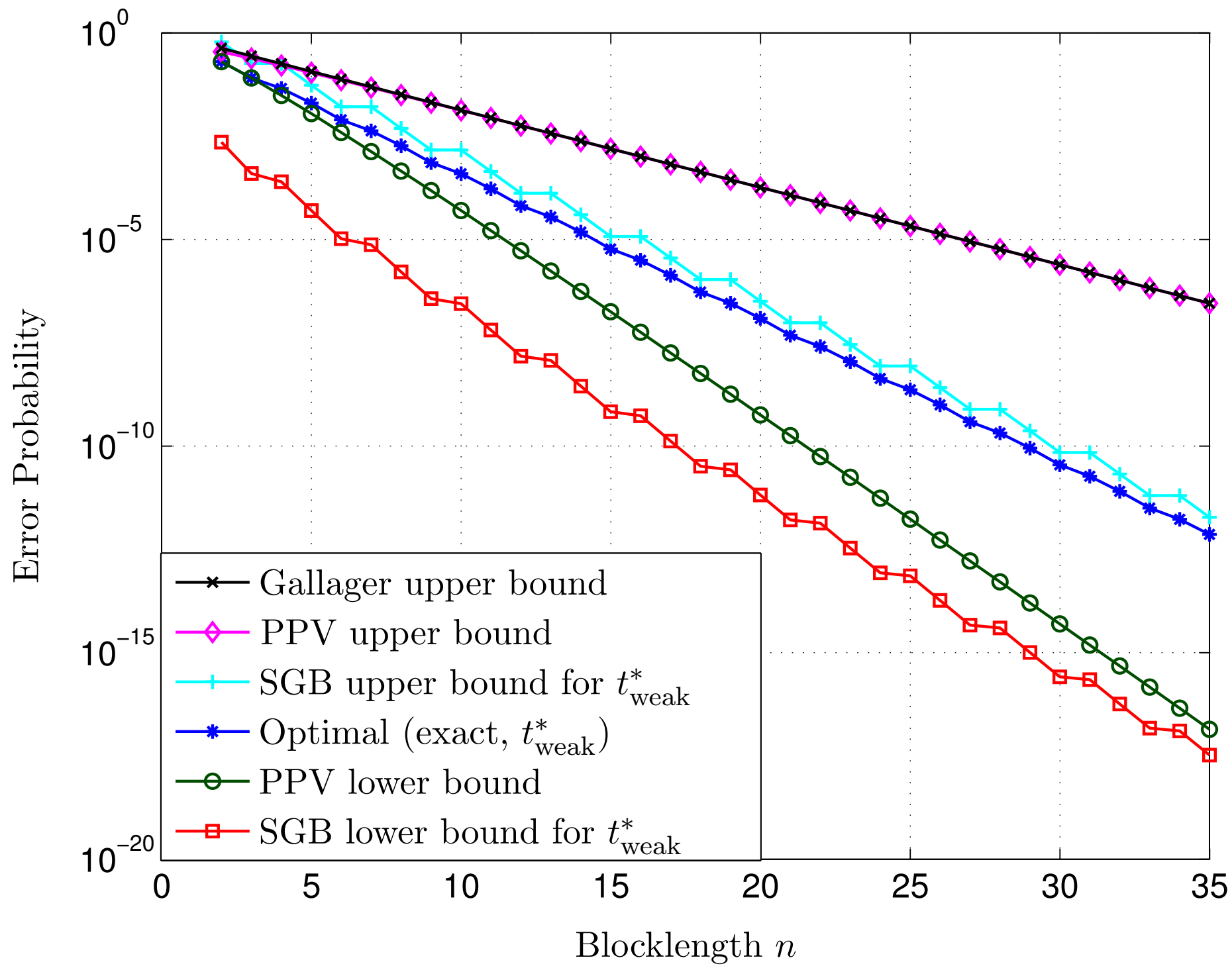

Figure 2: Exact value of, and bounds on, the performance of an optimal code with $\mathsf{M} = 3$ codewords on the BEC with $\delta = 0.3$ as a function of the blocklength $n$.

**Conjecture 55.** *Consider the BEC and a blocklength $n \geq 3$ that is not a multiple of $\mathsf{L} = 10$ (as the case of $n$ mod $10 = 0$ has been taken care in Conjecture 54), and define the shorthand*

$$\tau \triangleq \left\lfloor \frac{n}{10} \right\rfloor. \tag{138}$$

*An optimal code that minimizes the average error probability among all code designs*

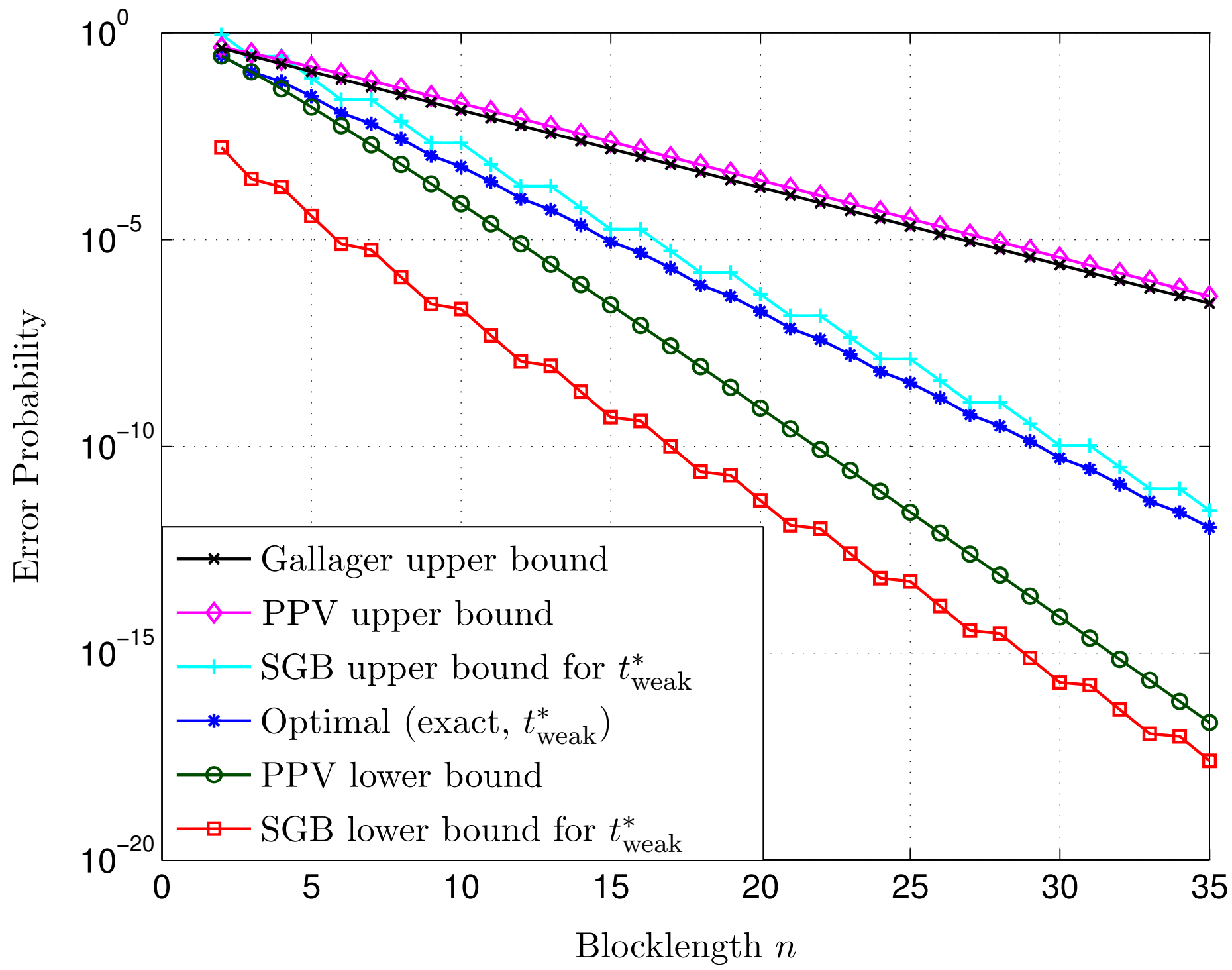

Figure 3: Exact value of, and bounds on, the performance of an optimal code with $\mathsf{M} = 4$ codewords on the BEC with $\delta = 0.3$ as a function of the blocklength $n$.

*with* $\mathsf{M} = 5$ *codewords is a weak flip code of type*

$$\mathbf{t}_{\text{weak}} = [t_3, t_5, t_6, t_7, t_9, t_{10}, t_{11}, t_{12}, t_{13}, t_{14}]$$
$$= \begin{cases} [\tau+1, \tau, \tau, \tau, \tau, \tau, \tau, \tau, \tau, \tau] & \text{if } n \bmod 10 = 1, \\ [\tau+1, \tau+1, \tau+1, \tau-1, \tau, \tau, \tau, \tau, \tau, \tau] & \text{if } n \bmod 10 = 2, \\ [\tau+1, \tau+1, \tau, \tau, \tau+1, \tau, \tau, \tau, \tau, \tau] & \text{if } n \bmod 10 = 3, \\ [\tau+1, \tau+1, \tau, \tau, \tau+1, \tau, \tau, \tau, \tau, \tau+1] & \text{if } n \bmod 10 = 4, \\ [\tau+1, \tau+1, \tau+1, \tau, \tau+1, \tau+1, \tau, \tau, \tau, \tau] & \text{if } n \bmod 10 = 5, \\ [\tau+1, \tau+1, \tau+1, \tau, \tau+1, \tau+1, \tau, \tau+1, \tau, \tau] & \text{if } n \bmod 10 = 6, \\ [\tau+1, \tau+1, \tau+1, \tau, \tau+1, \tau+1, \tau, \tau, \tau+1, \tau+1] & \text{if } n \bmod 10 = 7, \\ [\tau+2, \tau+1, \tau+1, \tau, \tau+1, \tau+1, \tau, \tau, \tau+1, \tau+1] & \text{if } n \bmod 10 = 8, \\ [\tau+1, \tau+1, \tau+1, \tau+1, \tau+1, \tau+1, \tau+1, \tau+1, \tau+1, \tau] & \\ & \text{if } n \bmod 10 = 9. \end{cases} \tag{139}$$

*Except for* $n \bmod 10 = 7$*, an optimal code that minimizes the average error probability*

*among all code designs with* $\mathsf{M} = 6$ *codewords is a weak flip code of type*

$$\begin{aligned}
\mathbf{t}_{\text{weak}} &= \bigl[t_7, t_{11}, t_{13}, t_{14}, t_{19}, t_{21}, t_{22}, t_{25}, t_{26}, t_{28}\bigr] \\
&= \begin{cases}
\bigl[\tau+1, \tau, \tau, \tau, \tau, \tau, \tau, \tau, \tau, \tau\bigr] & \text{if } n \bmod 10 = 1, \\
\bigl[\tau+1, \tau+1, \tau, \tau, \tau, \tau, \tau, \tau, \tau, \tau\bigr] & \text{if } n \bmod 10 = 2, \\
\bigl[\tau+1, \tau+1, \tau, \tau, \tau, \tau+1, \tau, \tau, \tau, \tau\bigr] & \text{if } n \bmod 10 = 3, \\
\bigl[\tau+1, \tau+1, \tau, \tau, \tau, \tau+1, \tau, \tau+1, \tau, \tau\bigr] & \text{if } n \bmod 10 = 4, \\
\bigl[\tau+1, \tau+1, \tau, \tau, \tau+1, \tau+1, \tau, \tau+1, \tau, \tau\bigr] & \text{if } n \bmod 10 = 5, \\
\bigl[\tau+1, \tau+1, \tau+1, \tau, \tau+1, \tau+1, \tau, \tau+1, \tau, \tau\bigr] & \text{if } n \bmod 10 = 6, \\
\bigl[\tau+1, \tau+1, \tau+1, \tau+1, \tau+1, \tau+1, \tau+1, \tau+1, \tau, \tau\bigr] & \\
& \text{if } n \bmod 10 = 8, \\
\bigl[\tau+1, \tau+1, \tau+1, \tau+1, \tau+1, \tau+1, \tau+1, \tau+1, \tau+1, \tau\bigr] & \\
& \text{if } n \bmod 10 = 9.
\end{cases}
\end{aligned} \tag{140}$$

*For* $n \bmod 10 = 7$ *and* $\mathsf{M} = 6$*, an optimal code that minimizes the average error probability among all code designs is actually not a weak flip code but a nonweak flip code of type* $\mathbf{t}$ *satisfying*

$$\begin{cases}
t_{14} = t_{22} = t_{26} = t_{28} = \tau, \\
t_7 = t_{11} = t_{13} = t_{19} = t_{21} = t_{25} = \tau + 1, \\
t_{30} = 1, \\
t_j = 0 \text{ for the remaining indices.}
\end{cases} \tag{141}$$

*Note that* $t_{30}$ *is the only nonweak flip column in this code.*

Surprisingly, the optimal code for $n \bmod 10 = 7$ and $\mathsf{M} = 6$ is not a weak flip code. We point out again that the exact average error probability expression for the BEC with $\mathsf{M} = 6$ is a function of the discrete multivariate nonnegative integers $t_1, t_2, \ldots, t_{31}$ under the constraint that their sum equals $n$. If we allow noninteger solutions, the minimizers are $t_j = n/\mathsf{L}$ for all $t_j$ belonging to weak flip columns. Yet, (141) shows that the nearest *integer* minimizer might be a only "nearly weak flip" code instead of a weak flip code.

Note that according to Conjecture 55, it is possible to recursively construct optimal codes with $\mathsf{M} = 5, 6$ codewords using a step size $\gamma < 10$.

For our quest of understanding the optimal code design for larger $\mathsf{M}$, we believe that it will be useful to substantiate these observations further.

### 5.6 Codes with Large $r$-Wise Hamming Distances for Arbitrary $\mathsf{M}$

We have already pointed out that a code having a large (or even maximum) pairwise Hamming distance is not necessarily an optimal code. It is crucial to look at all $r$-wise Hamming distances for $2 \le r \le \bar{\ell}$.

In the following theorem we will confirm this intuition once again.

**Theorem 56.** *Let the number of codewords be* $\mathsf{M} = 2\bar{\ell}$ *or* $2\bar{\ell} - 1$ *where* $\bar{\ell}$ *is an arbitrary positive even integer, and let the blocklength* $n$ *be such that* $n \bmod \mathsf{L} = 0$*. Then* *an $(\bar{\ell}-1)$-wise equidistant weak flip code that achieves the largest minimum $(\bar{\ell}-1)$-wise Hamming distance*[22] *but not the largest minimum $\bar{\ell}$-wise Hamming distance has a strictly worse performance on the BEC than the fair weak flip code.*

*Proof:* We will prove the theorem only for the case of $\mathsf{M} = 2\bar{\ell}-1$, the case of $\mathsf{M} = 2\bar{\ell}$ will be similar. So, let $\mathsf{M} = 2\bar{\ell}-1$ with $\bar{\ell}$ even and let the blocklength be $n = \mathsf{L}\tau$ for some $\tau \in \mathbb{N}$. Let $\mathscr{C}^{(\mathsf{M},n)}_{\mathbf{t}^\circ_{\text{weak}}}$ be an $(\bar{\ell}-1)$-wise equidistant weak flip code that achieves the largest minimum $(\bar{\ell}-1)$-wise Hamming distance, but does not achieve the largest minimum $\bar{\ell}$-wise Hamming distance, and let $\mathscr{C}^{(\mathsf{M},n)}_{\text{fair}}$ be the fair weak flip code that according to Corollary 46 is $\bar{\ell}$-wise equidistant and achieves the largest minimum $\bar{\ell}$-wise Hamming distance. Therefore, according to Theorem 45 and Theorem 49, we have

$$
\begin{aligned}
&P_{\text{e}}\left(\mathscr{C}^{(\mathsf{M},n)}_{\mathbf{t}^\circ_{\text{weak}}}\right) - P_{\text{e}}\left(\mathscr{C}^{(\mathsf{M},n)}_{\text{fair}}\right) \\
&\quad = \frac{1}{\mathsf{M}}\sum_{r=2}^{\mathsf{M}}(-1)^r \sum_{\substack{\mathcal{I}\subseteq\{1,\ldots,\mathsf{M}\}:\\ |\mathcal{I}|=r}} \delta^{d_{\mathcal{I}}\left(\mathscr{C}^{(\mathsf{M},n)}_{\mathbf{t}^\circ_{\text{weak}}}\right)} - \frac{1}{\mathsf{M}}\sum_{r=2}^{\mathsf{M}}(-1)^r \sum_{\substack{\mathcal{I}\subseteq\{1,\ldots,\mathsf{M}\}:\\ |\mathcal{I}|=r}} \delta^{d_{\mathcal{I}}\left(\mathscr{C}^{(\mathsf{M},n)}_{\text{fair}}\right)} && (142)\\
&\quad = \frac{(-1)^{\bar{\ell}}}{\mathsf{M}} \sum_{\substack{\mathcal{I}\subseteq\{1,\ldots,\mathsf{M}\}:\\ |\mathcal{I}|=\bar{\ell}}} \delta^{d_{\mathcal{I}}\left(\mathscr{C}^{(\mathsf{M},n)}_{\mathbf{t}^\circ_{\text{weak}}}\right)} - \frac{(-1)^{\bar{\ell}}}{\mathsf{M}} \sum_{\substack{\mathcal{I}\subseteq\{1,\ldots,\mathsf{M}\}:\\ |\mathcal{I}|=\bar{\ell}}} \delta^{d_{\mathcal{I}}\left(\mathscr{C}^{(\mathsf{M},n)}_{\text{fair}}\right)} && (143)\\
&\quad = \frac{1}{\mathsf{M}}\sum_{w=1}^{\mathsf{L}} \delta^{n-t^\circ_{j_w}} - \frac{1}{\mathsf{M}}\cdot\mathsf{L}\cdot\delta^{n-\frac{n}{\mathsf{L}}} && (144)\\
&\quad = \frac{\mathsf{L}}{\mathsf{M}}\delta^n\left[\frac{1}{\mathsf{L}}\sum_{w=1}^{\mathsf{L}}\delta^{-t^\circ_{j_w}} - \delta^{-\tau}\right] && (145)\\
&\quad > \frac{\mathsf{L}}{\mathsf{M}}\delta^n\left[\left(\prod_{w=1}^{\mathsf{L}}\delta^{-t^\circ_{j_w}}\right)^{\frac{1}{\mathsf{L}}} - \delta^{-\tau}\right] && (146)\\
&\quad = \frac{\mathsf{L}}{\mathsf{M}}\delta^n\left[\delta^{-\frac{1}{\mathsf{L}}\sum_{w=1}^{\mathsf{L}} t^\circ_{j_w}} - \delta^{-\tau}\right] && (147)\\
&\quad = \frac{\mathsf{L}}{\mathsf{M}}\delta^n\left[\delta^{-\frac{n}{\mathsf{L}}} - \delta^{-\tau}\right] && (148)\\
&\quad = 0. && (149)
\end{aligned}
$$

Here, the second equality follows because the distance structure of the two codes only differ in the case of $r=\bar{\ell}$ (and $\bar{\ell}$ must be even in order to make the difference positive); in the subsequent equality we use the type vector to express the $\bar{\ell}$-wise Hamming distances of both codes and also use the fact that $\bar{\ell}$ is even; the inequality holds because the arithmetic mean (AM) is strictly larger than the geometric mean (GM); and finally we note that $\sum_{w=1}^{\mathsf{L}} t^\circ_{j_w} = n$. Note that since we assume that $\mathscr{C}^{(\mathsf{M},n)}_{\mathbf{t}^\circ_{\text{weak}}}$ does *not* achieve the $\bar{\ell}$-wise Plotkin bound, it follows that there must exist some $t^\circ_{j_w} \neq \tau$ and therefore the inequality is strict. □

[22]By Theorem 45 such a weak flip code also is $s$-wise equidistant and maximizes the $s$-wise Hamming distances for all $2 \leq s \leq \bar{\ell}-1$.

## 5.7 Linear vs. Nonlinear Codes

In this work, we are not really interested in linear codes as our focus lies on optimality in the sense of smallest average error probability. Nevertheless it is important to show the superiority of our proposed weak flip codes. To that goal we will next compare linear codes with nonlinear weak flip codes for the case of $\mathsf{M} = 8$ and $\mathsf{M} = 16$. We will see that best linear codes are often strictly suboptimal.

### 5.7.1 Comparisons for $\mathsf{M} = 8$

The following example shows that the fair linear code with $\mathsf{M} = 8$ codewords, which only achieves the 2-wise Plotkin bound, is strictly suboptimal on the BEC.

**Example 57.** Consider the fair linear code and the (nonlinear) fair weak flip code for $\mathsf{M} = 2^3$ and $n = 35$. From Theorem 49 we obtain

$$P_e\Big(\mathscr{C}^{(8,35)}_{\text{lin,fair}}\Big) = \frac{1}{8}\Bigg(\binom{8}{2}\delta^{n-15} - \binom{8}{3}\delta^{n-5} + 14\delta^{n-5} + \bigg(\binom{8}{4} - 14\bigg)\delta^n - \binom{8}{5}\delta^n + \binom{8}{6}\delta^n - \binom{8}{7}\delta^n + \binom{8}{8}\delta^n\Bigg), \tag{150}$$

and from Corollary 40 and also Theorem 49, we get

$$P_e\Big(\mathscr{C}^{(8,35)}_{\text{fair}}\Big) = \frac{1}{8}\Bigg(\binom{8}{2}\delta^{n-15} - \binom{8}{3}\delta^{n-5} + \binom{8}{4}\delta^{n-1} - \binom{8}{5}\delta^n + \binom{8}{6}\delta^n - \binom{8}{7}\delta^n + \binom{8}{8}\delta^n\Bigg). \tag{151}$$

Thus,

$$P_e\Big(\mathscr{C}^{(8,35)}_{\text{lin, fair}}\Big) - P_e\Big(\mathscr{C}^{(8,35)}_{\text{fair}}\Big) = \frac{14}{8}\big(\delta^{n-5} + 4\delta^n - 5\delta^{n-1}\big), \tag{152}$$

which can be seen to be strictly positive using an argument similar to the proof of Theorem 56 (AM–GM inequality). Hence, the fair linear code with dimension $k = 3$ and blocklength $n = 35$ is not optimal. ♢

Actually, this example can be generalized to any blocklength being a multiple of 7 except $n = 7$. The derivation (which is given in Appendix C) is based on elaborately extracting $n$ columns from the codebook matrix of a fair weak flip code with blocklength larger than $n$ to form a new $(8, n)$ nonlinear code (that actually is a concatenation of several *nonlinear* Hadamard codes). The technique fails for $n = 7$ because taking any seven columns from the code matrix of the $(8, 35)$ fair weak flip code always results in a Hadamard *linear* code. Also note that since there are no Hadamard codes for any blocklength $n \bmod 7 \neq 0$, the technique fails again for $n \bmod 7 \neq 0$.[23]

**Theorem 58.** *For $n \bmod 7 = 0$ apart from $n = 7$, the fair linear code with $\mathsf{M} = 8$ codewords is strictly suboptimal over the BEC.*

[23]Hadamard codes allow for the exact computation of the complete $r$-wise Hamming distance structure. In the case of an arbitrary weak flip code this is rather involved.

*Proof:* Since fair weak flip codes are only defined for $n = 35\tau$, where $\tau \in \mathbb{N}$, we propose the so-called *generalized fair weak flip code* for all blocklengths $n \bmod 7 = 0$ apart from $n = 7$ and $n = 35\tau$. We then show that this nonlinear code and fair weak flip code have a better performance than the corresponding fair linear code over the BEC. Note that the minimum 4-wise Hamming distance of the generalized fair weak flip code and fair weak flip code are always larger than the minimum 4-wise Hamming distance of the fair linear code. The details are given in Appendix C. □

It is interesting that for $\mathsf{M} = 8$ and for all blocklengths $n \bmod 35 = 0$, the fair linear code and the fair weak flip code both are 2-wise and 3-wise equidistant and both achieve the 2-wise and the 3-wise Plotkin bounds. However, only the fair weak flip code is also 4-wise equidistant and achieves the 4-wise Plotkin bound. This is in agreement with Theorem 56 and explains why the fair linear code is outperformed on the BEC.

Based on these insights, we actually believe that the fair weak flip code is globally optimal and that the generalized fair weak flip codes outperform the best linear codes for $\mathsf{M} = 8$.

In general, for blocklengths $n \bmod \mathsf{L} \neq 0$, the situation is unclear because the optimal discrete solution to the "fair noninteger" distribution among all weak flip columns might even end up with nonweak flip columns (compare with Conjecture 55). Still, we have numerical evidence that the best found weak flip codes are superior to the best linear codes. We are next going to elaborate on this.

The best linear codes for $\mathsf{M} = 8$ and any blocklength $n \leq 35$ are found by an exhaustive search over all possible linear code parameters $\mathbf{t}_{\text{lin}}$ such that

$$\mathbf{t}^*_{\text{lin}} = \min_{\mathbf{t}_{\text{lin}}} \Big\{ P_{\text{e}}\Big(\mathscr{C}^{(8,n)}_{\mathbf{t}_{\text{lin}}}\Big)\Big\},$$

where $\sum_{\ell=1}^{7} t_{j_\ell} = n$. Unfortunately, the same approach does not work for the weak flip codes, because we need to choose from 35 weak flip columns, which results in a too high complexity for an exhaustive search. Instead, we use a simulated annealing algorithm [30] to determine a good weak flip code type $\mathbf{t}^{\diamond}_{\text{weak}}$ (which therefore is not guaranteed to be optimal). This simulated annealing algorithm is briefly summarized as follows.

**Step 1:** We randomly choose $n$ columns $\mathbf{c}^{(\mathsf{M})}_j \in \mathcal{C}^{(\mathsf{M})}_{\text{weak}}$ to form a weak flip code $\mathscr{C}^{(\mathsf{M},n)}_{\text{weak}} = \big[\mathbf{c}^{(\mathsf{M})}_{j_1}, \cdots, \mathbf{c}^{(\mathsf{M})}_{j_n}\big]$. We compute the corresponding $\mathbf{t}_{\text{weak}}$ and error probability $P_{\text{e}}\big(\mathscr{C}^{(\mathsf{M},n)}_{\mathbf{t}_{\text{weak}}}\big)$ according to (111). We set a temperature $\mathsf{T} \leftarrow \mathsf{T}_{\text{s}}$.

**Step 2:** We randomly select two distinct $w, w'$ such that $\mathbf{c}^{(\mathsf{M})}_{j_w} \in \mathscr{C}^{(\mathsf{M},n)}_{\text{weak}}$ and $\mathbf{c}^{(\mathsf{M})}_{j_{w'}} \in \mathcal{C}^{(\mathsf{M})}_{\text{weak}} \backslash \mathscr{C}^{(\mathsf{M},n)}_{\text{weak}}$, and obtain a new code $\mathscr{C}^{(\mathsf{M},n)'}_{\text{weak}}$ by replacing $\mathbf{c}^{(\mathsf{M})}_{j_w}$ with $\mathbf{c}^{(\mathsf{M})}_{j_{w'}}$. For this new code we compute the code type $\mathbf{t}'_{\text{weak}}$ and the difference in error probability $\Delta\mathsf{E} = P_{\text{e}}\big(\mathscr{C}^{(\mathsf{M},n)}_{\mathbf{t}'_{\text{weak}}}\big) - P_{\text{e}}\big(\mathscr{C}^{(\mathsf{M},n)}_{\mathbf{t}_{\text{weak}}}\big)$. If $\Delta\mathsf{E} < 0$, we replace $\mathbf{t}_{\text{weak}}$ by $\mathbf{t}'_{\text{weak}}$ for sure; otherwise, we replace $\mathbf{t}_{\text{weak}}$ by $\mathbf{t}'_{\text{weak}}$ with probability $e^{-\Delta\mathsf{E}/\mathsf{T}}$.

**Step 3:** We repeat **Step 2** until either the number of column replacements or the number of iterations exceeds some prescribed number.

**Step 4:** We lower the temperature $\mathsf{T} \leftarrow \alpha\mathsf{T}$ for some $\alpha < 1$, and return to **Step 2** until we either observe a stable code configuration or the temperature is lower than a freezing temperature $\mathsf{T}_{\text{f}}$.

Table 1 lists the resulting minimum $r$-wise Hamming distances for $r = 2, 3, 4$ for both $\mathbf{t}^*_{\text{lin}}$ and $\mathbf{t}^\diamond_{\text{weak}}$ for $8 \le n \le 34$ even and also for $n$ being a multiple of 7. Note that for $n \le 7$, $\mathbf{t}^\diamond_{\text{weak}}$ is equivalent to $\mathbf{t}^*_{\text{lin}}$.

We observe that the found best weak flip codes always have a larger 4-wise Hamming distance and that $d_{\min;4}$ increases as $n$ grows. This is consistent with Theorem 58.

### 5.7.2 Comparisons for $\mathsf{M} = 16$

If we increase the number of codewords to $\mathsf{M} = 16$, the number of weak flip columns increases to $\mathsf{L} = \binom{15}{8} = 6435$. This turns out to be too large even for the simulated annealing algorithm used in Section 5.7.1. So, we had to reduce complexity further. In the following we are going to explain an alternative method of searching for a well-performing weak flip code with large $r$-wise Hamming distances. The idea is to take a fair linear code with $\mathsf{M} = 16$ codewords and with the short blocklength $\mathsf{K} = 15$, and to concatenate $\kappa$ copies of this code with randomly permuted codewords. By numerically searching through many such codes and picking the best one, one obtains a good weak flip code. Note that this algorithm can be used to create nonlinear weak flip codes of any blocklength satisfying $n \bmod \mathsf{K} = 0$ (apart from $n = \mathsf{K}$, for which the code will be linear).

**Step 1:** We choose an initial fair linear code $\mathscr{C}^{(\mathsf{M},\mathsf{K})}_{\text{lin,fair}}$ of blocklength $n = \mathsf{K}$ (this can always be done in a fashion similar to Example 25). We fix some $\kappa \in \mathbb{N} \setminus \{1\}$ and set $p \leftarrow 1$.

**Step 2:** We create $\kappa - 1$ codebooks $\mathscr{C}^{(\mathsf{M},\mathsf{K})}_j$, $j = 2, \ldots, \kappa$, by randomly permuting the codewords of $\mathscr{C}^{(\mathsf{M},\mathsf{K})}_{\text{lin,fair}}$ except the all-zero codeword (which remains on first position). Then we concatenate $\mathscr{C}^{(\mathsf{M},\mathsf{K})}_{\text{lin,fair}}$ with these $\kappa - 1$ codebooks to obtain a length-$(\kappa\mathsf{K})$ code:

$$\mathscr{C}^{(\mathsf{M},\kappa\mathsf{K})}_{\text{weak}} = \bigl[\mathscr{C}^{(\mathsf{M},\mathsf{K})}, \mathscr{C}^{(\mathsf{M},\mathsf{K})}_2, \ldots, \mathscr{C}^{(\mathsf{M},\mathsf{K})}_\kappa\bigr].$$

We compute the corresponding $P_{\text{e}}\bigl(\mathscr{C}^{(\mathsf{M},\mathsf{K}\kappa)}_{\text{weak}}\bigr)$ (using (111)), and if $P_{\text{e}}\bigl(\mathscr{C}^{(\mathsf{M},\mathsf{K}\kappa)}_{\text{weak}}\bigr) < p$, we replace any previously stored code by this one and set $p \leftarrow P_{\text{e}}\bigl(\mathscr{C}^{(\mathsf{M},\mathsf{K}\kappa)}_{\text{weak}}\bigr)$.

**Step 3:** We repeat **Step 2** until a prescribed number of iterations has been performed.

Note that Proposition 21 guarantees that the created code $\mathscr{C}^{(\mathsf{M},\mathsf{K}\kappa)}_{\text{weak}}$ is a weak flip code. Moreover, since we fix the first $\mathsf{K}$ columns of $\mathscr{C}^{(\mathsf{M},\mathsf{K}\kappa)}_{\text{weak}}$, the resulting code is only linear if it is a fair linear code, which happens only with a very small probability equal to $\bigl(\frac{1}{\mathsf{M}!}\bigr)^{\kappa-1}$.

In order to find a good code, we choose as initial code a fair linear code that achieves the largest minimum pairwise Hamming distance. The results are summarized in Table 2(a).

For blocklengths $n < 30$ (for which $n \neq \kappa\mathsf{K}$ with $\kappa \in \mathbb{N} \setminus \{1\}$ and $\mathsf{K} = 15$, and hence the above algorithm does not work) we start with the weak flip columns taken from the best weak flip code $\mathscr{C}^{(16,30)}_{\text{weak}}$ obtained with the above algorithm and then apply a modified version of the simulated annealing algorithm from Section 5.7.1 (in Step 1 $\mathcal{C}^{(\mathsf{M})}_{\text{weak}}$ is replaced by the weak flip columns taken from $\mathscr{C}^{(16,30)}_{\text{weak}}$) to determine $\mathbf{t}^\diamond_{\text{weak}}$. For the best linear code we use simulated annealing to obtain the best punctured linear code by deleting $30 - n$ coordinates from the fair linear code $\mathscr{C}_{\text{lin,fair}}^{(16,30)}$. This yields Table 2(b).

Table 1: The minimum $r$-wise Hamming distances of the best found weak flip codes and the best linear codes with $\mathsf{M} = 8$ for $8 \le n \le 35$. Note that for any blocklength $n$, the performance of $\mathscr{C}^{(8,n)}_{\mathbf{t}^{\diamond}_{\text{weak}}}$ is always strictly better than $\mathscr{C}^{(8,n)}_{\mathbf{t}^{*}_{\text{lin}}}$.

| $\mathsf{M}$ | $n$ | | $d_{\text{min};2}$ | $d_{\text{min};3}$ | $d_{\text{min};4}$ |
|---|---|---|---|---|---|
| 8 | 8 | $\mathbf{t}^{\diamond}_{\text{weak}}$ | 4 | 6 | 7 |
| | | $\mathbf{t}^{*}_{\text{lin}}$ | 4 | 6 | 6 |
| | 10 | $\mathbf{t}^{\diamond}_{\text{weak}}$ | 5 | 8 | 9 |
| | | $\mathbf{t}^{*}_{\text{lin}}$ | 5 | 8 | 8 |
| | 12 | $\mathbf{t}^{\diamond}_{\text{weak}}$ | 6 | 10 | 11 |
| | | $\mathbf{t}^{*}_{\text{lin}}$ | 6 | 10 | 10 |
| | 14 | $\mathbf{t}^{\diamond}_{\text{weak}}$ | 8 | 12 | 13 |
| | | $\mathbf{t}^{*}_{\text{lin}}$ | 8 | 12 | 12 |
| | 16 | $\mathbf{t}^{\diamond}_{\text{weak}}$ | 8 | 13 | 15 |
| | | $\mathbf{t}^{*}_{\text{lin}}$ | 8 | 13 | 13 |
| | 18 | $\mathbf{t}^{\diamond}_{\text{weak}}$ | 10 | 15 | 17 |
| | | $\mathbf{t}^{*}_{\text{lin}}$ | 10 | 15 | 15 |
| | 20 | $\mathbf{t}^{\diamond}_{\text{weak}}$ | 11 | 17 | 19 |
| | | $\mathbf{t}^{*}_{\text{lin}}$ | 11 | 17 | 17 |
| | 21 | $\mathbf{t}^{\diamond}_{\text{weak}}$ | 12 | 18 | 20 |
| | | $\mathbf{t}^{*}_{\text{lin}}$ | 12 | 18 | 18 |
| | 22 | $\mathbf{t}^{\diamond}_{\text{weak}}$ | 12 | 18 | 21 |
| | | $\mathbf{t}^{*}_{\text{lin}}$ | 12 | 18 | 18 |
| | 24 | $\mathbf{t}^{\diamond}_{\text{weak}}$ | 13 | 20 | 23 |
| | | $\mathbf{t}^{*}_{\text{lin}}$ | 13 | 20 | 20 |
| | 26 | $\mathbf{t}^{\diamond}_{\text{weak}}$ | 14 | 22 | 25 |
| | | $\mathbf{t}^{*}_{\text{lin}}$ | 14 | 22 | 22 |
| | 28 | $\mathbf{t}^{\diamond}_{\text{weak}}$ | 16 | 24 | 27 |
| | | $\mathbf{t}^{*}_{\text{lin}}$ | 16 | 24 | 24 |
| | 30 | $\mathbf{t}^{\diamond}_{\text{weak}}$ | 16 | 25 | 29 |
| | | $\mathbf{t}^{*}_{\text{lin}}$ | 16 | 25 | 25 |
| | 32 | $\mathbf{t}^{\diamond}_{\text{weak}}$ | 18 | 27 | 31 |
| | | $\mathbf{t}^{*}_{\text{lin}}$ | 18 | 27 | 27 |
| | 34 | $\mathbf{t}^{\diamond}_{\text{weak}}$ | 19 | 29 | 33 |
| | | $\mathbf{t}^{*}_{\text{lin}}$ | 19 | 29 | 29 |
| | 35 | $\mathbf{t}^{\diamond}_{\text{weak}}$ | 20 | 30 | 34 |
| | | $\mathbf{t}^{*}_{\text{lin}}$ | 20 | 30 | 30 |

Table 2 again validates our quality criterion of good codes: large minimum $r$-wise Hamming distances. The found nonlinear weak flip codes are always superior to the corresponding best linear codes and they all have larger minimum $r$-wise Hamming distances for some $r > 2$ than the corresponding best linear codes. We can also see that for some $r \geq 4$, the difference between the $d_{\min;r}$ of the best weak flip code and the $d_{\min;r}$ of the best linear code increases when $n$ grows.

## 6 Conclusion

In this paper we have broken away from traditional coding theory that focuses on finding codes with sufficient structure (like linearity) to allow efficient encoding and decoding and that analyzes such codes' performance for *large* blocklengths. Instead we have put our emphasis on optimal design in the sense of minimizing the average error probability (under ML decoding) for any *finite* blocklength. To that goal we have proposed a column-wise approach to the codebook matrix that allows us to define families of codes with interesting properties. Also based on the column-wise analysis of codebooks, we have further proposed an extension to the pairwise Hamming distance, called *r-wise Hamming distance*, investigated its properties and proven that it is a key factor to determine the exact error probability of a binary code of arbitrary blocklength $n$ on a BEC.

We have introduced the *weak flip codes*, a new class of codes containing both the class of binary nonlinear Hadamard codes and the class of linear codes as special cases. We have shown that weak flip codes have many desirable properties; in particular, we have succeeded in proving that besides retaining many of the good Hamming distance properties of Hadamard codes, they are actually optimal with respect to the minimum error probability over a BEC for certain numbers of codewords $\mathsf{M}$ and many finite blocklengths $n$.

The family of *fair weak flip codes*—a subclass of the nonlinear weak flip codes—can be seen as a generalization of linear codes to arbitrary numbers of codewords $\mathsf{M}$. We have shown that fair weak flip codes are optimal with respect to the average error probability for the BEC for $\mathsf{M} \leq 4$ and a blocklength that is a multiple of $\mathsf{L}$ and we have conjectured that this result continues to hold also for $\mathsf{M} > 5$. Furthermore, we have also shown that the optimal code performance is really close to the upper bound of Shannon–Gallager–Berlekamp on the BEC for $\mathsf{M} \leq 4$, while for the BSC this is not the case.

Note that it has been known for quite some time that binary nonlinear Hadamard codes have good Hamming distance properties [12]; however, their behavior with respect to error probability for finite blocklength remained uninvestigated. In particular, while fair weak flip codes have been used before (although without being named) in the derivation of results related to error probability [21] and have been shown to be best-error-exponent achieving, their global (among all possible linear or nonlinear codes) optimality with respect to the error probability was not known so far.

In conclusion, this paper tries to build a bridge between coding theory, which usually is concerned with the design of codes with good Hamming distance properties (like, e.g., the binary nonlinear Hadamard codes), and information theory, which deals with error probability and with the existence of codes that have good or optimal error probability

Table 2: The minimum $r$-wise Hamming distances of the best found weak flip codes and the best linear codes with $\mathsf{M} = 16$ for certain values of $n$. Note that for any blocklength $n$, the performance of $\mathscr{C}^{(16,n)}_{\mathbf{t}^{\diamond}_{\text{weak}}}$ is always strictly better than $\mathscr{C}^{(16,n)}_{\mathbf{t}^{\diamond}_{\text{lin}}}$.

| $\mathsf{M}$ | $n$ | | $d_{\min;2}$ | $d_{\min;3}$ | $d_{\min;4}$ | $d_{\min;5}$ | $d_{\min;6}$ | $d_{\min;7}$ | $d_{\min;8}$ |
|---|---|---|---|---|---|---|---|---|---|
| 16 | 30 | $\mathbf{t}^{\diamond}_{\text{weak}}$ | 16 | 24 | 24 | 28 | 28 | 28 | 29 |
| | | $\mathbf{t}^{\diamond}_{\text{lin}}$ | 16 | 24 | 24 | 28 | 28 | 28 | 28 |
| | 45 | $\mathbf{t}^{\diamond}_{\text{weak}}$ | 24 | 36 | 38 | 42 | 42 | 44 | 44 |
| | | $\mathbf{t}^{\diamond}_{\text{lin}}$ | 24 | 36 | 36 | 42 | 42 | 42 | 42 |
| | 60 | $\mathbf{t}^{\diamond}_{\text{weak}}$ | 32 | 48 | 52 | 56 | 57 | 58 | 59 |
| | | $\mathbf{t}^{\diamond}_{\text{lin}}$ | 32 | 48 | 48 | 56 | 56 | 56 | 56 |
| | 75 | $\mathbf{t}^{\diamond}_{\text{weak}}$ | 40 | 60 | 64 | 70 | 72 | 73 | 74 |
| | | $\mathbf{t}^{\diamond}_{\text{lin}}$ | 40 | 60 | 60 | 70 | 70 | 70 | 70 |
| | 90 | $\mathbf{t}^{\diamond}_{\text{weak}}$ | 48 | 72 | 78 | 84 | 86 | 88 | 89 |
| | | $\mathbf{t}^{\diamond}_{\text{lin}}$ | 48 | 72 | 72 | 84 | 84 | 84 | 84 |
| | 105 | $\mathbf{t}^{\diamond}_{\text{weak}}$ | 56 | 84 | 92 | 98 | 101 | 102 | 104 |
| | | $\mathbf{t}^{\diamond}_{\text{lin}}$ | 56 | 84 | 84 | 98 | 98 | 98 | 98 |

(a) $n = \kappa\mathsf{K}$ with $\kappa \in \mathbb{N} \setminus \{1\}$ and $\mathsf{K} = 15$

| $\mathsf{M}$ | $n$ | | $d_{\min;2}$ | $d_{\min;3}$ | $d_{\min;4}$ | $d_{\min;5}$ | $d_{\min;6}$ | $d_{\min;7}$ | $d_{\min;8}$ |
|---|---|---|---|---|---|---|---|---|---|
| 16 | 16 | $\mathbf{t}^{\diamond}_{\text{weak}}$ | 8 | 12 | 12 | 14 | 14 | 15 | 15 |
| | | $\mathbf{t}^{\diamond}_{\text{lin}}$ | 8 | 12 | 12 | 14 | 14 | 14 | 14 |
| | 18 | $\mathbf{t}^{\diamond}_{\text{weak}}$ | 8 | 13 | 14 | 16 | 16 | 17 | 17 |
| | | $\mathbf{t}^{\diamond}_{\text{lin}}$ | 8 | 13 | 13 | 16 | 16 | 16 | 16 |
| | 20 | $\mathbf{t}^{\diamond}_{\text{weak}}$ | 10 | 15 | 15 | 18 | 18 | 19 | 19 |
| | | $\mathbf{t}^{\diamond}_{\text{lin}}$ | 10 | 15 | 15 | 18 | 18 | 18 | 18 |
| | 22 | $\mathbf{t}^{\diamond}_{\text{weak}}$ | 11 | 17 | 17 | 20 | 20 | 21 | 21 |
| | | $\mathbf{t}^{\diamond}_{\text{lin}}$ | 11 | 17 | 17 | 20 | 20 | 20 | 20 |
| | 24 | $\mathbf{t}^{\diamond}_{\text{weak}}$ | 12 | 18 | 19 | 22 | 22 | 22 | 23 |
| | | $\mathbf{t}^{\diamond}_{\text{lin}}$ | 12 | 18 | 18 | 22 | 22 | 22 | 22 |
| | 26 | $\mathbf{t}^{\diamond}_{\text{weak}}$ | 13 | 20 | 20 | 24 | 24 | 25 | 25 |
| | | $\mathbf{t}^{\diamond}_{\text{lin}}$ | 13 | 20 | 20 | 24 | 24 | 24 | 24 |
| | 28 | $\mathbf{t}^{\diamond}_{\text{weak}}$ | 14 | 22 | 22 | 26 | 26 | 26 | 27 |
| | | $\mathbf{t}^{\diamond}_{\text{lin}}$ | 14 | 22 | 22 | 26 | 26 | 26 | 26 |

(b) $16 \leq n \leq 28$

behavior (even though often in the asymptotic sense for very large blocklengths). Our results suggest that in order to have good performance in the finite blocklength regime for the BEC, one must find a code design with large minimum $r$-wise Hamming distances for all $r \in \{2, 3, \ldots, \bar{\ell}\}$.

# A Proof of Theorem 51

We refer to [19, Def. 33] and define

$$P_{\mathrm{c}}\big(\mathscr{C}^{(\mathsf{M},n+\gamma)}\big) = P_{\mathrm{c}}\big(\mathscr{C}^{(\mathsf{M},n)}\big) + \frac{1}{\mathsf{M}}\sum_{m=1}^{\mathsf{M}} \sum_{\substack{\mathbf{y}^{(n+\gamma)} \\ \text{s.t. } \mathbf{y}^{(n)}\in\mathcal{D}_m^{(\mathsf{M},n)} \\ \text{but } \mathbf{y}^{(n+\gamma)}\in\mathcal{D}_{m'}^{(\mathsf{M},n+\gamma)} \\ \text{for some } m'\neq m}} \Bigg( P_{\mathbf{Y}|\mathbf{X}}\Big(\mathbf{y}^{(n+\gamma)}\Big|\mathbf{x}_{m'}^{(n+\gamma)}\Big) - P_{\mathbf{Y}|\mathbf{X}}\big(\mathbf{y}^{(n+\gamma)}\big|\mathbf{x}_{m}^{(n+\gamma)}\big)\Bigg) \tag{153}$$

$$\triangleq P_{\mathrm{c}}\big(\mathscr{C}^{(\mathsf{M},n)}\big) + \Delta\Psi\big(\mathscr{C}^{(\mathsf{M},n+\gamma)}\big). \tag{154}$$

In the proof of Theorem 51, our goal is to maximize the total probability increase $\Delta\Psi\big(\mathscr{C}^{(\mathsf{M},n+\gamma)}\big)$ among all possible $\mathscr{C}^{(\mathsf{M},\gamma)}$ with $\gamma = 1$ for $\mathsf{M} = 3, 4$. Note that the codebook $\mathscr{C}^{(\mathsf{M},n+\gamma)}$ is formed by concatenating $\mathscr{C}^{(\mathsf{M},n)}$ with $\mathscr{C}^{(\mathsf{M},\gamma)}$. The proof is based on induction and follows along the same lines as in the proof for the BSC shown in [19, App. C.A] with some modifications that take into account the details of the decoding rule for the BEC. Similarly to [19, App. C.A], we need a case distinction depending on $n \bmod 3$. For space reason, we only outline the case from $n-1 = 3k-1$ to $n = 3k$. Moreover, we only consider the more complicated case of $\mathsf{M} = 4$. Similar arguments can be applied to $\mathsf{M} = 3$.

We start with the code $\mathscr{C}_{\mathbf{t}^{\diamond}_{\text{weak}}}^{(4,n-1)}$, whose type is as follows:

$$\mathbf{t}^{\diamond}_{\text{weak}} = [t^{\diamond}_3, t^{\diamond}_5, t^{\diamond}_6] = [k, k, k-1], \tag{155}$$

and need to pick a candidate columns from $\mathcal{C}^{(4)}$ to append to $\mathscr{C}_{\mathbf{t}^{\diamond}_{\text{weak}}}^{(4,n-1)}$. We require to show that appending $\mathbf{c}_6^{(4)}$ yields the largest total probability increase among all possible candidate columns in $\mathcal{C}^{(4)}$.

To that goal, we investigate how to extend the decoding regions of $\mathscr{C}_{\mathbf{t}^{\diamond}_{\text{weak}}}^{(4,n-1)}$. For each codeword, there are three possible extended decoding regions of blocklength $n$:

$$\big[\mathcal{D}_m^{(4,n-1)}\ 0\big], \quad \big[\mathcal{D}_m^{(4,n-1)}\ 1\big], \quad \big[\mathcal{D}_m^{(4,n-1)}\ 2\big], \quad m = 1, \ldots, 4. \tag{156}$$

Owing to the fact that for a BEC $P_{Y|X}(0|1) = P_{Y|X}(1|0) = 0$, and using $b \in \{0, 1\}$ to denote the value of the appended bit to the $m$th codeword, $x_{m,n} = b$, we see that the decoding region $\mathcal{D}_m^{(4,n)}$ should include both $\big[\mathcal{D}_m^{(4,n-1)}\ b\big]$ and $\big[\mathcal{D}_m^{(4,n-1)}\ 2\big]$, and that all the received vectors in $\big[\mathcal{D}_m^{(4,n-1)}\ \bar{b}\big]$ will be decoded to one of the other three codewords. Since

$$\psi_m\big(\mathscr{C}^{(4,n-1)}\big) = \psi_m\big(\mathscr{C}^{(4,n-1)}\big)\cdot(1-\delta+\delta) \tag{157}$$

$$= \Pr\Big(\mathcal{D}_m^{(4,n-1)}\Big|\mathbf{x}_m^{(n-1)}\Big)\big(P_{Y|X}(b|b) + P_{Y|X}(2|b)\big) \tag{158}$$

$$= \Pr\Big(\big[\mathcal{D}_m^{(4,n-1)}\ b\big]\Big|\big[\mathbf{x}_m^{(n-1)}\ b\big]\Big) + \Pr\Big(\big[\mathcal{D}_m^{(4,n-1)}\ 2\big]\Big|\big[\mathbf{x}_m^{(n-1)}\ b\big]\Big), \tag{159}$$

we obtain that $\left[\mathcal{D}_m^{(4,n-1)}\ b\right] \cup \left[\mathcal{D}_m^{(4,n-1)}\ 2\right]$ does not create any probability increase, i.e., the total probability increase for each codeword will be determined by how the received vectors in $\left[\mathcal{D}_m^{(4,n-1)}\ \bar{b}\right]$ are moved to one of decoding regions of the other three codewords.

We now investigate each possible appended column in a case-by-case fashion.

**Appending $\mathbf{c}_1^{(4)}$:** We build a new length-$n$ code $\mathscr{C}_{\mathbf{t}}^{(4,n)}$ from the given code $\mathscr{C}_{\mathbf{t}^\diamond_{\text{weak}}}^{(4,n-1)}$ by appending $\mathbf{c}_1^{(4)} = (0\ 0\ 0\ 1)^\mathsf{T}$. The type becomes

$$\mathbf{t}_1 = [1, 0, k, 0, k, k-1, 0]. \tag{160}$$

We now compute the total probability increase in this case. Because $x_{4,n} = 1$ and $x_{m,n} = 0$ for $m = 1, 2, 3$, some[24] of the vectors in the extended decoding regions $\left[\mathcal{D}_{\mathbf{t}^\diamond_{\text{weak}};m}^{(4,n-1)}\ 1\right]$ for $m = 1, 2, 3$ will be moved to $\mathcal{D}_{\mathbf{t}_1;4}^{(4,n)}$ (and some of the received vectors in the extended decoding region $\left[\mathcal{D}_{\mathbf{t}^\diamond_{\text{weak}};4}^{(4,n-1)}\ 0\right]$ will be moved to one of $\mathcal{D}_{\mathbf{t}_1;m}^{(4,n)}$, $m = 1, 2, 3$). The total probability increase $\Delta\Psi\big(\mathscr{C}_{\mathbf{t}_1}^{(4,n)}\big)$ is

$$\begin{aligned}
&\Delta\Psi\left(\mathscr{C}_{\mathbf{t}_1}^{(4,n)}\right) \\
&= \Pr\left(\left[\overline{\mathcal{D}}_4^{(4,n-1)}\ 1\right] \cap \left(\left[\overline{\mathcal{D}}_1^{(4,n-1)}\ 1\right] \cup \left[\overline{\mathcal{D}}_2^{(4,n-1)}\ 1\right] \cup \left[\overline{\mathcal{D}}_3^{(4,n-1)}\ 1\right]\right) \,\middle|\, \left[\mathbf{x}_4^{(n-1)}\ 1\right]\right) && (161)\\
&= \Pr\left(\overline{\mathcal{D}}_4^{(4,n-1)} \cap \left(\bigcup_{m=1}^{3} \overline{\mathcal{D}}_m^{(4,n-1)}\right) \,\middle|\, \mathbf{x}_4^{(n-1)}\right)(1-\delta) && (162)\\
&= \Pr\left(\bigcup_{m=1}^{3} \left(\overline{\mathcal{D}}_m^{(4,n-1)} \cap \overline{\mathcal{D}}_4^{(4,n-1)}\right) \,\middle|\, \mathbf{x}_4^{(n-1)}\right)(1-\delta) && (163)\\
&= \Bigg(\Pr\left(\overline{\mathcal{D}}_1^{(4,n-1)} \cap \overline{\mathcal{D}}_4^{(4,n-1)} \,\middle|\, \mathbf{x}_4^{(n-1)}\right) \\
&\qquad + \Pr\left(\overline{\mathcal{D}}_2^{(4,n-1)} \cap \overline{\mathcal{D}}_4^{(4,n-1)} \,\middle|\, \mathbf{x}_4^{(n-1)}\right) \\
&\qquad + \Pr\left(\overline{\mathcal{D}}_3^{(4,n-1)} \cap \overline{\mathcal{D}}_4^{(4,n-1)} \,\middle|\, \mathbf{x}_4^{(n-1)}\right) \\
&\qquad - \Pr\left(\overline{\mathcal{D}}_1^{(4,n-1)} \cap \overline{\mathcal{D}}_2^{(4,n-1)} \cap \overline{\mathcal{D}}_4^{(4,n-1)} \,\middle|\, \mathbf{x}_4^{(n-1)}\right) \\
&\qquad - \Pr\left(\overline{\mathcal{D}}_1^{(4,n-1)} \cap \overline{\mathcal{D}}_3^{(4,n-1)} \cap \overline{\mathcal{D}}_4^{(4,n-1)} \,\middle|\, \mathbf{x}_4^{(n-1)}\right) \\
&\qquad - \Pr\left(\overline{\mathcal{D}}_2^{(4,n-1)} \cap \overline{\mathcal{D}}_3^{(4,n-1)} \cap \overline{\mathcal{D}}_4^{(4,n-1)} \,\middle|\, \mathbf{x}_4^{(n-1)}\right) \\
&\qquad + \Pr\left(\overline{\mathcal{D}}_1^{(4,n-1)} \cap \overline{\mathcal{D}}_2^{(4,n-1)} \cap \overline{\mathcal{D}}_3^{(4,n-1)} \cap \overline{\mathcal{D}}_4^{(4,n-1)} \,\middle|\, \mathbf{x}_4^{(n-1)}\right)\Bigg)(1-\delta) && (164)\\
&= \left(\delta^{n-1-t_6^\diamond} + \delta^{n-1-t_5^\diamond} + \delta^{n-1-t_3^\diamond} - \delta^{n-1} - \delta^{n-1} - \delta^{n-1} + \delta^{n-1}\right)(1-\delta) && (165)\\
&= \left(\delta^{2k-1} + \delta^{2k-1} + \delta^{2k} - 2\delta^{n-1}\right)(1-\delta), && (166)
\end{aligned}$$

where (161) holds because of the definition of the closed decoding regions and because $\left[\overline{\mathcal{D}}_4^{(4,n-1)}\ 1\right] \cap \left[\overline{\mathcal{D}}_m^{(4,n-1)}\ 1\right]$, $m = 1, 2, 3$, are not empty; (162) is because the BEC is a DMC; (164) follows directly from applying the inclusion–exclusion principle; and finally, (165) follows from the same $r$-wise Hamming distances perspective as already used in the derivations of Theorem 49.

[24] The reason why we write "some" instead of "all" is that some vectors in $\left[\mathcal{D}_{\mathbf{t}^\diamond_{\text{weak}};m}^{(4,n-1)}\ 1\right]$ cannot occur and fall out of consideration.

**Appending $\mathbf{c}_2^{(4)}$:** The derivations here are similar to the first case (or, indeed, also for the cases of appending $\mathbf{c}_4^{(4)}$ or $\mathbf{c}_7^{(4)}$), so we omit the details and directly state the total probability increase:

$$\begin{aligned}
&\Delta\Psi\Big(\mathscr{C}_{\mathbf{t}_2}^{(4,n)}\Big) \\
&= \big(\delta^{n-1-t_5^\diamond} + \delta^{n-1-t_6^\diamond} + \delta^{n-1-t_3^\diamond} - \delta^{n-1} - \delta^{n-1} - \delta^{n-1} + \delta^{n-1}\big)(1-\delta) && (167)\\
&= \big(\delta^{2k-1} + \delta^{2k-1} + \delta^{2k} - 2\delta^{n-1}\big)(1-\delta). && (168)
\end{aligned}$$

**Appending $\mathbf{c}_3^{(4)}$:** If we append $\mathbf{c}_3^{(4)} = (0\ 0\ 1\ 1)^{\mathsf{T}}$, the new type for blocklength $n$ becomes

$$\mathbf{t}_3 = [0, 0, k+1, 0, k, k-1, 0]. \tag{169}$$

Since $x_{1,n} = x_{2,n} = 0$ and $x_{3,n} = x_{4,n} = 1$, again using an argument like in the first case, we find that some received vectors in the extended decoding regions $\big[\mathcal{D}_1^{(4,n-1)}\ 1\big]$ and $\big[\mathcal{D}_2^{(4,n-1)}\ 1\big]$ will be moved to either $\mathcal{D}_3^{(4,n)}$ or $\mathcal{D}_4^{(4,n)}$. We obtain a total probability increase

$$\begin{aligned}
&\Delta\Psi\Big(\mathscr{C}_{\mathbf{t}_3}^{(4,n)}\Big) \\
&= \Pr\Big(\Big(\big[\overline{\mathcal{D}}_1^{(4,n-1)}\ 1\big] \cup \big[\overline{\mathcal{D}}_2^{(4,n-1)}\ 1\big]\Big) \cap \big[\overline{\mathcal{D}}_3^{(4,n-1)}\ 1\big] \Big| \big[\mathbf{x}_3^{(n-1)}\ 1\big]\Big) \\
&\quad + \Pr\Big(\Big(\big[\overline{\mathcal{D}}_1^{(4,n-1)}\ 1\big] \cup \big[\overline{\mathcal{D}}_2^{(4,n-1)}\ 1\big]\Big) \cap \big[\overline{\mathcal{D}}_4^{(4,n-1)}\ 1\big] \Big| \big[\mathbf{x}_4^{(n-1)}\ 1\big]\Big) \\
&\quad - \Pr\Big(\Big(\big[\overline{\mathcal{D}}_1^{(4,n-1)}\ 1\big] \cup \big[\overline{\mathcal{D}}_2^{(4,n-1)}\ 1\big]\Big) \\
&\qquad\qquad \cap \Big(\big[\overline{\mathcal{D}}_3^{(4,n-1)}\ 1\big] \cap \big[\overline{\mathcal{D}}_4^{(4,n-1)}\ 1\big]\Big) \Big| \big[\mathbf{x}_{\ell,\ell\in\{3,4\}}^{(n-1)}\ 1\big]\Big) && (170)\\
&= \Big(\Pr\Big(\overline{\mathcal{D}}_1^{(4,n-1)} \cap \overline{\mathcal{D}}_3^{(4,n-1)} \Big| \mathbf{x}_3^{(n-1)}\Big) \\
&\quad + \Pr\Big(\overline{\mathcal{D}}_2^{(4,n-1)} \cap \overline{\mathcal{D}}_3^{(4,n-1)} \Big| \mathbf{x}_3^{(n-1)}\Big) \\
&\quad - \Pr\Big(\overline{\mathcal{D}}_1^{(4,n-1)} \cap \overline{\mathcal{D}}_2^{(4,n-1)} \cap \overline{\mathcal{D}}_3^{(4,n-1)} \Big| \mathbf{x}_3^{(n-1)}\Big) \\
&\quad + \Pr\Big(\overline{\mathcal{D}}_1^{(4,n-1)} \cap \overline{\mathcal{D}}_4^{(4,n-1)} \Big| \mathbf{x}_4^{(n-1)}\Big) \\
&\quad + \Pr\Big(\overline{\mathcal{D}}_2^{(4,n-1)} \cap \overline{\mathcal{D}}_4^{(4,n-1)} \Big| \mathbf{x}_4^{(n-1)}\Big) \\
&\quad - \Pr\Big(\overline{\mathcal{D}}_1^{(4,n-1)} \cap \overline{\mathcal{D}}_2^{(4,n-1)} \cap \overline{\mathcal{D}}_4^{(4,n-1)} \Big| \mathbf{x}_4^{(n-1)}\Big) \\
&\quad - \Pr\Big(\overline{\mathcal{D}}_1^{(4,n-1)} \cap \overline{\mathcal{D}}_3^{(4,n-1)} \cap \overline{\mathcal{D}}_4^{(4,n-1)} \Big| \mathbf{x}_{\ell,\ell\in\{3,4\}}^{(n-1)}\Big) \\
&\quad - \Pr\Big(\overline{\mathcal{D}}_2^{(4,n-1)} \cap \overline{\mathcal{D}}_3^{(4,n-1)} \cap \overline{\mathcal{D}}_4^{(4,n-1)} \Big| \mathbf{x}_{\ell,\ell\in\{3,4\}}^{(n-1)}\Big) \\
&\quad + \Pr\Big(\overline{\mathcal{D}}_1^{(4,n-1)} \cap \overline{\mathcal{D}}_2^{(4,n-1)} \cap \overline{\mathcal{D}}_3^{(4,n-1)} \cap \overline{\mathcal{D}}_4^{(4,n-1)} \Big| \mathbf{x}_{\ell,\ell\in\{3,4\}}^{(n-1)}\Big)\Big)(1-\delta) && (171)\\
&= \big(\delta^{n-1-t_5^\diamond} + \delta^{n-1-t_6^\diamond} - \delta^{n-1} + \delta^{n-1-t_6^\diamond} + \delta^{n-1-t_5^\diamond} - \delta^{n-1} \\
&\qquad - \delta^{n-1} - \delta^{n-1} + \delta^{n-1}\big)(1-\delta) && (172)\\
&= \big(\delta^{2k-1} + \delta^{2k} + \delta^{2k} + \delta^{2k-1} - 3\delta^{n-1}\big)(1-\delta), && (173)
\end{aligned}$$

where in (170) we use the rule of total probability;[25] in (171) we apply the inclusion–exclusion principle; and where (172) again follows from the $r$-wise Hamming distances perspective.

**Appending $\mathbf{c}_4^{(4)}$:** Using an argumentation similar to the case of appending $\mathbf{c}_1^{(4)}$, we have a total probability increase

$$\Delta\Psi\left(\mathscr{C}_{\mathbf{t}_4}^{(4,n)}\right) = \left(\delta^{n-1-t_3^\diamond} + \delta^{n-1-t_6^\diamond} + \delta^{n-1-t_5^\diamond} - \delta^{n-1} - \delta^{n-1} - \delta^{n-1} + \delta^{n-1}\right)(1-\delta) \quad (174)$$
$$= \left(\delta^{2k-1} + \delta^{2k} + \delta^{2k-1} - 2\delta^{n-1}\right)(1-\delta). \quad (175)$$

**Appending $\mathbf{c}_5^{(4)}$:** Using an argumentation similar to the case of appending $\mathbf{c}_3^{(4)}$, we have a total probability increase

$$\Delta\Psi\left(\mathscr{C}_{\mathbf{t}_5}^{(4,n)}\right) = \left(\delta^{n-1-t_3^\diamond} + \delta^{n-1-t_6^\diamond} - \delta^{n-1} + \delta^{n-1-t_6^\diamond} + \delta^{n-1-t_3^\diamond} - \delta^{n-1} - \delta^{n-1} - \delta^{n-1} + \delta^{n-1}\right)(1-\delta) \quad (176)$$
$$= \left(\delta^{2k-1} + \delta^{2k} + \delta^{2k} + \delta^{2k-1} - 3\delta^{n-1}\right)(1-\delta). \quad (177)$$

**Appending $\mathbf{c}_6^{(4)}$:** Using an argumentation similar to the case of appending $\mathbf{c}_3^{(4)}$, we have a total probability increase

$$\Delta\Psi\left(\mathscr{C}_{\mathbf{t}_6}^{(4,n)}\right) = \left(\delta^{n-1-t_3^\diamond} + \delta^{n-1-t_5^\diamond} - \delta^{n-1} + \delta^{n-1-t_3^\diamond} + \delta^{n-1-t_5^\diamond} - \delta^{n-1} - \delta^{n-1} - \delta^{n-1} + \delta^{n-1}\right)(1-\delta) \quad (178)$$
$$= \left(\delta^{2k-1} + \delta^{2k-1} + \delta^{2k-1} + \delta^{2k-1} - 3\delta^{n-1}\right)(1-\delta). \quad (179)$$

**Appending $\mathbf{c}_7^{(4)}$:** Using an argumentation similar to the case of appending $\mathbf{c}_1^{(4)}$, we have a total probability increase

$$\Delta\Psi\left(\mathscr{C}_{\mathbf{t}_7}^{(4,n)}\right) = \left(\delta^{n-1-t_3^\diamond} + \delta^{n-1-t_5^\diamond} + \delta^{n-1-t_6^\diamond} - \delta^{n-1} - \delta^{n-1} - \delta^{n-1} + \delta^{n-1}\right)(1-\delta) \quad (180)$$
$$= \left(\delta^{2k-1} + \delta^{2k-1} + \delta^{2k} - 2\delta^{n-1}\right)(1-\delta). \quad (181)$$

Using the fact that $\delta^d$ is strictly decreasing in $d$ for $0 < \delta < 1$, we can conclude that

$$\operatorname*{argmax}_{1\le j\le 7} \Delta\Psi\left(\mathscr{C}_{\mathbf{t}_j}^{(4,n)}\right) = 6. \quad (182)$$

This completes the proof. The proofs for $n \bmod 3 = 1$ or $2$ are similar and omitted.

# B Proof of Theorem 52

The proof of Theorem 52 is based on the exact average success probability for a BEC as a function of the type vector $\mathbf{t}$ with a blocklength $n = \sum_{j=1}^{J} t_j$. This problem is then transformed into a discrete multivariate constrained optimization problem.

[25] Note that $\left(\left[\overline{\mathcal{D}}_1^{(4,n-1)} \; 1\right] \cup \left[\overline{\mathcal{D}}_2^{(4,n-1)} \; 1\right]\right) \cap \left[\overline{\mathcal{D}}_3^{(4,n-1)} \; 1\right]$ and $\left(\left[\overline{\mathcal{D}}_1^{(4,n-1)} \; 1\right] \cup \left[\overline{\mathcal{D}}_2^{(4,n-1)} \; 1\right]\right) \cap \left[\overline{\mathcal{D}}_4^{(4,n-1)} \; 1\right]$ are not necessarily disjoint.

We define the region of all possible types $\mathbf{t}$ as

$$\mathcal{T}^{(\mathsf{M})} \triangleq \left\{ \mathbf{t} \in (\mathbb{N} \cup \{0\})^{\mathsf{J}} \colon \sum_{j=1}^{\mathsf{J}} t_j = n \right\}. \tag{183}$$

Our goal is to find the globally optimized type $\mathbf{t}^*$ that satisfies

$$\mathbf{t}^* = \operatorname*{argmin}_{\mathbf{t} \in \mathcal{T}^{(\mathsf{M})}} P_{\mathrm{e}}\Big(\mathscr{C}_{\mathbf{t}}^{(\mathsf{M},n)}\Big). \tag{184}$$

Applying Theorem 49 for $\mathsf{M} = 3$ or $\mathsf{M} = 4$, we have

$$P_{\mathrm{e}}\Big(\mathscr{C}_{\mathbf{t}}^{(3,n)}\Big) = \frac{1}{3}\big(\delta^{n-t_1} + \delta^{n-t_2} + \delta^{n-t_3} - \delta^n\big); \tag{185}$$

$$\begin{aligned} P_{\mathrm{e}}\Big(\mathscr{C}_{\mathbf{t}}^{(4,n)}\Big) = \frac{1}{4}\Big(&\delta^{n-(t_1+t_2+t_3)} + \delta^{n-(t_1+t_4+t_5)} + \delta^{n-(t_1+t_6+t_7)} \\ &+ \delta^{n-(t_2+t_4+t_6)} + \delta^{n-(t_2+t_5+t_7)} + \delta^{n-(t_3+t_4+t_7)} \\ &- \delta^{n-t_1} - \delta^{n-t_2} - \delta^{n-t_4} - \delta^{n-t_7} + \delta^n\Big). \end{aligned} \tag{186}$$

Since we consider the optimization problem for any fixed blocklength $n$ and hence $\delta^n$ is a constant, we can reformulate the discrete multivariate constrained minimization problem as follows:

$$\begin{aligned} &\text{minimize } f^{(\mathsf{M})}(\mathbf{t}) \triangleq \frac{\mathsf{M}}{\delta^n} P_{\mathrm{e}}\big(\mathscr{C}_{\mathbf{t}}^{(\mathsf{M},n)}\big) + (-1)^{\mathsf{M}+1} \\ &\text{subject to} \qquad \mathbf{t} \in \mathcal{T}^{(\mathsf{M})} \end{aligned} \tag{187}$$

where the minimization objective functions for $\mathsf{M} = 3$ or $\mathsf{M} = 4$ are

$$f^{(3)}(\mathbf{t}) = \delta^{-t_1} + \delta^{-t_2} + \delta^{-t_3} \tag{188}$$

and

$$\begin{aligned} f^{(4)}(\mathbf{t}) = {}& \delta^{-t_1-t_2-t_3} + \delta^{-t_1-t_4-t_5} + \delta^{-t_1-t_6-t_7} + \delta^{-t_2-t_4-t_6} + \delta^{-t_2-t_5-t_7} \\ &+ \delta^{-t_3-t_4-t_7} - \delta^{-t_1} - \delta^{-t_2} - \delta^{-t_4} - \delta^{-t_7}, \end{aligned} \tag{189}$$

respectively. Note that we add $(-1)^{\mathsf{M}+1}$ in (187) to simplify the expression of $f^{(\mathsf{M})}(\mathbf{t})$.

We firstly consider the easier case of $\mathsf{M} = 3$. Taking the locally optimal type $\mathbf{t}^\diamond$ from Theorem 51, we will now prove that it is actually globally optimal for (188). Using $t_3 = n - t_1 - t_2$, we have

$$\begin{aligned} f^{(3)}(\mathbf{t}) &= \delta^{-t_1} + \delta^{-t_2} + \delta^{t_1+t_2-n} &(190)\\ &\geq 2\sqrt{\delta^{-t_1}\delta^{-t_2}} + \delta^{t_1+t_2-n} &(191)\\ &\triangleq 2\delta^{-t} + \delta^{2t-n} &(192)\\ &\triangleq h(t), &(193) \end{aligned}$$

where (191) holds because the arithmetic mean (AM) is never smaller than the geometric mean (GM), and in (192) we define $t \triangleq (t_1 + t_2)/2$. It can be seen that the function $2\delta^{-t} + \delta^{n-2t}$ is convex in $t$. Hence, its global minimum $3\delta^{-n/3}$ is given for the $t$ satisfying

$$\frac{\partial}{\partial t}\big(2\delta^{-t} + \delta^{2t-n}\big) \overset{!}{=} 0, \tag{194}$$

where "$\stackrel{!}{=}$" means "should be equal to," i.e., the global minimizer of $h(t)$ is $t^* = \frac{n}{3}$. However, one must be aware that the minimizer of $f^{(3)}(\mathbf{t})$ must be a positive integer. So, if $n = 3k$, taking $t_1^* = t_2^* = t_3^* = t^*$ trivially achieves the global minimum of $h(t)$, i.e., $3\delta^{-n/3}$. In the following we will investigate the discrete minimizer $t^*$ for $h(t)$ for the case of $n = 3k + 1$. The case $n = 3k + 2$ is similar and omitted.

Since the function $h(t)$ is convex, the minimizer should be equal to $k$ or $k + 1$. Therefore,

$$\min\{h(k), h(k+1)\} = \min\left\{2\delta^{-k} + \delta^{-(k+1)}, 2\delta^{-(k+1)} + \delta^{-(k-1)}\right\} \tag{195}$$

$$= 2\delta^{-k} + \delta^{-(k+1)} \tag{196}$$

$$= h(k). \tag{197}$$

Here we again use the AM–GM inequality to show that $2\delta^{-k} < \delta^{-(k+1)} + \delta^{-(k-1)}$. Thus the discrete global minimizer for $h(t)$ is $t^* = k$. Finally, since the inequality of (191) is achievable by $[t_1, t_2, t_3] = [k, k, k+1]$, we can conclude that a discrete global minimizer for $f^{(3)}(\mathbf{t})$ is $\mathbf{t}^* = [k, k, k+1]$. Note that in Theorem 52, we state that the optimal type is $\mathbf{t}^* = [k+1, k, k]$. It is not difficult to show that the performance of these two codes is equivalent; so the optimal codes are not unique when $n = 3k + 1$.

In the case of $\mathsf{M} = 4$ we must first prove that the globally optimal type $\mathbf{t}^*$ must satisfy $t_1^* = t_2^* = t_4^* = t_7^* = 0$ for an arbitrary blocklength $n$. This turns out to be quite technical.

We reformulate the optimization problem in (187) as follows: introducing

$$u_j \triangleq \delta^{-t_j}, \quad 1 \leq j \leq \mathsf{J}, \tag{198}$$

and noting that $1 \leq u_j \leq \delta^{-n}$ for $0 < \delta < 1$, we rewrite (189) as

$$g^{(4)}(\mathbf{u}) \triangleq f^{(4)}(\mathbf{t}) \tag{199}$$

and the optimization region (183) as

$$\mathcal{U}^{(4)} \triangleq \left\{ \mathbf{u} \in \mathbb{R}^{\mathsf{J}} : u_j \geq 1 \text{ and } \prod_{j=1}^{\mathsf{J}} u_j = \delta^{-n} \right\}. \tag{200}$$

Note that while $\mathcal{T}^{(4)}$ is convex, $\mathcal{U}^{(4)}$ is not. We have

$$g^{(4)}(\mathbf{u})$$

$$= u_1u_2u_3 + u_1u_4u_5 + u_1u_6u_7 + u_2u_4u_6 + u_2u_5u_7 + u_3u_4u_7 - (u_1 + u_2 + u_4 + u_7) \tag{201}$$

$$= u_1(u_2u_3 + u_4u_5 + u_6u_7 - 1) + u_2u_4u_6 + u_2u_5u_7 + u_3u_4u_7 - (u_2 + u_4 + u_7) \tag{202}$$

$$\geq u_1\Big(3(u_2u_3u_4u_5u_6u_7)^{\frac{1}{3}} - 1\Big) + u_2u_4u_6 + u_2u_5u_7 + u_3u_4u_7 - (u_2 + u_4 + u_7) \tag{203}$$

$$= u_1\left(3\left(\frac{\delta^{-n}}{u_1}\right)^{\frac{1}{3}} - 1\right) + u_2u_4u_6 + u_2u_5u_7 + u_3u_4u_7 - (u_2 + u_4 + u_7) \tag{204}$$

$$= \left(3\delta^{-\frac{n}{3}}u_1^{\frac{2}{3}} - u_1\right) + u_2u_4u_6 + u_2u_5u_7 + u_3u_4u_7 - (u_2 + u_4 + u_7). \tag{205}$$

Here, (203) follows from the AM–GM inequality, where equality holds if

$$u_2u_3 = u_4u_5 = u_6u_7. \tag{206}$$

In (204), we use the fact that $\prod_{j=1}^{7} u_j = \delta^{-n}$. The first term in parentheses on the right-hand-side (RHS) of (205) is concave and nondecreasing in $u_1$ for $1 \leq u_1 \leq \delta^{-n}$, and independent of the other variables $u_2, \ldots, u_7$. This implies that if we want to minimize (205), we should have $u_1^* = 1$ and the minimization is irrelevant to $u_2^*, \ldots, u_7^*$. To achieve equality in (203), we only need to satisfy the condition (206), which means that $u_1^* = 1$ is both the discrete global minimizer of the RHS of (205) and $g^{(4)}(\mathbf{u})$. Using the same argument, we can also show that the discrete global optimizer $\mathbf{u}^*$ must satisfy that $u_1^* = u_2^* = u_4^* = u_7^* = 1$, i.e., $t_1^* = t_2^* = t_4^* = t_7^* = 0$.

So the discrete multivariate constrained optimization problem is reduced to

$$\min_{\mathbf{t}_{\text{weak}} \in \mathcal{T}_{\text{weak}}^{(4)}} f^{(4)}(\mathbf{t}_{\text{weak}}) = \min_{\mathbf{t}_{\text{weak}} \in \mathcal{T}_{\text{weak}}^{(4)}} \bigl(2\delta^{-t_3} + 2\delta^{-t_5} + 2\delta^{-t_6} - 4\bigr), \tag{207}$$

where

$$\mathcal{T}_{\text{weak}}^{(4)} \triangleq \left\{ \mathbf{t}_{\text{weak}} \in (\mathbb{N} \cup \{0\})^{\mathsf{L}} \colon t_j \geq 0,\ j \in \{3,5,6\},\ \text{and} \sum_{j \in \{3,5,6\}} t_j = n \right\}. \tag{208}$$

This problem can be solved in an analogous way as for $\mathsf{M} = 3$. We obtain

$$\mathbf{t}^* = \mathbf{t}_{\text{weak}}^* = \bigl[t_3^*, t_5^*, t_6^*\bigr] = \left[ \left\lfloor \frac{n+2}{3} \right\rfloor, \left\lfloor \frac{n+1}{3} \right\rfloor, \left\lfloor \frac{n}{3} \right\rfloor \right]. \tag{209}$$

# C Proof of Theorem 58

The proof is based on the exact average ML error probability formula expressed as a function of the linear type vector $\mathbf{t}_{\text{lin}}$. Applying Lemma 23 and Theorem 49 for the general three-dimensional linear code (whose corresponding $r$-wise Hamming distances can be derived from Example 25), we obtain

$$\begin{aligned} f^{(8)}(\mathbf{t}_{\text{lin}}) &\triangleq \frac{8}{\delta^n} P_{\text{e}}\Bigl(\mathscr{C}_{\mathbf{t}_{\text{lin}}}^{(8,n)}\Bigr) &(210)\\ &= 4\bigl(u_1u_2u_3 + u_1u_4u_5 + u_1u_6u_7 + u_2u_4u_6 + u_2u_5u_7 + u_3u_4u_7 + u_3u_5u_6\bigr)\\ &\quad - 8\bigl(u_1 + u_2 + u_3 + u_4 + u_5 + u_6 + u_7\bigr)\\ &\quad + 2\bigl(u_1 + u_2 + u_3 + u_4 + u_5 + u_6 + u_7\bigr)\\ &\quad + \binom{8}{4} - 14 - \binom{8}{5} + \binom{8}{6} - \binom{8}{7} + \binom{8}{8} &(211)\\ &= 4\bigl(u_1u_2u_3 + u_1u_4u_5 + u_1u_6u_7 + u_2u_4u_6 + u_2u_5u_7 + u_3u_4u_7 + u_3u_5u_6\bigr)\\ &\quad - 6\bigl(u_1 + u_2 + u_3 + u_4 + u_5 + u_6 + u_7\bigr) + 21, &(212) \end{aligned}$$

where for convenience we set

$$u_\ell \triangleq \delta^{-t_{j_\ell}}, \quad 1 \leq \ell \leq \mathsf{K} = 7. \tag{213}$$

For a blocklength $n = 7\kappa$, we know that the type of the fair linear code is

$$t_{j_1}^* = t_{j_2}^* = \cdots = t_{j_7}^* = \kappa. \tag{214}$$

Plugging this into (212), we obtain that a fair linear code with blocklength $n$ being a multiple of 7 has

$$f^{(8)}(\mathbf{t}_{\text{lin}}^*) = 28\delta^{-3\kappa} - 42\delta^{-\kappa} + 21. \tag{215}$$

To show that this fair linear code is strictly suboptimal, we start to find a code of identical size and blocklength that has better performance. According to Example 57, such a code can be constructed from the fair weak flip code $\mathscr{C}_{\text{fair}}^{(8,n)}$ of blocklength $n \bmod \mathsf{L} = 0$ (for $\mathsf{M} = 8$ we have $\mathsf{L} = 35$). By Corollary 40, a fair weak flip code with blocklength $n = 35\tau$ for $\tau \in \mathbb{N}$ and corresponding type $\mathbf{t}_{\text{fair}}$

$$t_{j_1} = t_{j_2} = \cdots = t_{j_{35}} = \tau \tag{216}$$

satisfies

$$f^{(8)}(\mathbf{t}_{\text{fair}}) = \binom{8}{2}\delta^{-15\tau} - \binom{8}{3}\delta^{-5\tau} + \binom{8}{4}\delta^{-\tau} - \binom{8}{5} + \binom{8}{6} - \binom{8}{7} \tag{217}$$

$$= 28\delta^{-15\tau} - 56\delta^{-5\tau} + 70\delta^{-\tau} - 36. \tag{218}$$

Because no fair weak flip codes are defined for $n \neq 35\tau$, we propose a so-called *generalized fair weak flip code* for $n = 35\tau + 7\eta = 7\kappa$ with $\kappa = 5\tau + \eta \geq 2$, $\tau \in \mathbb{N} \cup \{0\}$, $0 < \eta \leq 4$, by carefully choosing $n$ columns from the fair weak flip code with blocklength $35(\tau + 1) > n$ to form a new $(8, n)$ nonlinear weak flip code that is a concatenation of different $(8, 7)$ Hadamard codes. As such, $7\eta$ components of the corresponding type vector $\mathbf{t}_{\text{weak}}^{\diamond}$ are equal to $\tau + 1$, and the remaining $(35 - 7\eta)$ components are equal to $\tau$ (so $n = 7\eta(\tau + 1) + (35 - 7\eta)\tau = 35\tau + 7\eta$). With this generalization and together with the fair weak flip codes for $n \bmod 35 = 0$ (i.e., $\eta = 0$), we succeed in showing that there exist nonlinear codes with a blocklength $n = 7\kappa$ ($\kappa \geq 2$) that have a better performance over the BEC than the corresponding fair linear codes.

Note that while there are many different $(8, 7)$ Hadamard codes, they are all equivalent, i.e., they are only row- and column-permutations of (33). For each of these $(8, 7)$ Hadamard code, all the pairwise and three-wise Hamming matches are equal to 3 and 1, respectively; and there are 14 four-wise Hamming matches equal to 1 and $\binom{8}{4} - 14 = 56$ four-wise Hamming matches equal to 0. So, when we concatenate $\kappa$ different $(8, 7)$ Hadamard codes in order to construct the $(8, 7\kappa)$ generalized fair weak flip code, we will automatically achieve that all pairwise Hamming matches equal to $3\kappa$ and that all three-wise Hamming matches equal to $\kappa$. For the four-wise Hamming matches, we select the Hadamard carefully to minimize the resulting four-wise Hamming matches. Indeed, we repetitively append the $(8, 7)$ Hadamard code $\eta$ times to the fair weak flip code with $n = 35\tau$ to create an $(8, n = 35\tau + 7\eta = 7\kappa)$ generalized fair weak flip code such that $14\eta$ four-wise Hamming matches equal to $\tau + 1$ and $70 - 14\eta$ four-wise Hamming matches equal to $\tau$.

Hence, we see that

$$f^{(8)}(\mathbf{t}_{\text{weak}}^{\diamond}) = 28\delta^{-3\kappa} - 56\delta^{-\kappa} + 14\eta\delta^{-(\tau+1)} + (70 - 14\eta)\delta^{-\tau} - 36. \tag{219}$$

The proof is completed if one can show that except for $\kappa = 1$ (i.e., $\tau = 0$ and $\eta = 1$),

$$f^{(8)}(\mathbf{t}_{\text{lin}}^{*}) - f^{(8)}(\mathbf{t}_{\text{weak}}^{\diamond}) = 14\Big[(\delta^{-\kappa} + 4) - \big(\eta\delta^{-(\tau+1)} + (5 - \eta)\delta^{-\tau}\big)\Big] > 0. \tag{220}$$

To that goal define $u \triangleq \delta^{-1} > 1$, and rewrite the terms in the bracket on the RHS of (220) as

$$p(u) \triangleq u^{5\tau+\eta} + 4 - \eta u^{\tau+1} - (5 - \eta)u^{\tau}. \tag{221}$$

Observe that $p(1) = 0$ and that for $\tau = 0$,

$$\frac{\partial p(u)}{\partial u} = \eta u^{\eta-1} - \eta > 0, \quad \text{if } \eta \neq 1, \tag{222}$$

(where the inequality holds because $u > 1$) and for $\tau \geq 1$,

$$\frac{\partial p(u)}{\partial u} = (5\tau + \eta)u^{5\tau+\eta-1} - \eta(\tau+1)u^{\tau} - (5\tau - \eta\tau)u^{\tau-1} \tag{223}$$
$$> (5\tau + \eta)u^{5\tau+\eta-1} - \eta(\tau+1)u^{\tau-1} - (5\tau - \eta\tau)u^{\tau-1} \tag{224}$$
$$= (5\tau + \eta)u^{5\tau+\eta-1} - (5\tau + \eta)u^{\tau-1} \tag{225}$$
$$\geq 0 \tag{226}$$

(where the inequalities again hold because $u > 1$). This implies that $p(u)$ is strictly larger than zero unless $\kappa = 1$.

# References

[1] Claude E. Shannon, "A mathematical theory of communication," *Bell System Technical Journal*, vol. 27, pp. 379–423 and 623–656, July and October 1948.

[2] Shu Lin and Daniel J. Costello, Jr., *Error Control Coding*, 2nd ed. Upper Saddle River, NJ, USA: Prentice-Hall, 2004.

[3] Chia-Lung Wu, Po-Ning Chen, Yunghsiang S. Han, and Yan-Xiu Zheng, "On the coding scheme for joint channel estimation and error correction over block fading channels," in *Proceedings IEEE International Symposium on Personal, Indoor and Mobile Radio Communications (PIMRC)*, Tokyo, Japan, September 13–16, 2009, pp. 1272–1276.

[4] Giuseppe Durisi, Tobias Koch, and Petar Popovski, "Towards massive, ultra-reliable, and low-latency wireless communication with short packets," March 2016, to appear in *Proceedings of the IEEE*. Available: http://arxiv.org/abs/1504.06526

[5] Vitaly Skachek, "Batch and PIR codes and their connections to locally repairable codes," *arXiv*, November 2016. Available: https://arxiv.org/abs/1611.09914

[6] George M. Church, Yuan Gao, and Sriram Kosuri, "Next-generation digital information storage in DNA," *Science*, vol. 337, no. 6102, p. 1628, 2012.

[7] Nick Goldman, Paul Bertone, Siyuan Chen, Christophe Dessimoz, Emily M. LeProust, Botond Sipos, and Ewan Birney, "Towards practical, high-capacity, low-maintenance information storage in synthesized DNA," *Nature*, vol. 494, pp. 77–80, February 2013.

[8] Robert N. Grass, Reinhard Heckel, Michela Puddu, Daniela Paunescu, and Wendelin J. Stark, "Robust chemical preservation of digital information on DNA in silica with error-correcting codes," *Angewandte Chemie International Edition*, vol. 54, no. 8, pp. 2552–2555, 2015.

[9] S. M. Hossein Tabatabaei Yazdi, Han Mao Kiah, Eva Garcia-Ruiz, Jian Ma, Huimin Zhao, and Olgica Milenkovic, "DNA-based storage: Trends and methods," *IEEE Transactions on Molecular, Biological, and Multi-Scale Communications*, vol. 1, no. 3, pp. 230–248, September 2015.

[10] Arash Einolghozati and Faramarz Fekri, "Analysis of error-detection schemes in diffusion-based molecular communication," *IEEE Journal on Selected Areas in Communications*, vol. 34, no. 3, pp. 615–624, March 2016.

[11] A. Robert Calderbank, Eric M. Rains, P. W. Shor, and Neil J. A. Sloane, "Quantum error correction via codes over GF(4)," *IEEE Transactions on Information Theory*, vol. 44, no. 4, pp. 1369–1387, July 1998.

[12] F. Jessy MacWilliams and Neil J. A. Sloane, *The Theory of Error-Correcting Codes.* Amsterdam, The Netherlands: North-Holland, 1977.

[13] Victor K. Wei, "Generalized Hamming weights for linear codes," *IEEE Transactions on Information Theory*, vol. 37, no. 5, pp. 1412–1418, September 1991.

[14] Tor Helleseth, Torleiv Kløve, and Øyvind Ytrehus, "Generalized Hamming weights of linear codes," *IEEE Transactions on Information Theory*, vol. 38, no. 3, pp. 1133–1140, May 1992.

[15] Torleiv Kløve, "Minimum support weights of binary codes," *IEEE Transactions on Information Theory*, vol. 39, no. 2, pp. 648–654, March 1993.

[16] Tor Helleseth, Torleiv Kløve, Vladimir I. Levenshtein, and Øyvind Ytrehus, "Bounds on the minimum support weights," *IEEE Transactions on Information Theory*, vol. 41, no. 2, pp. 432–440, March 1995.

[17] Po-Ning Chen, Hsuan-Yin Lin, and Stefan M. Moser, "Nonlinear codes outperform the best linear codes on the binary erasure channel," in *Proceedings IEEE International Symposium on Information Theory (ISIT)*, Hong Kong, China, June 14–19, 2015, pp. 1751–1755.

[18] Po-Ning Chen, Hsuan-Yin Lin, and Stefan M. Moser, "Weak flip codes and applications to optimal code design on the binary erasure channel," in *Proceedings Fiftieth Allerton Conference on Communication, Control and Computing*, Allerton House, Monticello, IL, USA, October 1–5, 2012, pp. 160–167.

[19] Po-Ning Chen, Hsuan-Yin Lin, and Stefan M. Moser, "Optimal ultrasmall block-codes for binary discrete memoryless channels," *IEEE Transactions on Information Theory*, vol. 59, no. 11, pp. 7346–7378, November 2013.

[20] Robert G. Gallager, *Information Theory and Reliable Communication.* New York, NY, USA: John Wiley & Sons, 1968.

[21] Claude E. Shannon, Robert G. Gallager, and Elwyn R. Berlekamp, "Lower bounds to error probability for coding on discrete memoryless channels," *Information and Control*, pp. 65–103, February 1967, part I.

[22] A. B. Fontaine and W. W. Peterson, "Group code equivalence and optimum codes," *IRE Transactions on Information Theory*, vol. 5, no. 5, pp. 60–70, May 1959.

[23] Yury Polyanskiy, H. Vincent Poor, and Sergio Verdú, "Channel coding rate in the finite blocklength regime," *IEEE Transactions on Information Theory*, vol. 56, no. 5, pp. 2307–2359, May 2010.

[24] Yury Polyanskiy, “Saddle point in the minimax converse for channel coding,” *IEEE Transactions on Information Theory*, vol. 59, no. 5, pp. 2576–2595, May 2013.

[25] Po-Ning Chen, Hsuan-Yin Lin, and Stefan M. Moser, “Equidistant codes meeting the Plotkin bound are not optimal on the binary symmetric channel,” in *Proceedings IEEE International Symposium on Information Theory (ISIT)*, Istanbul, Turkey, July 7–13, 2013, pp. 3015–3019.

[26] Hsuan-Yin Lin, “Proof of minimum $r$-wise distance for linear codes,” February 2016, personal notes.

[27] Christine Bachoc and Gilles Zémor, “Bounds for binary codes relative to pseudo-distances of $k$ points,” *Advances in Mathematics of Communications*, vol. 4, no. 4, pp. 547–565, 2010.

[28] Douglas R. Stinson, *Combinatorial Designs: Constructions and Analysis.* Springer Verlag, 2003, ISBN: 0–387–95487–2.

[29] Richard A. Brualdi, *Introductory Combinatorics*, 5th ed. Upper Saddle River, NJ, USA: Prentice-Hall, 2010.

[30] Abbas A. El Gamal, Lane A. Hemachandra, Itzhak Shperling, and Victor K. Wei, “Using simulated annealing to design good codes,” *IEEE Transactions on Information Theory*, vol. 33, no. 1, pp. 116–123, January 1987.